\documentclass[iop]{emulateapj}

\def\Zsol{\hbox{Z$_{\odot}$}}

\usepackage{epsfig,natbib}

\usepackage{soul}

\usepackage{pdflscape}

\usepackage{rotating}

\newcommand{\kms}{km\,s$^{-1}$}
\newcommand{\hi}{H\,{\sc i}}
\newcommand{\hii}{H~{\sc ii}}

\newcommand{\foi}{[O~{\sc i}]}

\newcommand{\pii}{P~{\sc ii}}
\newcommand{\piii}{P~{\sc iii}}
\newcommand{\sii}{S~{\sc ii}}
\newcommand{\siii}{S~{\sc iii}}
\newcommand{\sIii}{Si~{\sc ii}}
\newcommand{\sIiii}{Si~{\sc iii}}
\newcommand{\sIiv}{Si~{\sc iv}}
\newcommand{\feii}{Fe~{\sc ii}}
\newcommand{\feiii}{Fe~{\sc iii}}
\newcommand{\Ni}{N~{\sc i}}
\newcommand{\Nii}{N~{\sc ii}}

\newcommand{\Nv}{N~{\sc v}}
\newcommand{\nIii}{Ni~{\sc ii}}
\newcommand{\nIiii}{Ni~{\sc iii}}
\newcommand{\ci}{C~{\sc i}}
\newcommand{\cii}{C~{\sc ii}}

\newcommand{\oi}{O~{\sc i}}
\newcommand{\oii}{O~{\sc ii}}
\newcommand{\oiv}{O~{\sc iv}}

\newcommand{\mgii}{Mg~{\sc ii}}

\newcommand{\mnii}{Mn~{\sc ii}}

\newcommand{\ciii}{C~{\sc iii}}
\newcommand{\civ}{C~{\sc iv}}

\newcommand{\Lya}{Ly$\alpha$}

\shorttitle{Investigating Nearby SFGs with HST/COS -- Paper I}
\shortauthors{James et al.}

\begin{document}


\title{Investigating Nearby Star-Forming Galaxies in the Ultraviolet with HST/COS Spectroscopy. I: Spectral Analysis and Interstellar Abundance Determinations}


\author{{B.~James\altaffilmark{1,2}}}\email{bjames@ast.cam.ac.uk}
\altaffiltext{1}{Space Telescope Science Institute, Baltimore, MD 21218, USA}
\altaffiltext{2}{Institute of Astronomy, University of Cambridge, Madingely Road, Cambridge, CB3 0HA, UK}
\author{A.~Aloisi\altaffilmark{1}}
\author{T.~Heckman\altaffilmark{3}}
\altaffiltext{3}{The Johns Hopkins University, Baltimore, MD 21218, USA }


\author{{S.~T.~Sohn}\altaffilmark{1}}


\author{M.~Wolfe\altaffilmark{1}}




\begin{abstract}

This is the first in a series of three papers describing a project with the Cosmic Origins Spectrograph on the Hubble Space Telescope to measure abundances of the neutral interstellar medium (ISM) in a sample of 9 nearby star-forming galaxies. The goal is to assess the (in)homogeneities of the multiphase ISM in galaxies where the bulk of metals can be hidden in the neutral phase, yet the metallicity is inferred from the ionized gas in the H\,{\sc ii} regions. The sample, spanning a wide range in physical properties, is to date the best suited to investigate the metallicity behavior of the neutral gas at redshift z\,=\,0. ISM absorption lines were detected against the far-ultraviolet spectra of the brightest star-forming region(s) within each galaxy. Here we report on the observations, data reduction, and analysis of these spectra. Column densities were measured by a multi-component line-profile fitting technique, and neutral-gas abundances were obtained for a wide range of elements. Several caveats were considered including line saturation, ionization corrections, and dust depletion. Ionization effects were quantified with `ad-hoc' CLOUDY models reproducing the complex photoionization structure of the ionized and neutral gas surrounding the UV-bright sources. An  `average spectrum of a redshift z\,=\,0 star-forming galaxy' was obtained from the average column densities of unsaturated profiles of neutral-gas species. This template can be used as a powerful tool for studies of the neutral ISM at both low and high redshift.

\end{abstract}

\keywords{galaxies: individual (I~Zw~18, M83, NGC~3690, NGC~4214, NGC~4449, NGC~4670,  NGC~5253, SBS~0335-052, SBS~1415+437) --- galaxies: ISM --- galaxies: starburst ---  ISM: abundances --- ultraviolet: ISM}

\section{Introduction}
This is the first in a series of three papers in which we present the study of neutral-gas abundances in a sample of nearby star-forming galaxies (SFGs), as obtained from absorption-line spectroscopy in the ultraviolet (UV) with the Hubble Space Telescope (HST). The present paper discusses the spectroscopic observations with the HST Cosmic Origins Spectrograph (COS), the spectral data reduction, the measurements of the column densities from line-profile fitting of absorption lines arising from ions of several species, and the interstellar abundance determinations. The second paper \citep[][hereafter Paper II]{Aloisi:2014a} derives the metallicity offset between neutral and ionized gas (as inferred from the \hii~regions) within each galaxy of the sample and discusses the implications for the chemical (in)homogeneity of the multiphase ISM in these SFGs of the local Universe. Finally, the third paper \citep[][hereafter Paper III]{Aloisi:2014b} compares the interstellar abundances in the galaxies of our sample with those observed in the local and high-redshift Universe and discusses the implications of our findings within a cosmological context. 

Abundances in the interstellar medium (ISM) of star-forming galaxies (SFGs) are typically determined using optical and NIR emission-line spectroscopy of the \hii\, regions around the star-formation sites. Oxygen is the element that is probed directly and most reliably \citep[e.g., ][]{Pagel:1997}, and galaxy metallicities are usually quoted in terms of 12+log(O/H) \citep[8.69 corresponds to a solar abundance;][]{Asplund:2009}. However, \hii\, regions are associated with recent star-formation and may suffer from additional enrichment compared to the ISM \citep[see, e.g.][]{Kunth:1986}. In addition to this, SFGs are characterized by the presence of a large reservoir of \hi, since this is a fundamental component for the star-formation onset and this reservoir may amount to as much as 90-95\% of the baryonic matter in the so-called blue compact dwarf (BCD) galaxies \citep[e.g.,][]{Kniazev:2000,Begum:2008}. As the dominant component of the baryonic mass, the neutral gas can hide the bulk of the metals even if much less enriched than the \hii\, regions. It is therefore important to directly study abundances in the neutral ISM in order to infer a more complete picture of the metal content within SFGs.

The metal content in the neutral ISM of a galaxy can be assessed via (far) ultraviolet (UV) absorption lines arising from the most common heavy elements. This technique requires bright UV sources to use as background spectra that get absorbed along the sightline, similarly to the UV/optical studies of absorbing systems along the sightline to quasars \citep[e.g.,][]{Lu:1996}. Such bright UV sources tend to be present in galaxies with strong star-formation (SF), such as starbursts, BCD galaxies, and merging/interacting systems. While the neutral ISM within a SFG could be probed in principle by using a background quasar, quasars that are bright enough in the UV for this kind of study are quite rare. Furthermore, the sightlines sampled by quasars are different from the ones sampled by star-forming regions, the former being usually much farther away into the halo of a galaxy than the latter, and this can bias the quasar results towards lower abundances due to the well-known metallicity gradients within galaxies \citep[e.g.,][]{Kobulnicky:1998}.

\citet{Aloisi:2003} pioneered a methodology to study neutral ISM abundances in nearby SFGs using FUSE \citep[Far Ultraviolet Spectroscopic Explorer;][]{Moos:2000} in the FUV spectral range $\sim$ 900--1200~\AA, and applied this technique with success to the most metal-poor galaxy known in the local Universe, the BCD I~Zw~18 (12+log(O/H)=7.2, Z$\sim$1/30~\Zsol). The analysis of the FUSE spectra of I~Zw~18 yielded very interesting results, indicating that the abundances of the alpha elements O, Ar, and Si, as well as N are $\sim$0.5 dex lower in the neutral ISM than in the \hii\, regions, while Fe is instead the same \citep[for a different result on O from the same FUSE data see also][]{Lecavelier:2004}.

The result of an abundance offset between the nebular gas and the neutral gas through which FUSE sightlines pass, has been confirmed by several other studies of nearby SFGs published in the refereed literature \citep[][and references therein]{Lebouteiller:2009}. A compilation of N, O, Ar, and Fe abundance measurements of the nebular and neutral ISM suggest that there is a systematic underabundance by a factor $\sim$10 of the neutral gas compared to the ionized gas in the \hii\, regions \citep[but see also the HST/STIS study of SBS 1543+593 by][] {Bowen:2005}. A number of theories have been put forward to explain this observed offset. \citet{Lebouteiller:2009} hypothesise that this offset can be achieved and maintained if most of the metals released by star-formation episodes mix with the much larger amount of \hi\, gas, increasing only slightly its metallicity, while even a fraction as small as $\sim$ 1\%  that mixes locally can greatly enrich the ionized gas. \citet{Heckman:2001} suggested that abundances measured from neutral-species absorptions from small dense clumps within a superbubble interior, with the same metallicity as the ionized gas, may be getting diluted by more metal-poor gas lying further away.  Alternatively, \citet{Cannon:2005} suggested that the absorptions arise in neutral, metal-poor \hi\, halos, whilst metal-rich ionized gas lies close to regions of star-formation.  However, since the offset seems to be independent of \textbf{the current ISM chemical abundances,} this would only work under the assumption that star-formation remains localized over the galaxy's history.
 
There are, however, several issues associated with these FUSE studies. Firstly, the FUSE large aperture ($30''\times30''$) implies that abundances can be ill-defined averages over regions that may have quite different properties. Secondly, the neutral-gas abundance determination of N, O, Ar, and Fe each come with its own caveat:  O may be affected by hidden saturation, Ar and N by ionization corrections, and Fe by dust depletion \citep[see, e.g., ][]{Aloisi:2003}. And finally, the sample of SFGs investigated so far is still small and heavily biased toward very metal-poor BCDs which may have a peculiar SF and evolutionary history. As a result, very little is known about the behavior of the neutral gas metallicity at zero redshift as a function of the galaxy properties \citep[see, e.g., ][]{Lilly:2003}.  It is these three factors that form the motivation behind the COS study presented here (Paper I) and in Papers II and III.

COS onboard HST is a relatively new UV spectrograph designed to perform high-sensitivity spectroscopy between 1150--3200~\AA. With an aperture of only 2.5$''$ in diameter, it enables us to sample single \hii\, regions or young clusters within our sample of SFGs. As summarized in Table~\ref{tab:sample}, this sample spans a wide range in galaxy type (dwarf, spiral, interacting/merger system), metallicity (as inferred from O in the \hii\, regions), and SFR (indicated by log(L$_{UV}$ +L$_{FIR}$)). It allows us to increase and diversify the sample of nearby studied objects and investigate the metallicity behavior of the neutral ISM as a function of the galaxy properties at redshift $z = 0$.

As the first publication of our study, this paper presents the data calibration and analysis of COS spectra from HST Cycle 17 program 11579 (PI A.~Aloisi), and is organized as follows. In \S~\ref{sec:observations},  we describe the galaxy sample, the COS target selection from ACS pre-imaging, and the various stages of the calibration of the COS spectra, including removal of geocoronal contamination.  In \S~\ref{sec:analysis} we describe the analysis of the COS spectra, including continuum-fitting and line-fitting techniques.  The measurements of \hi\, and heavy-element column densities are presented in \S~\ref{sec:hi_colDens} and \ref{sec:abundances}, respectively, along with a detailed discussion of line saturation and ionization correction issues.  The results and spectra for each individual galaxy are described in \S~\ref{sec:results} and \S~\ref{sec:objects}, respectively.  Finally, in \S~\ref{sec:avg_spec} we present the so-called `average' star-forming galaxy spectrum, a synthetic spectrum obtained by considering the 'average' column density of each ion as obtained from the single measurements on the individual spectra of the SFG sample. Conclusions are given in \S~\ref{sec:discussion}.

\section{Observations}\label{sec:observations}
\subsection{Sample Selection}
\label{sec:targets}

The galaxies of our sample were selected on the basis of the following criteria. The most important one was the existence of archival FUSE data \citep[e.g., I~Zw~18 whose FUSE spectra have already been analyzed by][]{Aloisi:2003}.  Selecting galaxies with FUSE spectra provided three advantages: (i.) it ensured that the sample galaxies are strong emitters in the FUV, (ii.) it guaranteed the detection of metal absorption lines, and (iii.) the FUSE data provided a direct estimate of the flux in the FUV spectral region used for the COS observations. While our sample of 9 nearby SFGs is by no means a rigorously defined, statistically complete one, it does span a broad range in parameter space. More specifically, the sample covers a range of galaxy spectral types, metallicities as inferred from O in the \hii\, regions, and SFRs  as indicated by log(L$_{UV}+$L$_{FIR}$).  

The sample ordered by increasing metallicity and its basic properties are listed in Table \ref{tab:sample}. The galaxy right ascension and declination, morphological type, systemic velocity, and foreground Galactic extinction E(B-V) were all taken from the NASA Extragalactic Database (NED). The metallicities (Z) are given in terms of the oxygen abundance by number \citep[ Z = 12 + log (O/H), where $Z_{\odot} = 8.69$, ][]{Asplund:2009}. Further information about these metallicities (including references) are given in the table and its associated notes. The majority of distances were compiled from the HyperLeda database\footnote{http://leda.univ-lyon1.fr} \citep{Paturel:2003} and represent those calculated from the tip of the red giant branch (TRGB) obtained from resolved stellar population studies with \textit{HST}. For galaxies where this method cannot be applied, distances adopted were calculated using the standard linear Virgo-centric infall model of \citet{Mould:2000}, as listed in NED.

The far infrared (FIR) luminosity (L$_{FIR}$) was calculated from the ``FIR " parameter (the flux between 60 and 100 $\micron$m) published in the IRAS survey (Fullmer \& Lonsdale 1989).  The UV luminosity (L$_{UV}$) was calculated from the flux at 1900~\AA~corrected for reddening using the E(B-V) values listed in the table. See \citet{Heckman:1998} for more details on how these luminosities were calculated.

\subsection{ACS/SBC Pre-Imaging \& Target Selection}
\label{sec:ACSimaging}

\begin{figure*}
\center
\includegraphics[scale=0.6]{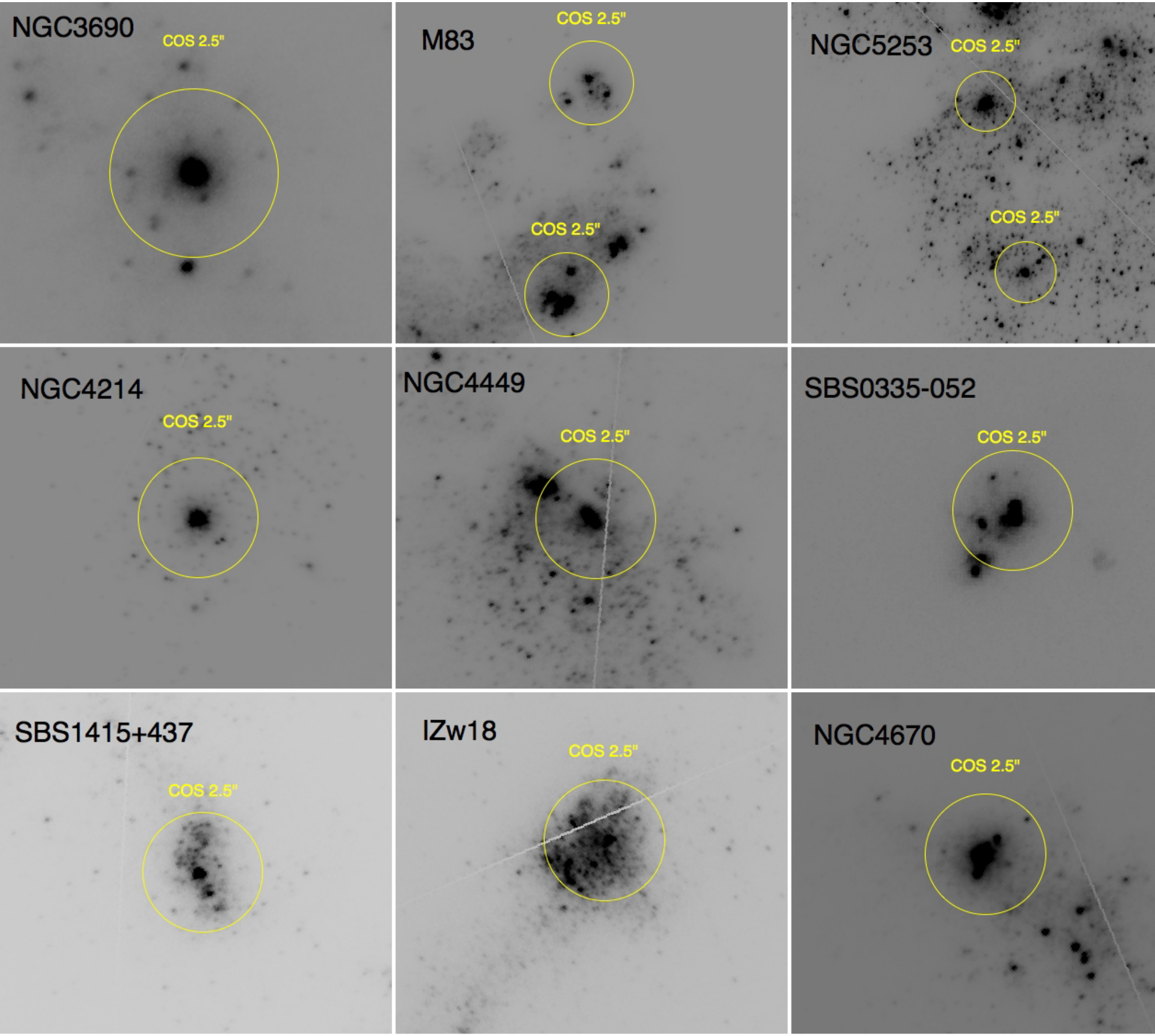}
\vspace{0.2cm}
\caption{Tiled ACS/SBC images of the galaxies detailed in Table~\ref{tab:sample}, with the COS 2.5$''$ PSA aperture overlaid on the selected UV bright ``point-like" sources (as detailed in Section~\ref{sec:ACSimaging}).} 
\label{fig:COS_targets}
\end{figure*}

As the primary motivation for this work was to study the abundances in the neutral ISM of nearby star-forming galaxies by using absorption lines arising from the COS spectra of UV background sources, it was necessary to select within each galaxy the optimal UV ``point-like'' sources to target with COS. In order to do this, we performed ACS/SBC FUV imaging of each galaxy in the F125LP filter (Fig.~\ref{fig:COS_targets}). The only exception was SBS~0335-052 for which an ACS/SBC F140LP archival image already existed and was used for this purpose. With $\lambda_{eff}=1440$~\AA, the F125LP filter has a bandwidth that covers as much as possible of the wavelength range of the COS G130M grating ($\sim$1150--1450~\AA), and at the same time avoids contamination from Ly$\alpha$. Table~\ref{tab:obs_ACS} lists the details of the ACS/SBC observations for all galaxies within our sample.  Reduction of the images was performed via the standard ACS reduction pipeline \textsc{CALACS} v5.0.5, which is part of the STSDAS (Space Telescope Science Data Analysis System) software package within IRAF (Image Reduction and Analysis Facility).    

Along with acting as $30''\times30''$ early-acquisition images for the spectroscopy, these images were crucial in guiding the selection of the COS targets.  The point sources were selected based on the following criteria. Firstly, in order to optimize the S/N within the allocated amount of time, we needed to select the brightest point sources within the ACS/SBC images. Secondly, since the 2.5$''$ COS aperture is larger than the extension of a point source, in order to avoid resolution degradation we had to consider the spatial extent of these point source. Based on the spectral resolution of FUSE observations of very similar targets \citep[e.g. I~Zw~18,][]{Aloisi:2003}, which was sufficient in performing the ISM analysis we present here, we found that a value as high as 60 km/s would work.  Combining this with the line-spread function of the COS spectrograph (which is linearly related to the spatial extent of the source within the aperture), this lower limit implied that we could observe sources that are $\sim$4 times wider than a point source, i.e. with a maximum diameter of $\sim$0.5~arcsec compared to the angular extent of 0.13~arcsec in the dispersion direction for a point source\footnote{As detailed in the COS instrument handbook: http://www.stsci.edu/hst/cos/documents/handbooks/current/ cos\_cover.html}. Thirdly, to avoid selecting red targets instead of UV targets as a result of the ACS/SBC `red leak', the optical color of each point source was derived using multi-band images of the sample galaxies in the $HST$ archive. The final consideration was to only select sources that were in relatively uncrowded regions, to allow the search procedure in the COS target acquisition to correctly select the desired point source.  Thus the selected target sources were as compact as possible, no larger than $\sim$0.5~arcsec in diameter, bright in FUV (i.e. $\leq$16~STMAG) and in uncrowded regions.  The coordinates of the UV point-like sources observed within each galaxy are listed in Table~\ref{tab:obs_COS}.

For two galaxies within the sample (NGC~5253 and M83), we obtained observations along two separate sightlines.  This was done with an aim to test for spatial variations of the ISM metal abundances within the galaxy.

\subsection{COS Observations \& Data Reduction}\label{COSdata_reduction}
We obtained \textit{HST}/COS FUV observations of 11 point-like sources in 9 nearby star-forming galaxies using the G130M medium-resolution grating with the $\lambda1291$ central-wavelength setting.  Each observation utilized the four FP-POS positions (small dithers in the dispersion direction) to get the best data quality achievable without a flat-field correction. This corresponded to a final wavelength range of $\sim$1130--1420~\AA.  Full details of each COS observation are given in Table~\ref{tab:obs_COS}.  For each COS observation, the telescope was initially pointed at the coordinates listed in Table~\ref{tab:obs_COS} and NUV imaging target acquisition (TA) was performed to proper center each target within the COS aperture.  A TA configuration of PSA and Mirror A was used in most cases, except for the brightest targets where Mirror B was instead adopted.
 
After retrieval from the archive, all data were reduced locally using the COS data reduction package {\sc CALCOS v.2.14.4}.  This calibration pipeline consists of three main components that calibrate COS data by (1) correcting for instrumental effects such as thermal drifts, geometric distortion corrections, Doppler corrections, and pixel-to-pixel variations in sensitivity (this latter step is not performed for FUV data due to a lack of a good 2D flat field), (2) generating an exposure-specific wavelength-calibrated scale, and (3) extracting and producing a final (one-dimensional) flux-calibrated (summed) spectrum for the entire observation \citep{COSDHB}.

In consideration of the absence of a 2D flat-field correction for COS FUV observations within the {\sc CALCOS} pipeline, it was necessary to investigate whether the quality of the data would be improved by applying something similar to such a correction.  To do this, we derived two sets of 1D flat-field templates using an iterative procedure which selects regular features in pixel space (e.g., the shadows of the ion-repeller grid wires of the COS FUV detector) within the four FP-POS positions of a spectroscopic observation\footnote{As detailed in http://www.stsci.edu/hst/cos/documents/ isrs/ISR2011\_03.pdf}. We used both our spectra and high S/N data of white dwarf stars for this purpose.  In both cases, due to the moderately high S/N of our data ($S/N \sim$\,10-30 per 6-pixel resolution element in the final combined spectra; see Table~\ref{tab:obs_COS}), the imperfect 1D flat-field correction introduced noise rather than removing any spurious detector features.  We therefore decided that in the case of our observations taken with all four FP-POS positions available for the setting selected, removing the grid wires by flagging them through the bad-pixel table used by {\sc CALCOS} during the data reduction was sufficient in removing the major flat-field features.

Each spectrum was then inspected in pixel space for any erroneous features due to effects of gain sag around pixels 7200 and 9100, (the position on the FUV detector where the geocoronal \Lya\, emission line falls when the COS spectrograph is used in two of its most popular observing modes).  If these features were present, the data were reduced with pulse height amplitude settings of 2-31 (instead of the default 4-31) to recover any flux lost due to this gain-sag effect.  The data were finally binned by a factor of 6 in order to improve the identification of spectral features without degrading the spectral resolution.

\begin{figure*}
\center
\includegraphics[angle=90,scale=0.75]{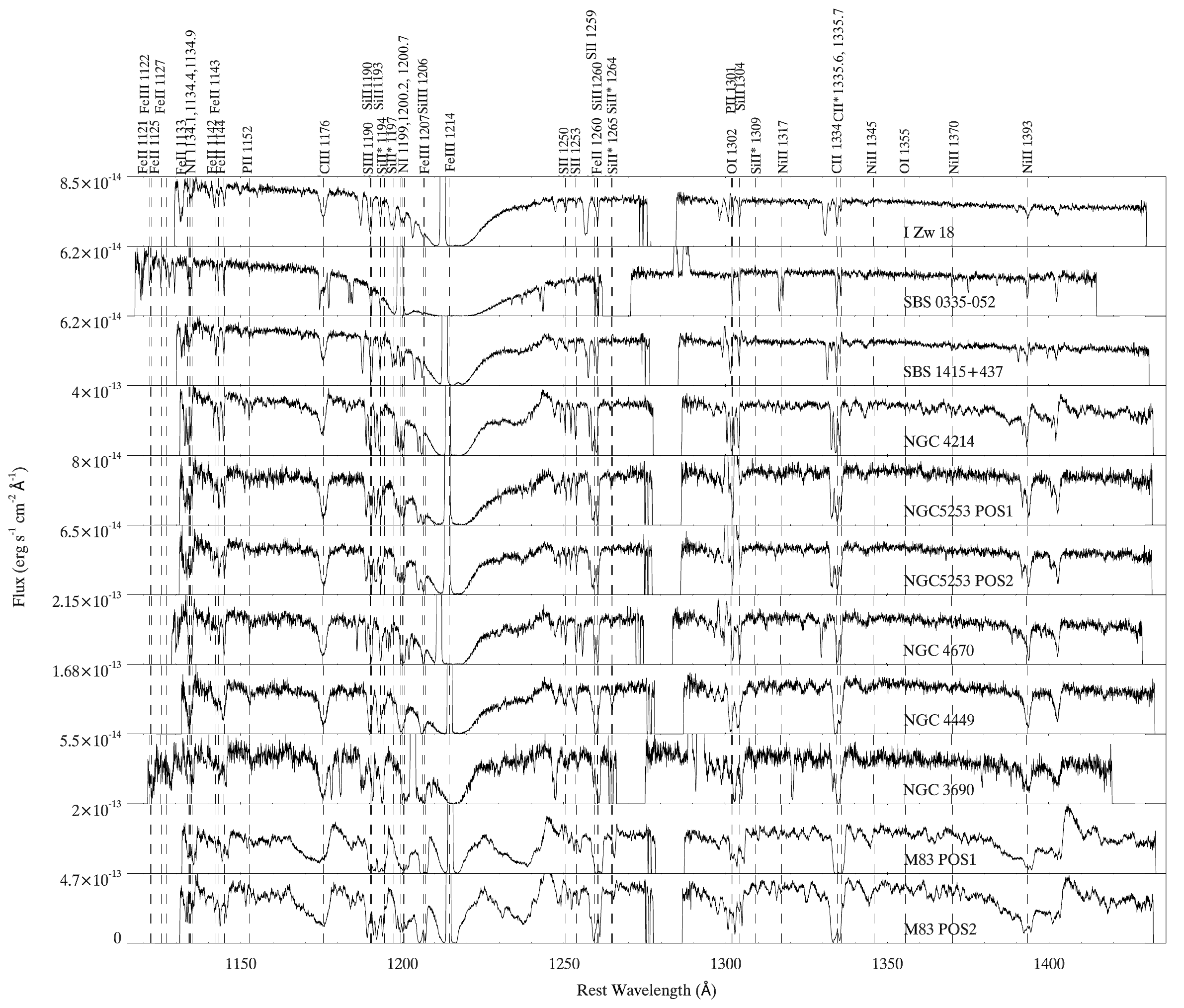}
\vspace{0.2cm}
\caption{HST/COS G130M spectra of all UV sources in the sample of nearby SFGs listed in Table~\ref{tab:sample}, each shown in their respective rest-frame wavelengths. The spectra are ordered by increasing metallicity from top to bottom. All absorption lines intrinsic to each galaxy analyzed within this study are labelled in the top panel. Notice that \ciii~$\lambda$1176 is the only stellar absorption line considered in this study from the UV background source within the galaxy. The radial velocity inferred from this line has been used to obtain the rest-frame wavelengths for each observed spectrum (see Table~\ref{tab:HIdensities}).}  
\label{fig:all_spec}
\end{figure*}

\begin{figure*}
\center
\includegraphics[angle=90,scale=0.75]{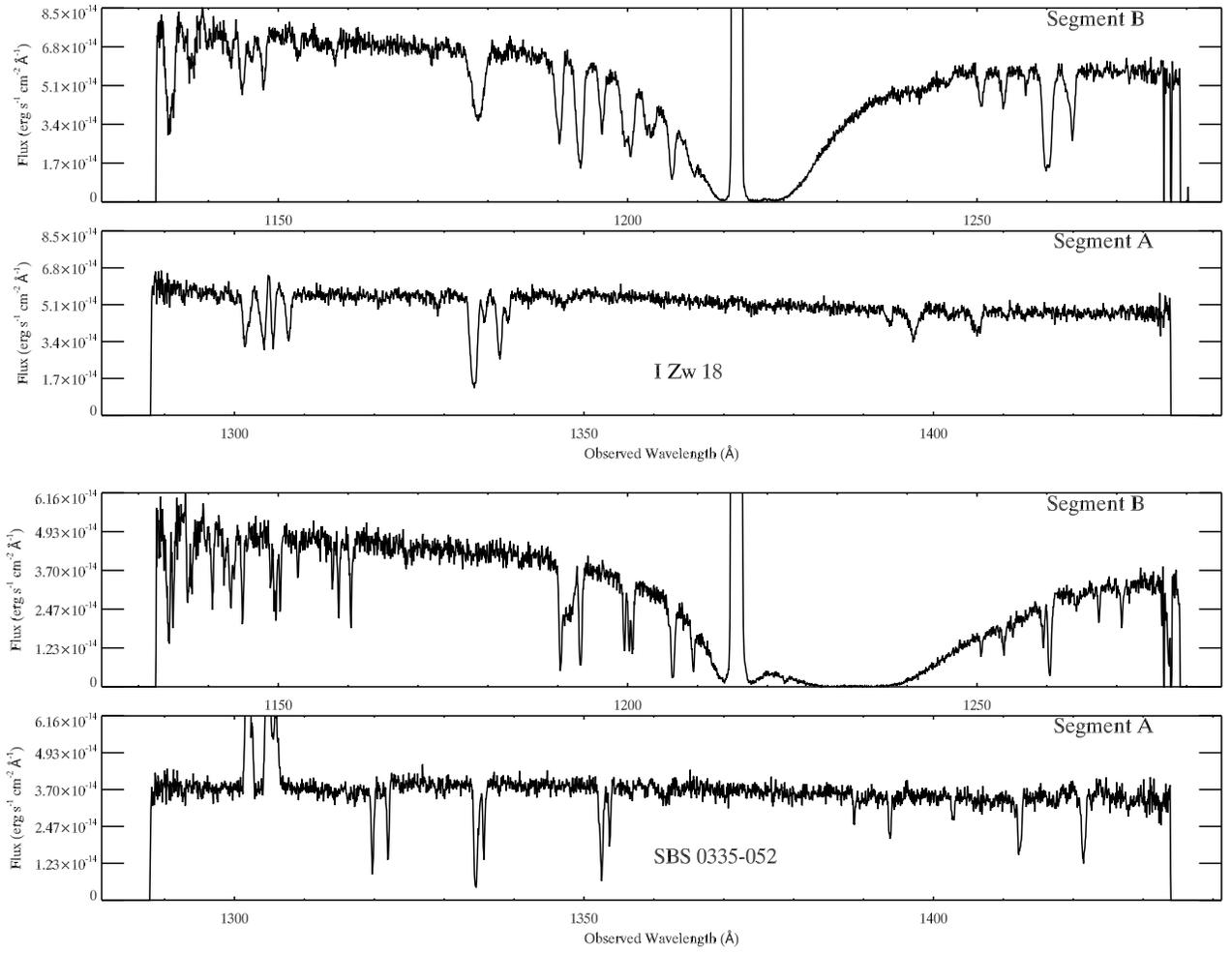}
\vspace{0.2cm}
\caption{HST/COS G130M spectra of I~Zw~18 (top two panels) and SBS~0335-052 (bottom two panels).} 
\label{fig:izw18_sbs0335_spec}
\end{figure*}

\begin{figure*}
\center
\includegraphics[angle=90,scale=0.75]{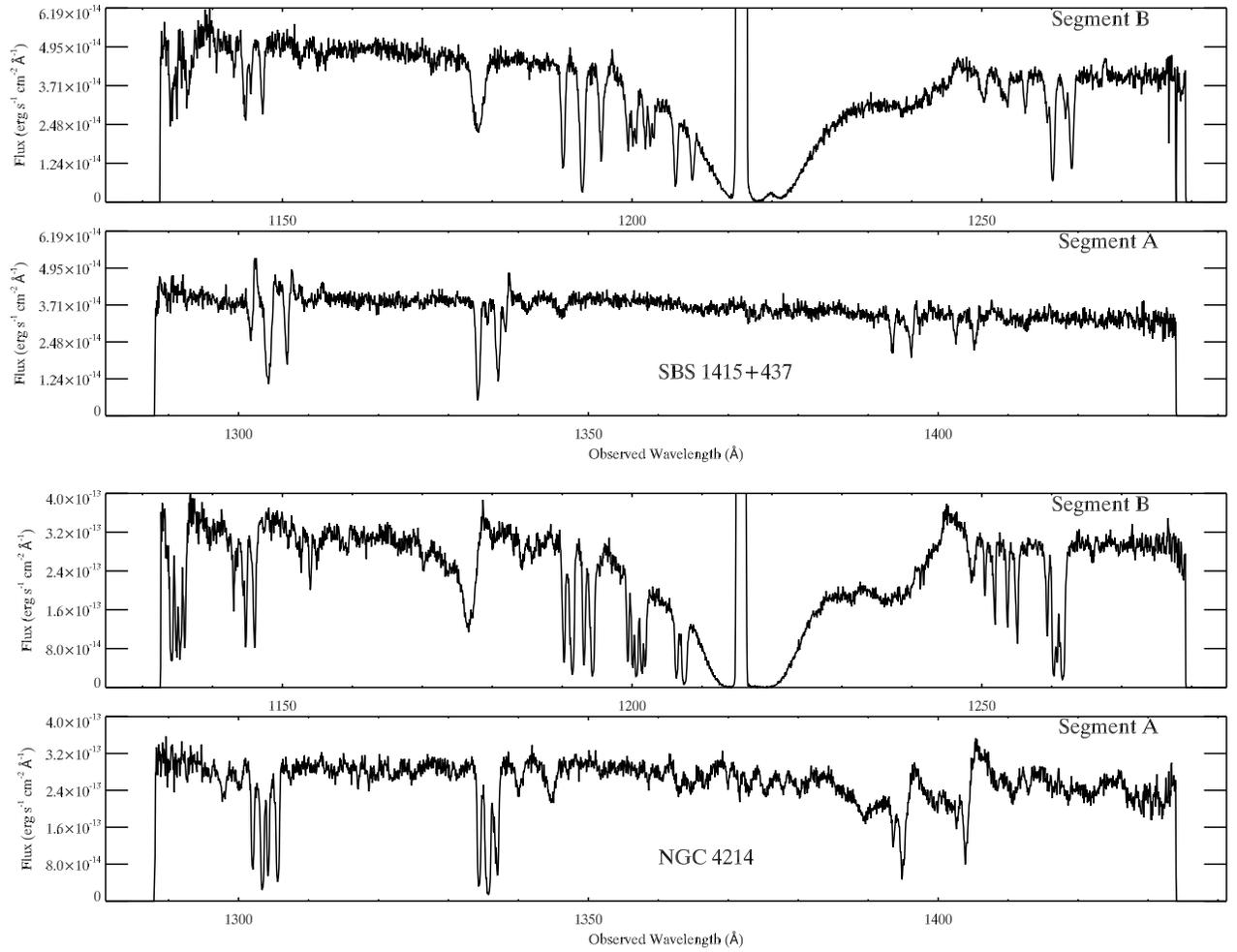}
\vspace{0.2cm}
\caption{HST/COS G130M spectra of SBS~1415+437 (top two panels) and NGC~4214 (bottom two panels).} 
\label{fig:sbs1415_ngc4214_spec}
\end{figure*}

\begin{figure*}
\center
\includegraphics[angle=90,scale=0.75]{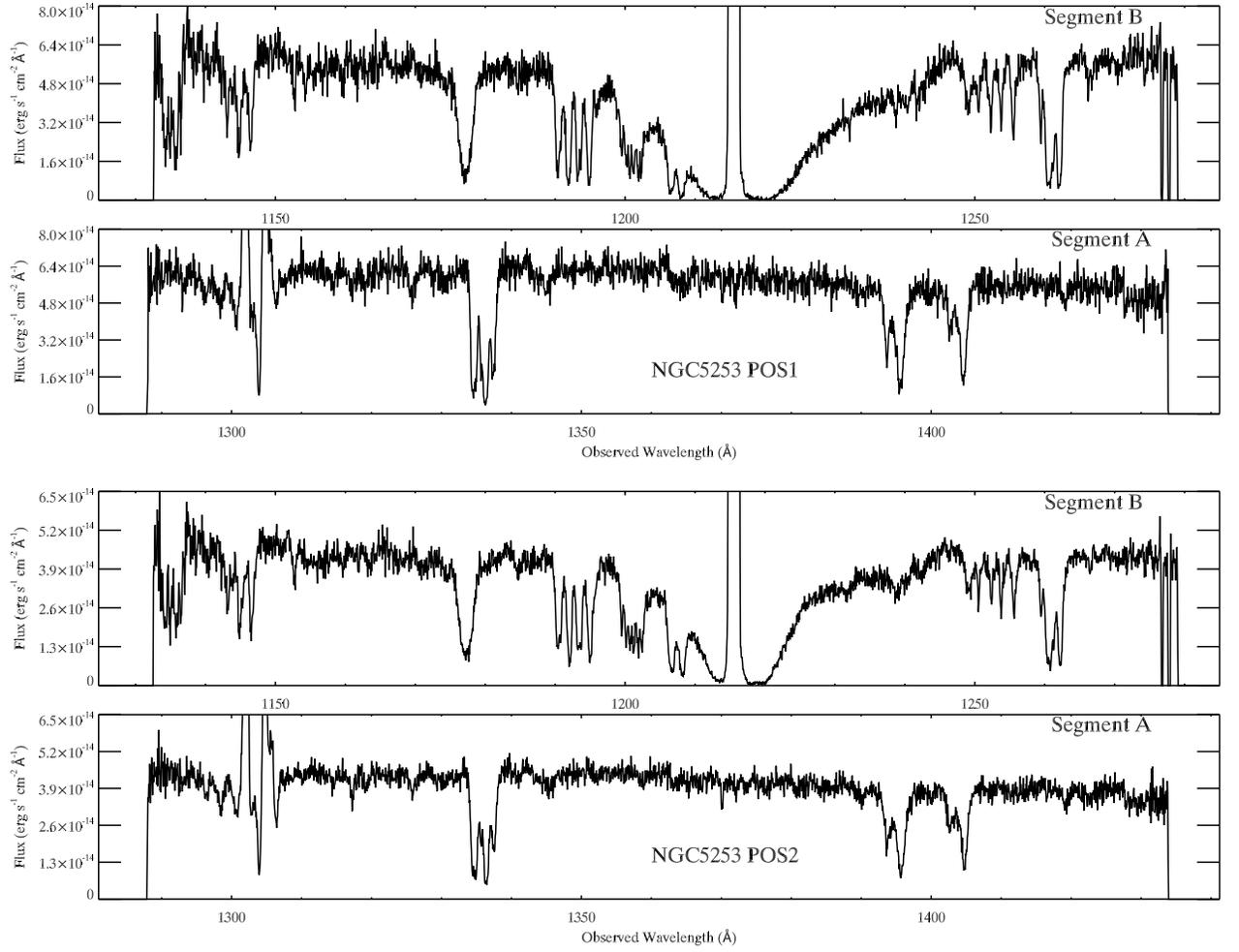}
\vspace{0.2cm}
\caption{HST/COS G130M spectra of NGC~5253-Pos$^n$~1 (top two panels) and NGC~5253-Pos$^n$~2 (bottom two panels).} 
\label{fig:ngc5253_spec}
\end{figure*}

\begin{figure*}
\center
\includegraphics[angle=90,scale=0.75]{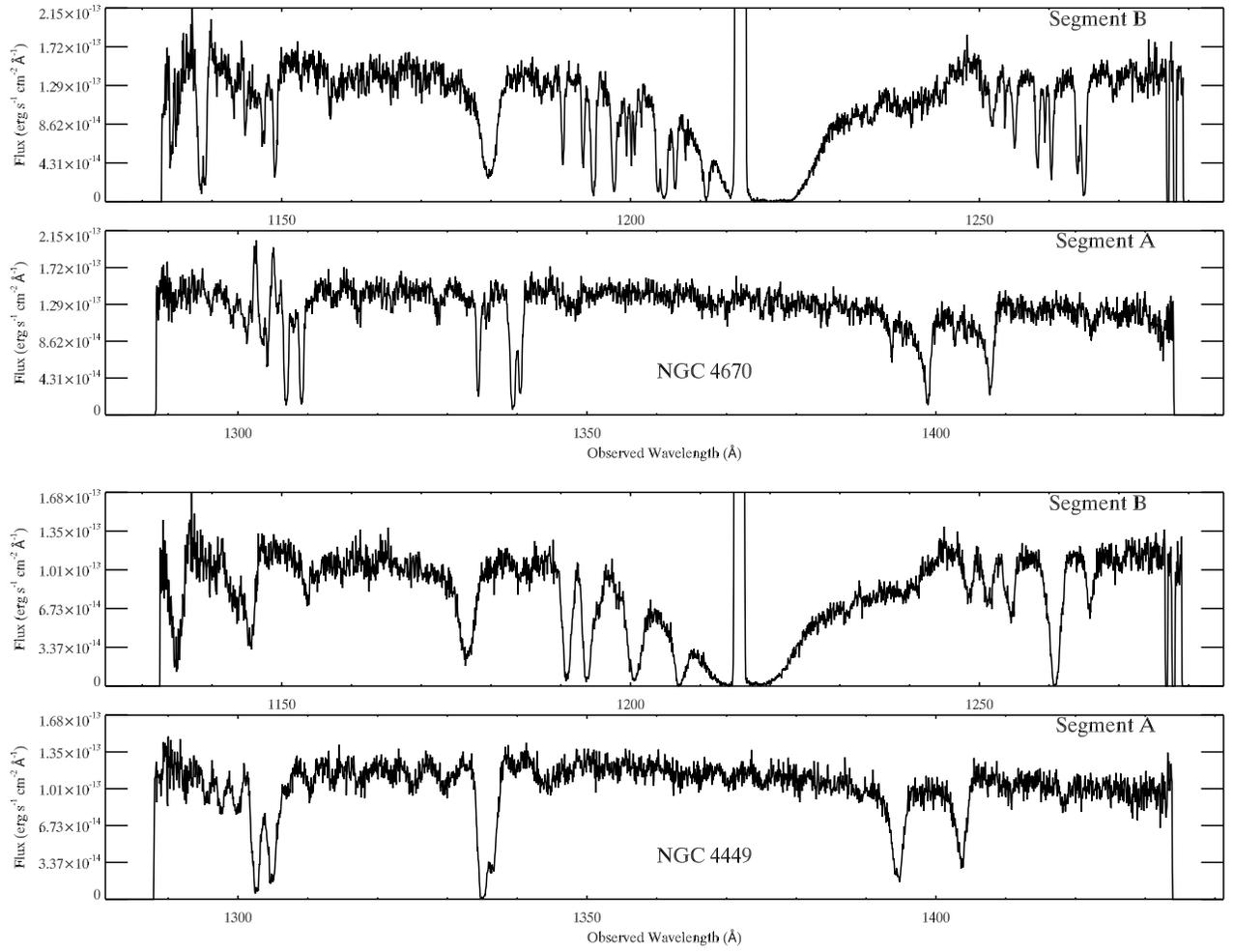}
\vspace{0.2cm}
\caption{HST/COS G130M spectra of NGC~4670 (top two panels) and NGC~4449 (bottom two panels).} 
\label{fig:ngc470_ngc4449_spec}
\end{figure*}

\begin{figure*}
\center
\includegraphics[angle=90,scale=0.75]{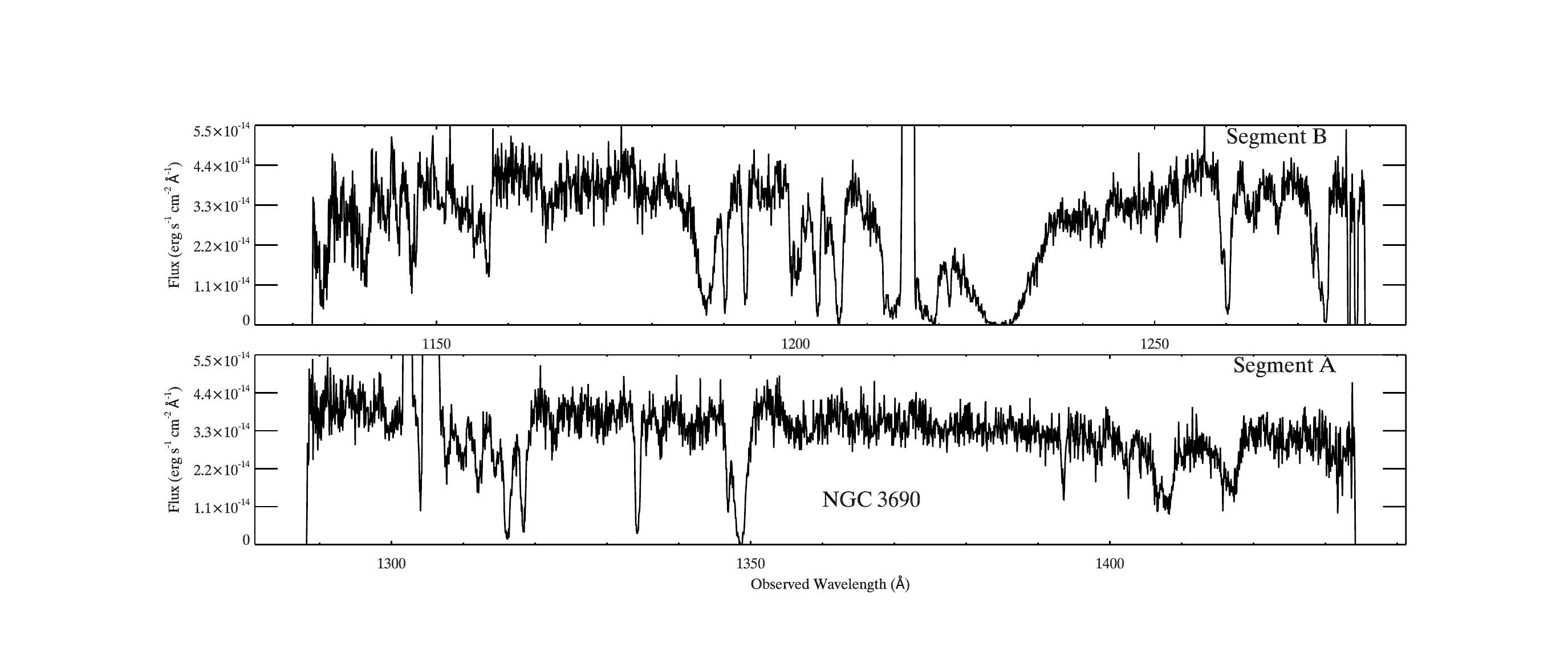}
\vspace{0.2cm}
\caption{HST/COS G130M spectra of NGC~3690.} 
\label{fig:ngc3690_m831_spec}
\end{figure*}

\begin{figure*}
\center
\includegraphics[angle=90,scale=0.75]{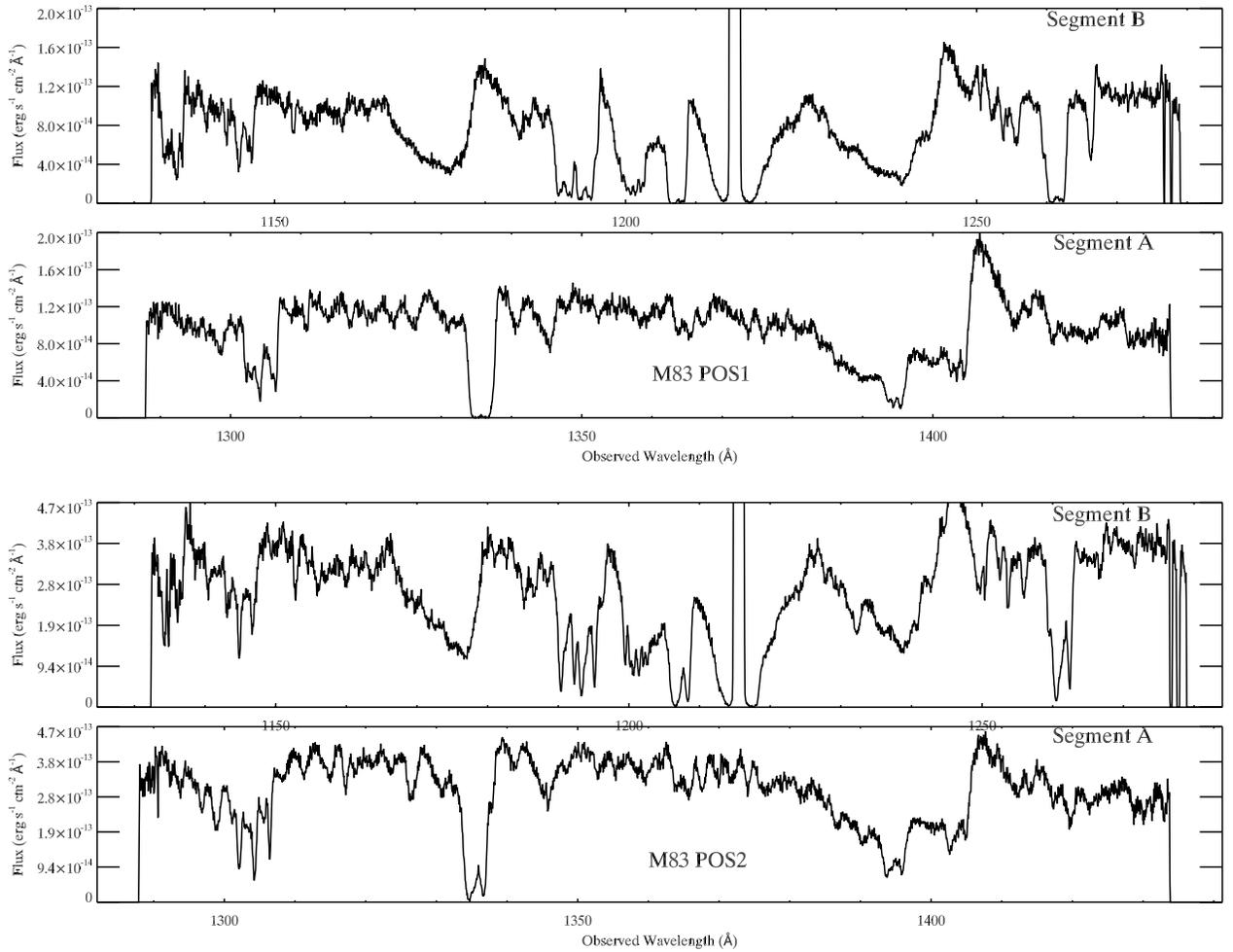}
\vspace{0.2cm}
\caption{HST/COS G130M spectra of M83-Pos$^n$~1(top two panels) and M83-Pos$^n$~2 (bottom two panels).} 
\label{fig:ngcm832_spec}
\end{figure*}

The resultant 11 COS spectra of the 9 targets in the rest-frame wavelengths are shown in Fig.~\ref{fig:all_spec}.  The radial velocity inferred from the \ciii~$\lambda$1176 stellar absorption line arising in the spectrum of the UV background source within each galaxy, has been used to convert observed into rest-frame wavelengths (see Table~\ref{tab:HIdensities}). Up to three absorprtion line systems at three different radial velocities are present in each spectrum: the radial velocity of the galaxy, the Milky Way (MW) and, if present, a high-velocity cloud (HVC).  Zoomed-in spectra for each of the galaxies in the observed wavelengths are shown separately in Figures~\ref{fig:izw18_sbs0335_spec}--\ref{fig:ngcm832_spec}.

\subsection{Removal of Geocoronal Contamination}
\label{sec:night}
Due to the passage of sunlight through the Earth's atmosphere, each observation is subject to contaminating geocoronal emission or `airglow'.  The magnitude of this contamination depends highly on the limb-angle and the altitude of the sun during each observation.  The emission spectrum is made up of two major components; the \Lya\, emission at 1215.67 \AA\, and a triplet of oxygen lines at 1302 \AA\, (\oi), 1304 \AA\, (\oi*) and 1306 \AA\, (\oi**).  The removal of the oxygen geocoronal emission is paramount for the low-redshift absorption line spectra presented here, where \oi~$\lambda$1302 galaxy absorption and \oi\, geocoronal emission coincide with one another. 

Whilst the observer cannot completely remove \Lya\, emission (due to it being in a very extended layer in the Earth's atmosphere), the strong dependence of the \oi\, geocoronal lines on limb-angle enables the observer to essentially select times when such emission is absent.   We were therefore able to remove the \oi\, emission contamination in each of our spectra by creating `night' spectra.  This was done in a three-stage process.  Firstly, the number of counts in both the 1215.67~\AA\, and 1302~\AA\, regions were analyzed as a function of time from each FPPOS \textsc{corrtag} file and compared against counts within $\sim$20~\AA\, regions either side of these wavelengths.  This essentially allowed us to determine when the emission was minimal/absent.  Secondly, we plotted the altitude of the sun as a function of time throughout each FP-POS observation.  Finally, we correlated these two pieces of information to determine the angle at which geocoronal contamination was absent and filtered the observations by angle, i.e. flagging observations above a certain angle as `bad'.

By only selecting the observations that were taken below this `airglow angle', where there was no increase in counts either side of the OI region, we were confident that the airglow had been removed rather than minimised. After re-combining the data, this left us with a `night' spectrum, upon which we were able to measure the \oi~$\lambda$1302 absorption line.  In some cases, where either the redshift of the galaxy was sufficiently high to shift this line away from the airglow emission, or where the limb-angle was large enough not to contaminate the spectrum, the creation of `night' spectra was not necessary.  

\section{Analysis of COS Data}\label{sec:analysis}

\subsection{Continuum Fitting}
The fit of the continuum was performed by interpolating between points of the observed flux free of apparent absorptions (referred to as `nodes' hereafter). Each node was chosen by carefully inspecting by eye the spectra of each target, an average of the flux within a $\sim$ 0.2 \AA~ region either side of each node was calculated, and a spline was used for the interpolation of the flux among these nodes.

Systematic errors on the measured column densities ($\sigma_{N(X)}$) due to the uncertainty in the continuum placement were considered as follows. Firstly, the total uncertainty on the continuum placement ($\sigma_{F_C}$) was considered as the error on the node flux (indicated as $F_c$), i.e., the variance on the mean flux within a 0.4~\AA\, region centered on the node.  $\sigma_{F_C}$ was calculated for each node across each spectrum, and was found to be of the order of $\sim$3\% at most. Secondly, we needed to assess the effect that $\sigma_{F_C}$ would have on the equivalent width ($W$) of the absorption line from which column densities, $N$(X), are derived. Following the calculations of \citet{Vollmann:2006}, the error on $W$ can be stated as:
\begin{equation}
\sigma_W^2 = [-\frac{\Delta\lambda}{F_c}\sigma_F]^2+[\frac{1}{F_c}(\Delta\lambda-W)\,\sigma_{F_c}]^2
\end{equation}
Whilst the first term is dealt with by our line-profile fitting software (which utilizes $\sigma_F$, the uncertainty in the flux at wavelength $\lambda$, and propagates it to derive $\sigma_{N(X)}$), the second term is not and needs to be considered independently. Thirdly, we needed to propagate $\sigma_W$ in terms of error on $N$(X).  Since the relationship between $W$ and $N$(X) varies according to the strength of the line, this had to be assessed for three different line profiles: unsaturated, saturated and damped.   We therefore calculated the magnitude of $\sigma_{N(X)}$ taking into account also the contribution of $\sigma_{F_C}$ to $\sigma_W$, for lines of varying oscillator strength and column density, in each of these regimes. The newly estimated $\sigma_{N(X)}$ was found to differ by much less than 1\% from the value given by our line-profile fitting software. To confirm this, an additional test was performed by measuring with our software a selection of lines on a spectrum whose continuum had been placed $\pm\,1\,\sigma_{F_c}$ above and below the original continuum nodes.  Column densities were found to vary within the uncertainties resulting from $\sigma_F$ alone.  We are therefore confident that $\sigma_{N(X)}$ is dominated by $\sigma_F$, which is fully taken into account by the line-profile fitting software (described below).

\subsection{Spectral Resolution}
\label{sec:resel}
Fitting of the absorption-line profiles requires the spectral resolution of each spectrum as a function of wavelength (FWHM$_{spec}$). This was derived using the FWHM of the COS instrument spectral response for a point source in the G130M grating, as measured from the G130M line-spread function (FWHM$_{lsf}$, $\sim 8$ pixels corresponding to $\sim 0.18$ arcsec)\footnote{The COS LSF experiences a wavelength-dependent non-Gaussianity due to the introduction of mid-frequency wave-front errors from the HST mirrors.  Tabulated profiles used for the calculation of the G130M $FWHM_{lsf}$ were obtained from http://www.stsci.edu/hst/cos/documents/isrs/.}, combined in quadrature with the intrinsic FWHM (FWHM$_{int}$) of the UV source.  FWHM$_{int}$ in arcseconds was calculated from the FWHM of the image profile measured in the NUV TA images (FWHM$_{imag}$) using the formula FWHM$_{int}^2$=\,\,FWHM$_{imag}^2-\,$FWHM$_{psf}^2$ where FWHM$_{psf}$ is the FWHM of the point-spread function in the NUV TA images ($\sim 2$ pixels corresponding to $\sim 0.05$ arcsec). The values of FWHM$_{spec}$ at $\sim$ 1150 \AA~for each UV point-like source targeted with COS are reported in Table~~\ref{tab:obs_COS}. Notice that most of the sources in our sample are only slightly extended (FWHM$_{spec}$ $\sim 19$ km\,s$^{-1}$ for a point-like source).

\subsection{Absorption-Line Fitting}
\label{sec:fitting}
We derived column densities of \hi\, and heavy elements from the COS spectra of each galaxy by line-profile fitting of the observed absorption lines.  Standard theoretical Voigt profiles were convolved with the spectral resolution inferred in \S~\ref{sec:resel} and fitted to the data.  The package \textsc{FITLYMAN} \citep{Fontana:1995} in MIDAS for multi-component fitting was used for this purpose.  This method is more powerful than a simple curve-of-growth analysis (based on the equivalent width of the absorption lines) because it allows for (1) the de-blending of multiple components along the line of sight contributing to the same absorption and (2) the simultaneous and independent fit of contaminating absorption components.  We were able to fit absorption lines of all the ions considered in each galaxy with very simple velocity-component models (i.e., one component, except for NGC~4214, NGC~3690 and M83-Pos\,1 where 2 components were necessary, and NGC~4449 where 3 components were necessary).

The real situation described by the far-UV spectra of these galaxies is, however, more complex.  COS detects a non-linear average absorption over the full extent of the stellar background sources covered by its 2.5\arcsec~aperture.  This implies that the observed lines arise from a combination of many unresolved velocity components from different absorbing clouds along the many sightlines within such aperture.  Moreover, as discussed in Section~\ref{sec:hidden_sat}, some lines of sight may have saturated absorption, even if the composite profile does not go to zero intensity \citep[e.g., ][]{Savage:1991}.

\citet{Jenkins:1986} has demonstrated that the single-velocity approximation applied to complex blends of features gives nearly the correct answer (the simulated-to-true column density ratio rarely goes below 0.8) if the distribution function of the line characteristics is not irregular.  This result holds also if different lines have different saturation levels or Doppler parameters $b$.  Since our COS data sample many sightlines towards each galaxy over the 2.5\arcsec\, aperture, we expect a quite regular distribution of the kinematical properties of the single absorbing components.  We thus believe that we amply fall within the regime where the single-velocity approximation is valid.

In light of these considerations, we preferred to maintain a simple approach in the determination of the column densities with the line-profile fitting method.  We thus avoided the introduction of additional free parameters, i.e., the number of intervening clouds and their velocity distribution, since we believe the resolution of the data does not allow to correctly constrain the real physical situation represented by this type of observation.  In the single-velocity approximation the fitting parameter $b$ has no precise physical meaning but is rather the result of the combination of both the various line Doppler widths present (due, e.g., to turbulent and/or thermal broadening) and the various velocity separations among the different line components  \citep[]{Hobbs:1974}.  On the other hand, according to \citet{Jenkins:1986} the column density is well constrained.  In addition, the column density of a certain ion is even better constrained if several lines with different values of $f\lambda$ ($\lambda$ is the rest-frame wavelength, and $f$ is the oscillator strength of the absorption) are available for simultaneous fit, and the results are independent of saturation problems affecting the strongest lines (see Section~\ref{sec:saturation} for a detailed discussion of saturation issues).

However, in order to check the reliability of the line-profile fitting results and the single-velocity approximation, following the method outlined in \citet{Aloisi:2003} we also determined total column densities of heavy elements by applying the apparent optical-depth method \citep[AODM,][]{Savage:1991}.  This method is very powerful for determining column densities.  Its strength lies in the fact that no assumption needs to be made concerning the velocity distribution of absorbers, i.e., it does not depend on the number of intervening clouds along the single or multiple lines of sight.  However, it does not provide a correct answer for quite strong lines that are not fully resolved.  In the case of our resolution, $\sim$18~\kms, the line-profile fitting should give a more accurate measure of the ionic column densities compared with the apparent optical depth method, especially for stronger lines.  
As mentioned previously, since AODM can only be applied to un-blended lines, we were limited to comparing only a few lines within each spectrum and only lines known to be free from saturation issues (Section~\ref{sec:saturation}).  For the handful of lines that were suitable, column densities resulting from the two methods are in excellent agreement, with an average difference of $\sim$0.1~dex.  

In our analysis, the similarities between the column densities obtained with the line-profile fitting and AODM for (unsaturated) lines strengthen our results and justify the single-velocity approximation.

\section{\hi\, Column Densities}\label{sec:hi_colDens}

The COS FUV spectral range covers only one absorption line of the \hi\, Lyman series, i.e., \Lya\, at $\lambda=1215.67$~\AA.  Measurements of the \hi\, column density ($N$(\hi)) for each target are listed in Table~\ref{tab:HIdensities}.  For each object, we also list the foreground \hi\, column densities from the Milky Way ($N$(\hi)$_{MW}$). Due to the nearby proximity of the galaxies within our sample, the foreground \hi\, column density in the direction of each object was often not clearly separated from that intrinsic to the object.  In such cases, the foreground+object absorption profiles were fitted simultaneously by considering a fixed MW velocity component at the average velocity of the MW lines, and by using the blue wing of \Lya\, to constrain the fit of the MW absorption and the red wing to constrain the fit of the intrinsic absorption.  If the blue wing of the MW  Lya\, was heavily blended with that of the target, e.g. for very low-redshift targets or where multiple velocity components exist, we inferred $N$(\hi)$_{MW}$ using values measured by the composite all-sky map of neutral hydrogen column density ($N$(\hi)$_{map}$, also given in Table~\ref{tab:HIdensities}), formed from the Leiden/Dwingeloo survey data \citep{Hartmann:1997} and the composite $N$(\hi) map of \citet{Dickey:1990}.  In such cases $N$(\hi)$_{MW}$ was fixed at the map value (along with a fixed central wavelength; the b parameter does not matter in this case, since the line is in the damped part of the curve of growth) and the source $N$(\hi) was allowed to vary freely.  These cases can be identified in Table~\ref{tab:HIdensities} as having no error on $N$(\hi)$_{MW}$.  A case-by-case description of the fitting of each \Lya\, profile is described Section~\ref{sec:objects}.

The \Lya\, absorption lines of the galaxies in our sample appear to be interstellar in origin according to their typical damped profile.  However, UV spectra of Galactic early-B stars \citep[e.g., ][]{Pellerin:2002} clearly show the presence of a photospheric component.  This implies that when early-B stars start to dominate the integrated spectrum of a stellar population (e.g., a burst with an age greater than $\sim$10~Myr), the wings of a large photospheric contribution are superposed on the cores of the damped interstellar absorption.  However, the contamination is minimal if the stellar temperature is below $\sim$35,000~K \citep[see Figures A1 and A2 of ][]{Lebouteiller:2009}.

In order to assess the age of the stellar population dominating the UV spectrum, each spectrum was fitted to model stellar spectra created by \textsc{TLUSTY} \citep{Hubeny:1995}.  Models were prepared for a range of temperatures and metallicities and include stellar rotation at approximately 200~\kms (Lebouteiller~V., private communication).  The main lines used to estimate the age and metallicity of the stellar population were \civ~$\lambda$1169, \ciii~$\lambda$1175 and the \oiv\, doublet at $\lambda\lambda$1338, 1343.  The \civ\, and \ciii\, profiles provide a good constraint on the upper limit of the stellar temperature, increasing slowly in strength with decreasing temperature up to the B0 type \citep{Pellerin:2002}.  At the same time, the \oiv\, doublet decreases with strength and thus provides us with a lower limit on the temperature.  Models of various metallicities ($0.001<$ \Zsol $<2$) and temperatures ($32,500<T~(K)<55,000$) were overlaid on each spectrum and a Z, T grid of best-fitting models was determined for each galaxy's stellar population. 

For each galaxy, the \Lya\, stellar absorption profile for each set of best-fitting stellar models spanning a wide range in metallicities and temperatures, was fitted using the same procedure adopted for the interstellar absorption lines. The fitted parameters of the strongest stellar absorption were then included as an additional component when measuring the target \Lya\, absorption profile.  Minimal changes in \Lya\, column densities were found  with the addition of this worst-case scenario for the stellar absorption ($N$(\hi) values decreased by 0.0--0.07~dex). Since this is well within the errors on $N$(\hi) from the fit of the \Lya\, absorptions with the interstellar component only, we did not consider it further.\\

\section{Heavy-Element Column Densities and Abundances} 
\label{sec:abundances}
We cover several transitions of neutral, singly and doubly ionized atoms of heavy elements.  Table~\ref{tab:fvals} lists theoretical parameters of each ion measured within our sample of COS data; the vacuum rest wavelengths $\lambda_{lab}$ of the transitions are from the compilation by \citet{Morton:1991} and oscillator strengths, $f$, are from the references indicated.

We derived values of column density for ions of interest by using the line-profile fitting technique described in Section~\ref{sec:analysis}.  The line profiles of most ions appear symmetric and were simultaneously fitted by a single-velocity component.  The only exceptions are NGC~4214, NGC~3690, M83-Pos~1 and NGC~4449, where the multi-component structure of their profiles required the introduction of additional components in order to get an acceptable value for $\chi^2$ on the fit.  In Table~\ref{tab:Nvals} we list the best-fit results for the ions considered within each spectrum, for each target.  In the multi-component cases listed above, we separately list the best-fit parameters of each of the velocity components.  Column densities affected by either `hidden' or `classical' saturation, as discussed in the following sections, should be considered as lower limits and are identified with a $\dagger$ symbol in Table~\ref{tab:Nvals}.  

\subsection{Saturation}\label{sec:saturation}
Absorption line profiles can suffer from two types of saturation; `classical' saturation, where a line is optically thick and line width increases logarithmically with column density, and `hidden' saturation, where multiple lines of sight containing both large and small column densities combine to make a line \textit{appear}, e.g., unsaturated, when in fact it is saturated.  In the following sections, we describe the methods used to check for, assess, and treat both types of saturation.

\subsubsection{Hidden Saturation}\label{sec:hidden_sat}

\begin{figure*}
\center
\includegraphics[scale=0.4]{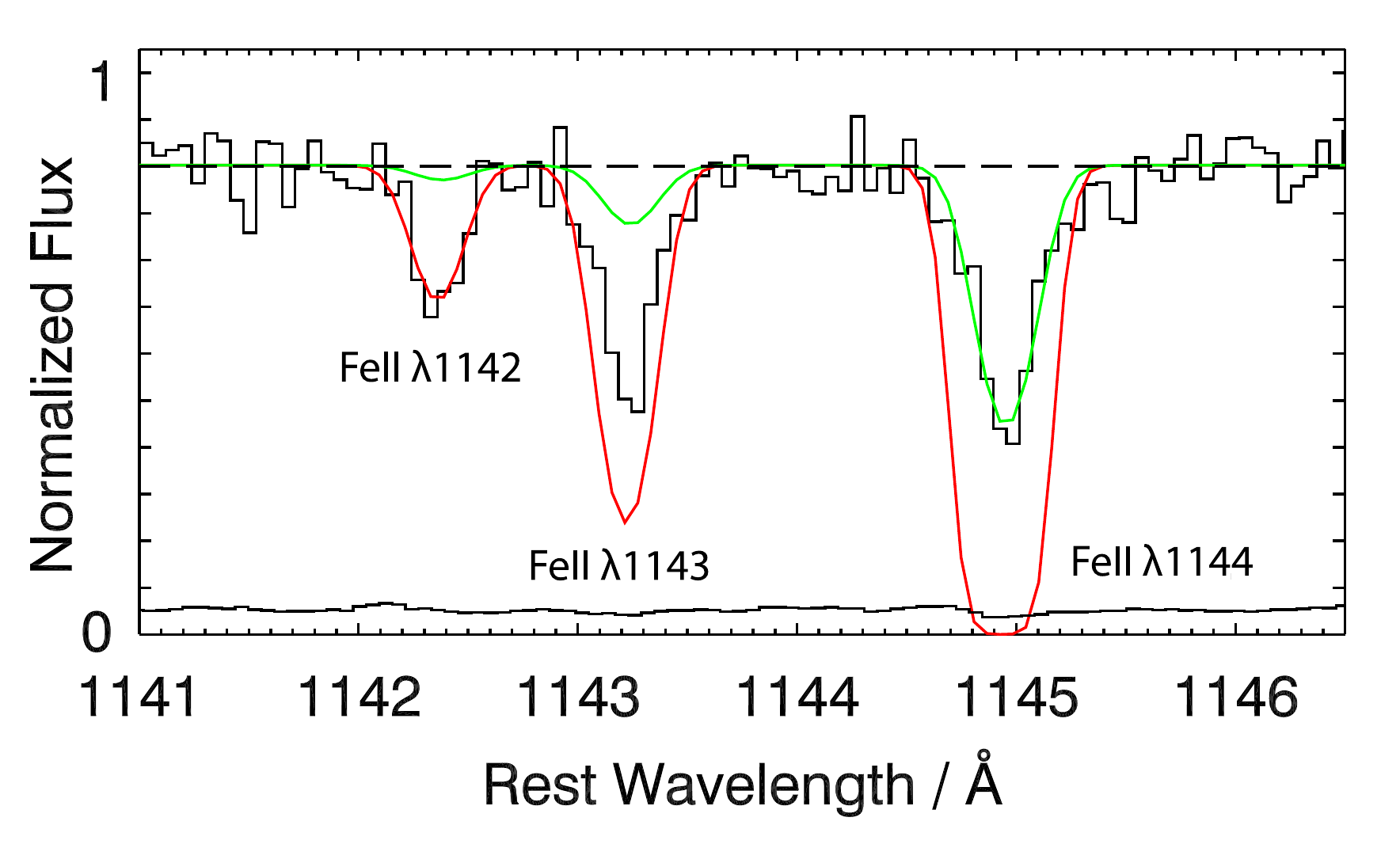}
\caption{An example of hidden saturation affecting our \feii\, column densities.  Here we show the spectra of SBS~0335-052 around the \feii\, lines at 1142~\AA, 1143~\AA, and 1144~\AA. The \feii\, column density obtained by fitting the strongest 1144~\AA\, line (green solid line) clearly under-predicts the strength of the weakest line at 1142~\AA. Similarly, the \feii\, column density inferred by fitting the weakest 1142~\AA\, line (red solid line) is overestimating the strength of the other strongest two lines. This is an indication that hidden saturation is at play in at least the 1143~\AA\,  and 1144~\AA lines. In this case, the column density from the 1144~\AA\, line is underestimated by up to $\sim$1~dex if the weakest 1142~\AA\, is used to infer the column density.} 
\label{fig:feii_saturation}
\end{figure*}

Whilst there are strong lines that are clearly affected by saturation (namely \cii~$\lambda$1134 and \oii~$\lambda$1302, see below), cases of hidden saturation are also present within the COS spectra.  This effect can be recognized when fitting multiple lines of the same ion, since the strongest lines are too weak (i.e., saturated) compared to the other weaker lines, as illustrated in Figure~\ref{fig:feii_saturation}.  For the spectra in our study, this effect was evident in the \feii\, triplet (at 1142, 1143 and 1144\,\AA) and \Ni\, triplet (at 1134.1, 1134.4 and 1134.9\,\AA), since in both cases several lines are available of varying oscillator strength.  Here we discuss the variable amounts of hidden saturation observed within our spectra and assess the roles that various parameters have in causing this effect.

Table~\ref{tab:saturation} lists the difference in inferred column density between the weakest and strongest lines within the above mentioned \feii\, and \Ni\, triplets for the majority of galaxies within our sample (in some cases we were unable to constrain column densities from the weaker lines only, due to blending).  In the case of \feii\, we constrained the column density using \feii~$\lambda$1133, $\lambda$1142 and/or $\lambda$1143 (the weakest lines available) and excluded the strongest line, \feii~$\lambda$1144, from the fit.  On average, we found that \feii~$\lambda$1144 under-predicts $N$(\feii) by 0.6~dex compared to using only the weakest lines.  The difference was less extreme for \Ni, where \Ni~$\lambda$1134.9 under-predicts $N$(\Ni) by $\sim$0.2~dex compared to the $\lambda\lambda$1134.1, 1134.4 lines (the \Ni~$\lambda$1200 triplet was not used as these lines are very strong and almost certainly suffer from `classical' saturation).  

\begin{figure*}
\center
\includegraphics[angle=90,scale=0.5]{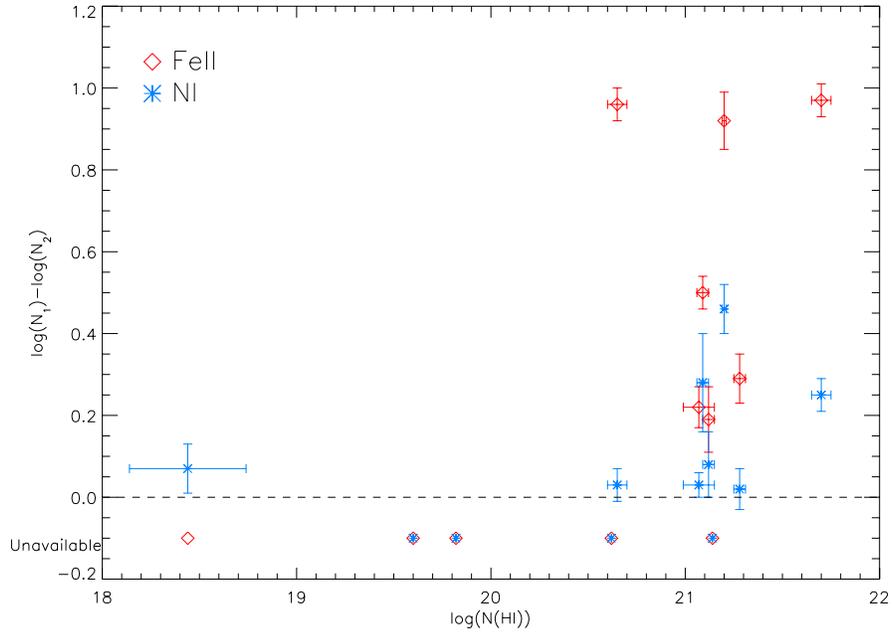}
\caption{Difference in the column density of the strongest and weakest lines in the \feii\, (log[$N$(1142 or 1143)]$-$log[$N$(1144)], red open diamonds) and \Ni\, (log[$N$(1134.1,1134.4)]$-$log[$N$(1134.9)], blue asterisks) triplets as a function of \hi~column density. Symbols at $y = -0.1$ represent galaxies for which we were unable to constrain column densities due to blending.} 
\label{fig:saturation_trends}
\end{figure*}

Figure~\ref{fig:saturation_trends} shows the values listed in Table~\ref{tab:saturation} as a function of neutral hydrogen column density, log[$N$(\hi)/cm$^{-2}$]. It can be seen that overall, the amount of hidden saturation is greatest for galaxies with log[$N$(\hi)/cm$^{-2}$]$\,>\,$20.6.  For \Ni\, alone, saturation of the 1134.9~\AA\, line only occurs when log[$N$(\hi)/cm$^{-2}$]$\,>\,$21.1. Unfortunately, in the case of \feii\, we are unable to provide a definitive limit to the amount of $N$(\hi) where hidden saturation becomes an issue because several of the galaxies with log[$N$(\hi)/cm$^{-2}$]$\,<\,$20.6 have multiple velocity components and column densities from the \feii~$\lambda$1142 or \feii~$\lambda$1143 absorption lines could not be constrained.

\begin{figure*}
\center
\includegraphics[angle=90,scale=0.5]{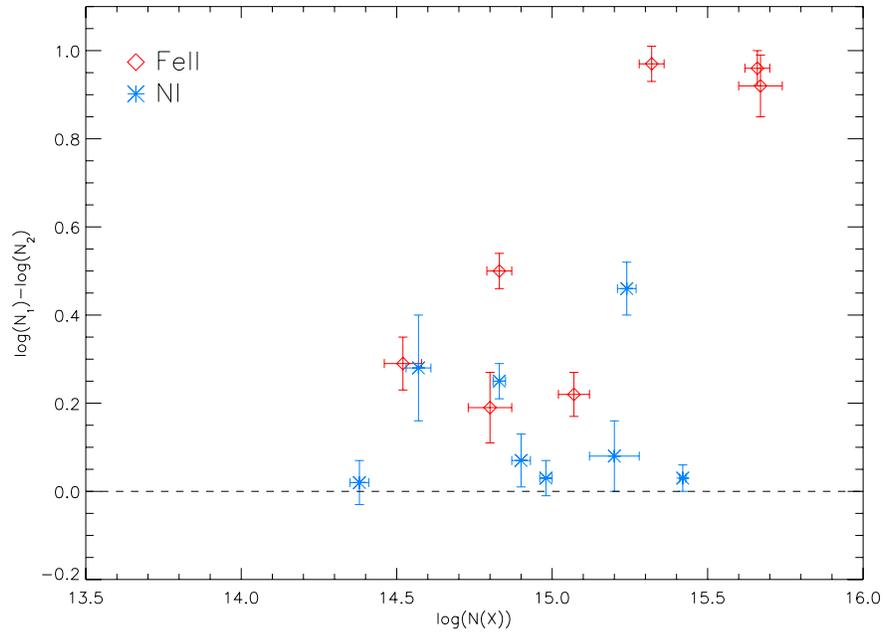}
\caption{Magnitude of under-estimation in column density, due to hidden saturation, as a function of column density of the weakest (i.e., most reliable) absorption line(s).  Red open diamonds are for \feii~and blue asterisks for \Ni. Whilst the effects in \Ni\, are small and practically negligible in about half of the cases, a clear correlation can be seen between the physical amount of \feii\, within the gas and the extent of hidden saturation.} 
\label{fig:saturation_ndens}
\end{figure*}

However, this is a multi-parameter problem, affected by not only the amount of gas within the sightline, but also the physical amount of that species within the gas.  Figure~\ref{fig:saturation_ndens} shows the magnitude of $N$(X) under-estimation as a function of $N$(X).  Whilst the effects on nitrogen are smaller and practically negligible in about half of the cases due to the lower abundance of this species in our sample, a clear increase in the amount of hidden saturation can be seen with increasing $N$(\feii), where the \feii~$\lambda$1144 line is very strong compared to the others and hidden saturation can be easily assessed.  

\begin{figure*}
\center
\includegraphics[angle=90,scale=0.8]{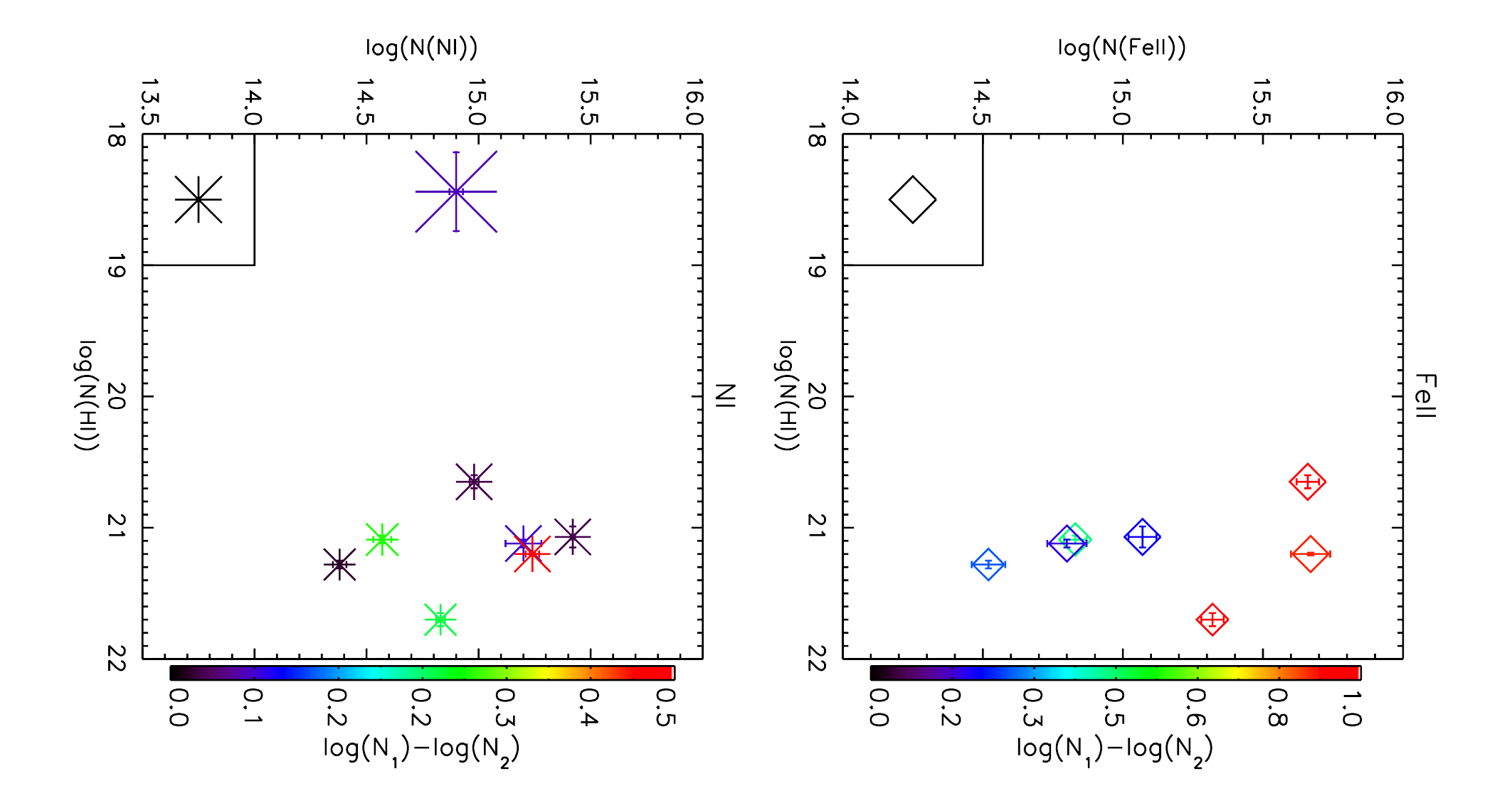}
\caption{Factors that contribute to the amount of hidden saturation, i.e., column density of the species (i.e., $N$(\feii), top panel, or $N$(\Ni), bottom panel) versus \hi\,column density, where colors indicate the amount of hidden saturation and symbol size scales with metallicity of the \hii\, (nebular) gas. For reference, the solar metallicity is represented in the little square at the bottom left of each plot.} 
\label{fig:saturation_all}
\end{figure*}

The main contributing parameters to hidden saturation are combined in Figure~\ref{fig:saturation_all}, which shows the amount of saturation as a function of $N$(X), $N$(\hi) and metallicity as inferred from the nebular gas (see Table~\ref{tab:sample}).  For \feii\, (top panel of Fig.~\ref{fig:saturation_all}) it can be seen that, as expected, the maximum amount of hidden saturation occurs within the galaxies with highest $N$(\feii), metallicity and $N$(\hi).  Whilst there are galaxies with $N$(\hi) comparable to those showing maximum saturation, the physical amount of \feii~as assessed from $N$(\feii) in the neutral ISM (and the galaxy metallicity to a lesser extent) is lower, effectively resulting in weaker lines and decreasing the overall effect of hidden saturation. For nitrogen (lower panel of Fig.~\ref{fig:saturation_all}), we can see the effects of low $N$(\hi) counter-acting the metallicity of the galaxy, where the highest metallicity galaxy displays a negligible amount of saturation, despite having $N$(\Ni) comparable to galaxies with log[$N$(\hi)/cm$^{-2}$]$\,>\,$20.6. This illustrates that it is a combination of the amount of the ion and the amount of the absorbing gas which result in hidden saturation.  

To conclude, the amount of hidden saturation of a certain ion X from its absorption lines within a spectrum is a function of two main parameters, $N$(X) and $N$(\hi).  For both \feii\, and \Ni\, we find that hidden saturation effects occur when log[$N$(X)/cm$^{-2}$]$\,>\,$14.5, but only when log[$N$(\hi)/cm$^{-2}$]$\,>\,$20.6 for \feii\,, whilst for \Ni\, saturation effects are only experienced at higher \hi\, column densities, i.e., when log[$N$(\hi)/cm$^{-2}$]$\,>\,$21.1, corresponding to abundance limits of [Fe/H]$\,>\,-6.1$ and [N/H]$\,>\,-6.6$.

As a result of these findings, and for the purpose of this paper, column density measurements of ions that have multiple lines within our wavelength range (e.g., \feii, \Ni, \sii, \sIii, \sIii*), were made using all the ``weak" lines that were consistent with the weakest line available and avoiding strong lines that may be affected by hidden saturation.  For galaxies where multiple velocity components exist in their absorption lines, we were often unable to isolate the weakest lines as they were merged with velocity components from the stronger lines.  For these cases, column densities are listed as lower limits in Table~\ref{tab:Nvals}.

\subsubsection{Classical Saturation \& Curve-of-Growth Analysis}\label{sec:COGs}
In order to determine whether our measurements were affected by `classical' saturation, for each ion in each galaxy we checked the column density measurements from the line-profile fitting against the curve of growth (COG) suited for that ion.  This method is particularly useful for those ions where only one absorption line exists preventing us from identifying this kind of saturation by simply performing a line-profile fitting. In fact, the fitting always gives the lowest column density possible in the ``saturated" COG regime, despite the same equivalent width may correspond to a wide range of column densities and $b$ parameters. Each ion has a specific COG which expresses how the equivalent width, log($W/\lambda$), changes with column density, log($fN$), where $f$ is the oscillator strength for the transition undertaken to produce a photon of wavelength $\lambda$.  As the strength of the absorption line profile increases, the curve takes on three separate parts: (1) a `linear part', where W grows linearly with $N$ and is independent of the Doppler parameter $b$; (2) a `saturated part', where the wings continue to grow deeper and broader, but the core gets flatter ($W$ increases much more slowly, as $(\ln N)^{1/2}$, so it is only weakly dependent on $N$ but is strongly dependent on $b$); and (3) a `square-root part', where the wings grow even deeper, $W$ increases as $N^{1/2}$ and is independent of $b$.  

In order to compare our measurements with the best suited COG, we had to adopt the appropriate $b$ parameter for each ion considered in the spectra of each galaxy. Due to the degeneracy between $b$ and $N$ in the saturated part of the COG (this degeneracy implies that for a certain $W$ value of an absorption line the column density $N$ cannot be univocally determined, and only an interval of values can be given, unless the $b$ parameter is known), we could not merely rely on our $b$ measurements from the line-profile fitting. This is particularly true for those cases where only one absorption line was available. We thus tried to estimate the most likely $b$ value based on the following considerations. For each galaxy the $b$ parameters of all the ions considered were plotted as a function of atomic mass to assess whether the broadening of the lines was due to thermal rather than turbulent motion of the atoms.  In the case of thermal motion, the $b$ parameter would decrease with the atomic mass of the ion, and we would be able to infer the $b$ value specific to the ion under consideration by taking this trend into account. Alternatively, for turbulent motions the $b$ parameters would be relatively constant compared to the atomic mass of the ions, and an average $b$ parameter would suffice. The latter case was found to be true for all the galaxies within our sample. Note that the reasoning above implies that we fall within the regime where the single-velocity approximation is valid (see \S~\ref{sec:fitting}).

For each species in each target, we generated the appropriate COG by using both the $b$ parameter inferred from the simultaneous line-profile fitting of all the available lines of that species (even if this value may not be reliable as the line(s) may be saturated) and the average $b$ parameter estimated from the analysis of the trends vs atomic mass (different velocity components were analyzed separately, under the assumption that each of them is real and arises from a separate volume of gas along the line of sight).  We then placed on this curve each line of that species by taking into account its $W$ and $N$ as inferred from the fitting. Since $W$ is just the integral of the fitted line, it is a good approximation of the real equivalent width. In this way we could easily determine whether or not `classical' saturation was affecting our measurements, e.g, because all the lines used were in the saturated part of the COG.

Figure~\ref{fig:COG_sample} shows a selection of COGs illustrating each of the line strength regimes encountered, each of which are discussed below.

\subsubsection{Assessing Saturation Effects}\label{sec:saturation_results}
The number of transitions available for each species in each target, the strength of each transition, the possibility to fit all transitions with one single fit or not, and the location of these transitions on the COG, have all been used to assess if the column density estimates reported in Table~\ref{tab:Nvals} are robust or should instead be considered as lower limits due to saturation effects. The following paragraphs discuss the level of saturation affecting the column density estimate of each ion in each target. 

\begin{figure*}
\center
\includegraphics{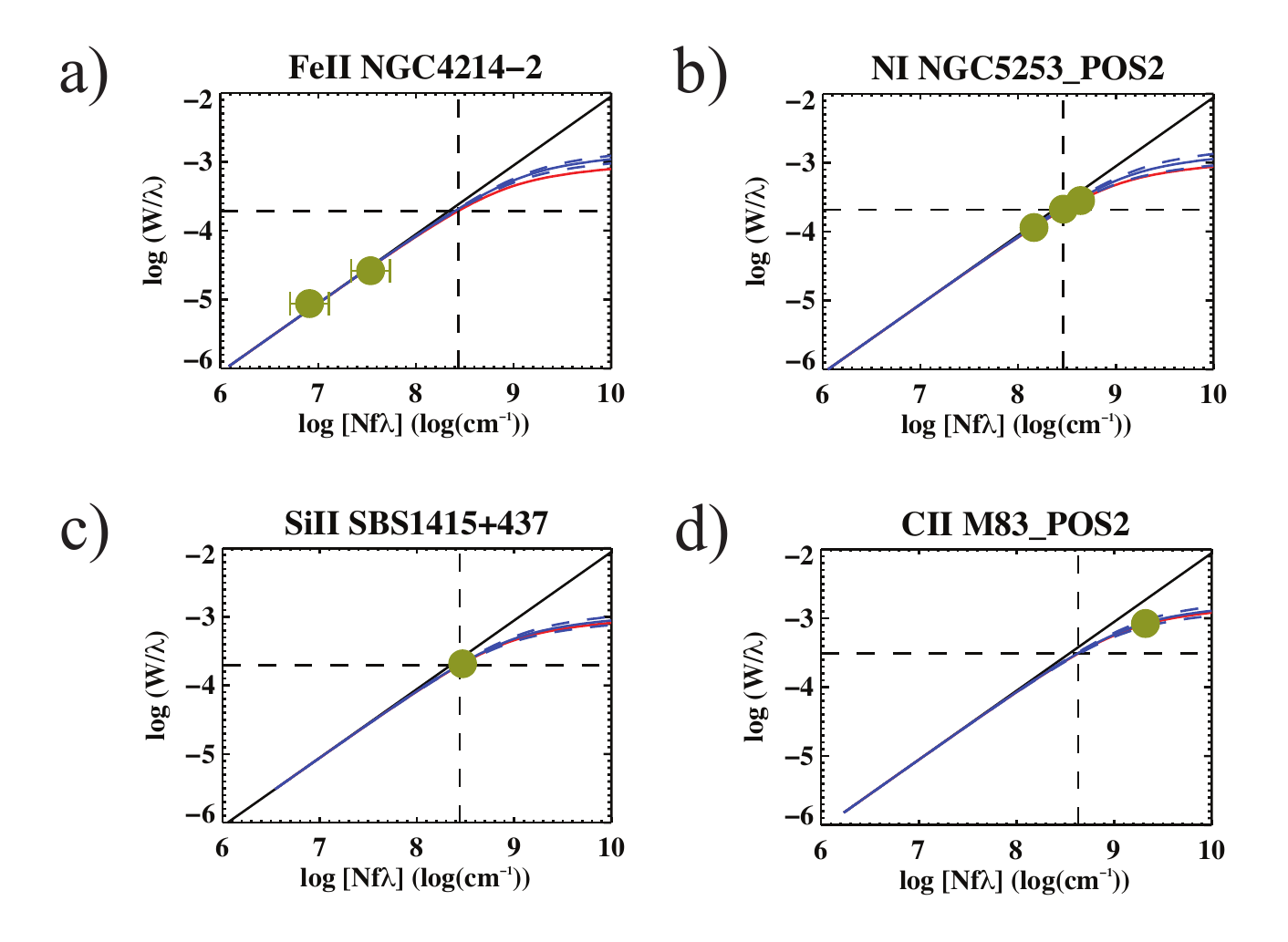}
\vspace{0.2cm}
\caption{A selection of curves of growth showing the linear and saturated (\textit{curved}) parts of this curve for a $b$ value corresponding to (1) the average $b$ parameter of all the species with absorption lines fitted in the spectra of the galaxy (or galaxy velocity component, in the case of multiple components; \textit{blue line}), and (2) the fitted $b$ parameter specific only to the species considered in the plot (\textit{red line}).  Each curve illustrates a different line strength regime encountered within this study: (a) multiple transitions showing un-saturated lines; (b) multiple transitions ruling out the possibility of hidden saturation; (c) a single transition, border-line case where saturation may exist; (d) single transition, saturated line.  The 1\,$\sigma$ errors on the $b$ parameters are also included as blue/red dashed lines, but being very small they are practically undistinguishable from the parameter values plotted. Overlaid brown points correspond to the equivalent width $W$ and column density $N$ of each line as derived from our line-profile fitting. The error bars on these two quantities are also included, but are in most cases  smaller than the size of the symbols plotted.} 
\label{fig:COG_sample}
\end{figure*}

\textit{\cii, \oi, and \sIiii:} As expected, when only one strong line such as \cii~$\lambda$1334, \oi~$\lambda$1302, or \sIiii~$\lambda$1206 is available for a certain ion, this line ends up being located in the saturated part of the COG  (see Fig.~\ref{fig:COG_sample} for an example), usually at the beginning of this part because \textsc{fitlyman} automatically resorts to the solution containing the lowest column density $N$ and largest $b$ parameter.  With only one line available, it is not possible to estimate the amount of  hidden saturation that can be present even when the line appears to be at the boundary between linear and saturated part of the COG. We have thus been conservative and assumed that all \cii, \oi, and \sIiii\, column densities are lower limits in Table~\ref{tab:Nvals}.

\textit{\Ni:}  Since in most cases there are several \Ni\, lines with different oscillator strengths available within our spectral range, the assessment of  \Ni\, saturation is straightforward.  In general the weakest lines fall within the linear regime on the COG (see Fig.~\ref{fig:COG_sample} for an example). For those cases with multiple lines where the weakest ones begin to approach the saturated part of the COG (i.e., SBS0335$-$052, velocity component 1 in NGC~4214, NGC~4670, and position 1 in NGC~5253), since we were able to fit more than one line with the same solution, we are confident that we are not dominated by hidden or classical saturation. 
The only exception is velocity component 1 in position 1 of M83. Since only the strongest \Ni\, $\lambda$1134.9 line was fitted in this target and this line falls in the saturated part of the COG, the corresponding \Ni\, column density is indicated as a lower limit in Table~\ref{tab:Nvals}.

\textit{\sIii:}  Despite the many transitions available for \sIii\, in our wavelength range, only the weakest (but still quite strong) \sIii~$\lambda$1304 line was used to estimate the column density, the other stronger lines being contaminated and/or clearly affected by hidden saturation. From the fitting we cannot determine if this line is also suffering from hidden saturation, independently of its location on the unsaturated or saturated part of the COG (which can unveil classical saturation;  see Fig.~\ref{fig:COG_sample} for an example). To be conservative we have thus considered all the \sIii\, column densities as lower limits in Table~\ref{tab:Nvals}. 

\textit{\cii*, \sIii*, \pii, \sii, and \nIii:} The column density measurements of these ions make use in most cases of few lines that are likely too weak to suffer from hidden saturation (if present). In addition, the majority of these lines lie well within the linear part or just at the intersection with the saturated part of the COG, thus excluding classical saturation. The only exception is the two \sii\, lines in NGC~4670. These lines are very close to each other at the beginning of the saturated part of the COG, and to be conservative  the corresponding \sii\, column density is considered as lower limit in Table~\ref{tab:Nvals}.

\textit{\feii:}  Several \feii\, transitions with different oscillator strengths are available in our wavelength range to address at some level both hidden and classical saturation. E.g., the simultaneous fit of multiple lines, including in some cases the strongest  \feii\,$\lambda$1144 line, gives us confidence that hidden saturation is not an issue. In addition, if these multiple lines have quite different oscillator strengths, with the weakest lying in the linear part of the COG, we can also exclude classical saturation effects. This is the case, e.g., in both velocity components of NGC~4214 (see, e.g., Fig.~\ref{fig:COG_sample}) and NGC~3690. In all other cases, only one \feii\, line was fitted and a careful analysis of the information available was necessary to assess saturation effects on a case by case basis. For SBS~0335$-$052 all lines of the \feii\, triplet around $\sim 1143$ \AA, in addition to \feii~$\lambda$1133, were available. Measurements of  the weak \feii~$\lambda$1142 line versus the strongest \feii~$\lambda$1144 line of the triplet indicate a hidden saturation effect of $\sim 1$ dex. This effect drops to $\sim 0.25$ dex if the $\sim 6$ times weaker \feii~$\lambda$1143 line is considered instead of \feii~$\lambda$1144. Since \feii~$\lambda$1142 (as well as \feii~$\lambda$1133, the other line used in the fit) is another factor of $\sim 4 (3)$ weaker compared to \feii~$\lambda$1143, we are confident that hidden saturation is not an issue. In this object, the two lines fitted are also still in the linear part of the COG, excluding classical saturation effects. SBS~0335$-$052 is the worst case of hidden saturation as indicated by \feii~$\lambda$1142 or \feii~$\lambda$1143 versus \feii~$\lambda$1144 measurements. If this case is taken as reference, we can confidently say that hidden saturation is not an issue for the two positions of NGC~5253, where only \feii~$\lambda$1142 was fitted, giving column densities $\sim 1$ dex lower than \feii~$\lambda$1144. The COG analysis for these pointings suggests that classical saturation is not an issue either. In I~Zw~18 and NGC~4670, only the \feii~$\lambda$1143 line was used for the fit, giving a $\sim$ 0.2-0.3 dex lower column density compared to \feii~$\lambda$1144. This is a factor of 2-3 lower than in SBS~0335$-$052 ($\sim 0.75$ dex) and similar to what found there as difference between \feii~$\lambda$1142 and \feii~$\lambda$1143. We thus assumed that in these two galaxies hidden saturation is not affecting the measurements. As suggested by the COG, classical saturation is not an issue either. \feii~$\lambda$1143 is the only line used also in SBS~1415+437. However, the column density inferred from this line differs by $\sim 0.5$ dex if compared to \feii~$\lambda$1144, indicating more severe hidden saturation effects than in the previous targets. Despite the COG suggesting that classical saturation is not a problem for this object, due to the uncertainties in the strength of hidden saturation, the \feii\, column density in Table~\ref{tab:Nvals} is considered as lower limit. Only \feii~$\lambda$1144 was used in all velocity components of NGC~4449, and all velocity components and positions in M83. It is extremely likely that these lines are affected by hidden saturation (see Section~\ref{sec:hidden_sat} and Fig.~\ref{fig:saturation_trends}), thus the \feii\,column density inferred for these targets is listed as lower limit in Table~\ref{tab:Nvals}.

\subsection{Corrections}
\label{sec:corrections}
Another major concern in abundance determinations from UV absorption-line analyses is represented by ionization and dust-depletion effects.  Our approach to each of these issues is described in the following sections.

\subsubsection{Ionization Corrections}
\label{sec:ioncorr}

Ionization effects are usually negligible in studies of the neutral ISM, and abundances of a certain element are inferred from the amount of that element in the primary ionization state at the temperature of the neutral gas. From Galactic interstellar studies it is well known that the singly ionized stage is the dominant one for most elements because their first ionization potential is below 13.6~eV (the H$^0$ ionization threshold) and their second one is above it.  The neutral stage instead prevails for those elements having the first ionization potential above 13.6~eV.  The reason for this is that the bulk of the \hi\, gas with $N_{HI}\gtrsim10^{19}$~cm$^{-2}$ is self-shielded from $hv>13.6$~eV photons, but transparent to $hv<13.6$~eV photons.  This means that \cii, \sIii, \pii, \sii, \feii, and \nIii, as well as \oi\, and \Ni\, will be the dominant ionization stages of these elements in the neutral gas of our sample of nearby star-forming galaxies. However, our sample also includes a few targets with $N_{HI}\lesssim10^{19}$~cm$^{-2}$ for which ionization effects can be noticeable and need to be properly taken into account in order to avoid an underestimate of the real amount of a certain element in the cold gas. 

There is also another kind of ionization effect that needs to be considered for our sample. Some of the ions that are dominant ionization states in \hi\, regions may also be produced in photoionized clouds (or \hii\, regions) where \hi\, is a small fraction of the total hydrogen content. These clouds are likely photoionized by the same UV background sources that we used to detect absorption lines in the neutral ISM of our targets, or could be simply along the same sightline. The formation of metal absorption lines in both ionized and neutral regions can have a significant impact on element abundance determinations of the neutral ISM, resulting in an overestimate of the real amount of a certain element in the cold gas. The size of the impact depends on the ionizing source and the geometry of the system.  For example, the ionization corrections are in general smaller when a stellar spectrum dominates over an external UV background, i.e., for damped \Lya\, systems \citep[see, e.g.,][]{Howk:1999}.  Fortunately, this is the case for our sample of SFGs, where the radiation field is predominantly made up of young massive stars within the galaxies themselves.

The ionization correction factors (ICFs) due to contaminating ionized gas along the line of sight of the COS targets in our sample, are usually larger than the so-called classical ionization correction factors needed to properly take into account all the ions of a certain element present in the neutral ISM.  From now onwards we will refer to these two corrections as ICF$_{ionized}$ and ICF$_{neutral}$, respectively, and to the total ionization correction as:

\begin{equation}
ICF_{total}  = ICF_{ionized} - ICF_{neutral} \label{eq:icftot}
\end{equation}

\noindent
(see Table~\ref{tab:ICFs} for a list of the adopted values for these corrections in each SFG of our sample). The final column density of each element X, corrected for both ionization effects, is defined as follows:

\begin{equation}
log[N(X)_{ICF}]=log[N(X)] - ICF_{total}. \label{eq:icf0}
\end{equation}

Column density ratios relative to H and abundances derived from ionization-corrected column densities are listed in Table~\ref{tab:abundances} as log\,(X/H)$_{ICF}$ and [X/H]$_{ICF}$, respectively.

\paragraph{Ionization Corrections due to Contaminating Ionized Gas} 

To estimate the magnitude of ionized gas contributing to our absorption spectra, we used the \textsc{CLOUDY} photoionization code \citep{Ferland:1998} to model the star-forming galaxies as spherically symmetric gas surrounding single ionizing sources (the real geometry of each system may be more complex than this simple assumption).  Our aim is to model both the ionized and neutral gas in order to understand the ionization structure along a single line of sight towards the galaxy.  Models were run over a grid of metallicities (0.02, 0.5, 1.0, and 4.0~Z/\Zsol) and neutral hydrogen column densities (log[$N$(\hi)/cm$^{-2}$] = 18.2, 18.7, 19.1, 19.7, 20.2,  20.7, 21.1, and 21.7) which encompass the range of metallicities and neutral hydrogen column densities measured within our sample.  For each combination of Z and $N$(\hi) over the grid, we modeled the galaxy as a spherical volume of constant density gas, over a large range of volume densities ($-6\,<\,$log\,[$n$(H)/cm$^{-3}]\,<\,6$).  The gas is ionized by a central star at a certain distance from the cloud with an effective temperature of 36,000~K (the average estimated temperature for the stellar population targeted in our sample - see Section~\ref{sec:hi_colDens}) and a UV luminosity of log[L$_\star$]\,=\,40.05 erg/s (derived from the average number of counts measured within the COS aperture on the ACS/SBC images; see Fig.~\ref{fig:COS_targets}).  The magnitude and shape of the ionizing spectrum was found to have a minimal effect on the relative mixture of neutral and ionized gas within our models, when compared to the effects of differing metallicity and column density of neutral hydrogen.  Models were also run for a variety of radii from the central ionizing source (0.1-100 pc).  Significant differences were only seen when the radii were large enough for the cloud to be considered as a plane-parallel slab, i.e., consisting of only neutral gas and therefore not applicable to the galaxies within our sample. A radius of 10 pc was ultimately selected for our modeling.

The modeling process consisted of two phases.  In the first phase, for each volume density considered and for each combination of Z and $N$(\hi) over the grid, the models are run and stopped once the desired \hi\, column density is reached, at which point we output the simulated ion column densities of the sphere.  By stopping the models at the desired $N$(\hi) (rather than a fixed radius) we ensure that they properly account for the correct amount of neutral gas along the line of sight, including all neutral gas outside the ionized gas volume.  

In order to proceed with the second phase, one needs to pinpoint the appropriate volume density of the gas. This is routinely done by using the ratio of successive ion stages.  One of the best indicators in this sense is the \sIiii/\sIii\, ratio. However, this ratio cannot be used for the galaxies in our sample due to saturation effects affecting both \sIii\, and \sIiii\, determinations. Another indicator of the gas volume density is the \feiii/\feii\, ratio, but due to the low redshift of our galaxies, the \feiii~$\lambda$1122 absorption line is only available in the spectrum of SBS~0335-052, giving in this case a ratio log[$N$(\feiii)/$N$(\feii)]\,=\,$-1.08\pm0.08$. In light of the above, we decided to take a conservative approach. We assumed the maximum \feiii/\feii\, ratio possible as determined by the preliminary modeling in the first phase, and run models which represent the `worst-case scenario' for a maximum amount of ionized gas along the line of sight for each combination of $N$(\hi) and metallicity. The results of this modelling are the column density of each element associated with both the ionized and neutral gas, i.e., $N$(X)$_{HI+HII}$, and the determination of the location of the photodissociation region (PDR), the latter obtained by plotting the ionization stages as a function of depth through the cloud and assessing the depth at which the ionic fraction of \hii\, is equal to that of \hi.

The second phase of the modeling process consisted of running the `worst-case scenario' model again using the same input parameters, but stopping the depth of the model at the location of the PDR for each combination of N(\hi) and metallicity.  This allowed us to estimate for each element the column density associated with the ionized gas only, i.e., $N($X$)_{HII}$. Using Equation~\ref{eq:1}, we can then assess the magnitude of the column densities arising from the neutral gas only, $N$(X)$_{HI}$: 

\begin{equation}
N(X)_{HI} = {N(X)_{HI+HII}}-{N(X)_{HII}} . \label{eq:1}
\end{equation}

The relative amount in logarithmic scale of the column density of the total (neutral+ionized) gas $N$(X)$_{HI+HII}$ (as seen from the outer edge of the gas cloud) compared to the column density of the neutral gas only $N$(X)$_{HI}$, gives the ionization correction factor ICF$_{ionized}$. This is the amount of column density $N(X)_{HII}$ that is within the ionized gas as seen from the edge of the PDR: 

\begin{equation}
ICF(X)_{ionized} = log[N(X)_{HI+HII}] - log[N(X)_{HI}] . \label{eq:icf1}
\end{equation}

\begin{figure*}
\center
\includegraphics[scale=1]{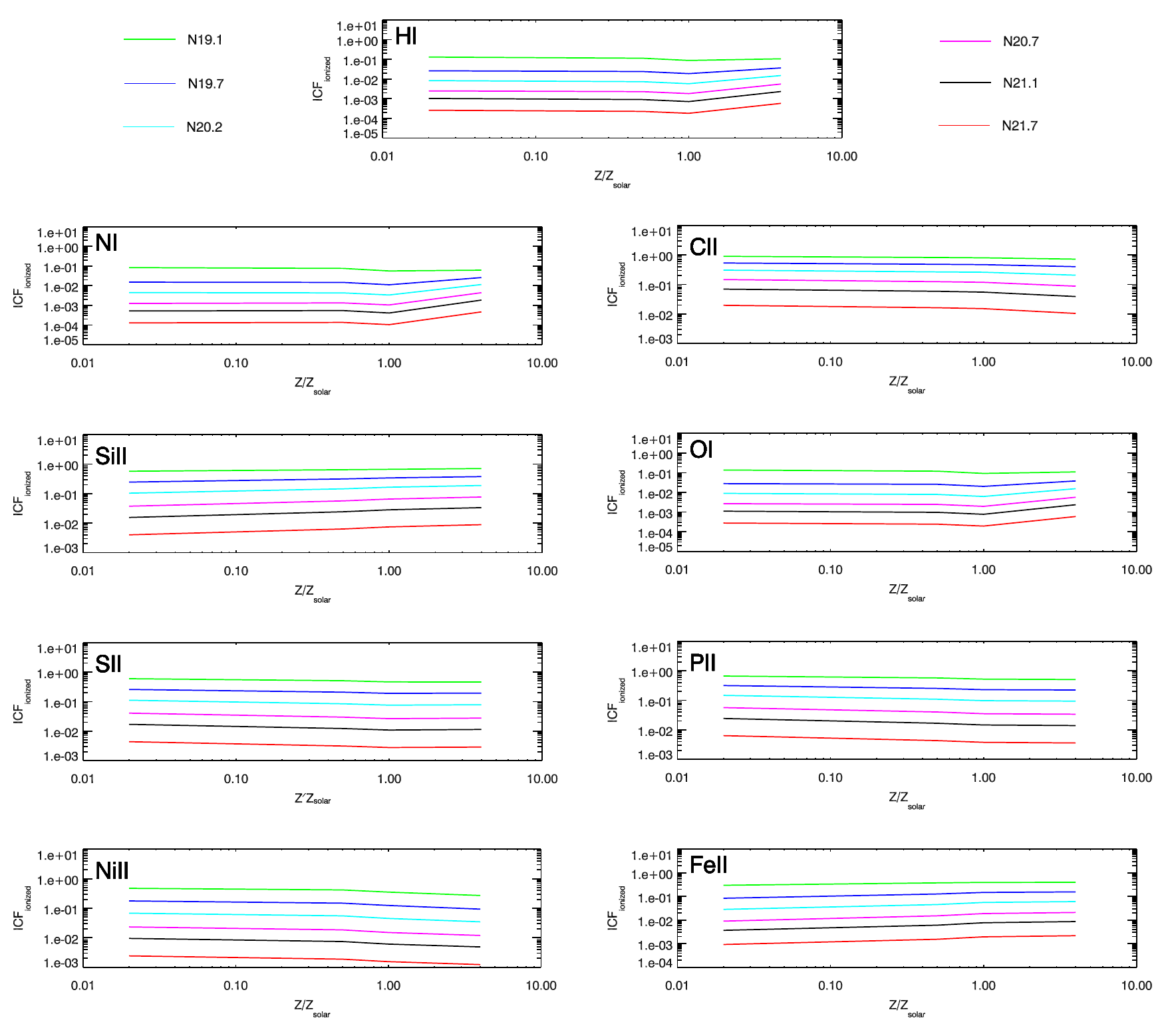}
\vspace{0.2cm}
\caption{Ionization corrections due to contaminating ionized gas along the line of sight for the dominant ions within the neutral gas of our sample of SFGs. The ICF$_{ionized}$ values represent the amount of those ions locked into the ionized gas in logarithmic scale. These values were calculated as the average between the `worst-case scenario' of maximum amount of ionized gas and the `best-case scenario' of no ionization along the line of sight for each combination of $N$(\hi) and metallicity by using the \textsc{CLOUDY} models described in Section~\ref{sec:ioncorr} and equation~\ref{eq:icf1}.Results are shown for models with metallicities of 0.02, 0.5, 1.0, and 4.0~Z/\Zsol\, and neutral hydrogen column densities of log[$N$(\hi)/cm$^{-2}]\,$= 19.1, 19.7, 20.2, 20.7, 21.1, and 21.7 (the models with log[$N$(\hi)/cm$^{-2}$] = 18.2 and 18.7 were fully ionized in the `worst-case scenario' and were not considered further).}
\label{fig:cloudy_models_ionized}
\end{figure*}

The ionized-gas ICFs obtained in this way are upper limits associated to the `maximum' \feiii/\feii\ ratio possible as suggested by the modelling itself. 
In order to estimate the final ICFs, we averaged these upper limits with what would be the `best-case scenario' of no ionization along the line of sight (no ICF corrections).  Fig.~\ref{fig:cloudy_models_ionized} shows the trends of these `average' ICF$_{ionized}$ values as a function of metallicity and for specific values of log[$N$(\hi)]. Only the dominant ions within the neutral gas phase are considered in these plots. By interpolating between the grid of `average' ICFs in metallicity and $N$(\hi) space, we derived the ICFs specific to each galaxy. In this case, the errors on the final ICFs are given by the difference between the `average' and the `maximum' ionization ICFs and are equal to the average. The final ionized-gas ICFs and associated errors adopted in our analysis are listed in the first part of Table~\ref{tab:ICFs}.

Our models show that the amount of correction required is small and comparable to the 1\,$\sigma$ errors for galaxies with log\,[$N$(\hi)/cm$^{-2}$]\,$\,\gtrsim\,$\,21 if Z/\Zsol\,$\lesssim\,0.5$, a finding in agreement with models of metal-poor DLAs by \cite{Cooke:2011}.  This correction becomes more important as the column density gets lower and the metallicity higher, i.e., for log\,[$N$(\hi)/cm$^{-2}$]$\,\lesssim\,$21 and  Z/\Zsol\,$\gtrsim\,0.5$,  to the point  that for the lowest \hi\, column densities considered, our `worst-case scenario' models predict that only ionized gas remains within the system. This is the reason why the models at log[$N$(\hi)/cm$^{-2}$] = 18.2 and 18.7 are not plotted in Fig.~\ref{fig:cloudy_models_ionized} and no ionized-gas ICFs are given for position 2 of M83 in Table~\ref{tab:ICFs}. It should be noted, however, that for this sightline we observe \Lya\ and many other neutral species in absorption in the COS spectrum, so the lack of neutral gas in the `worst-case scenario' model is just the result of our cautious approach.

There are two main caveats to the approach presented here.  Firstly, the models assume that the same abundances apply to both the ionized and neutral gas.  Secondly, the models assume that the solar abundance ratios apply. Both of these assumptions are routinely made in these types of studies and, despite the complexity of our models, we are aware that they may not be fully representative of the real situation which may be far more complex.  These two assumptions may not hold if, for example, chemical inhomogeneities exist where the galaxy halo has a different chemical imprint due to inefficient mixing.  

Our models are analogous to those adopted in the study of DLAs with the exception that in the latter case the ionized gas (clearly present in our study since the background source is a hot UV-bright target within our star-forming galaxies) is not considered during the modeling phase. This is because the DLAs are considered as a plane-parallel slab of purely neutral gas illuminated by a background source (e.g., QSO). The existence and amount of ionized gas (including ICF corrections) is then inferred by comparing the model input abundances to the observed abundances.

An alternative method to estimate the amount of ionized gas contaminating the absorption spectra, is to assess the ionization fractions within the ionized gas, i.e. $X^j:X^i$ (where $i>j$) for each element $X$ with $X^i$ only present in the ionized gas phase and $X^j$ present in both the neutral and ionized gas phases, and use $N(X^i)$ to calculate the magnitude $N(X^j)$ contaminating the neutral gas column densities \citep[see][]{Semback:2000}.  However, in our case we do not have any element where we detect the ions at the higher ionization state needed for this kind of correction (except for \feiii~and \feii~in SBS~0335-052).  This is the reason why we adopted the approach of using \textsc{CLOUDY} modeling to assess the ionization corrections needed for \textit{all} the ions measured in our spectra.  Despite the differences in approach, both methods yield similar answers.  For example, \citet{Sembach:2000} predicts that the $N$(\sii)$_{HII}$/$N$(\sii)$_{HI}$ ratio can be 0.3--0.8 for log[$N$(\hi)/cm$^{-2}$]\,=\,18.8--19.7, which is not inconsistent with our predictions of $\gtrsim$\,0.2 for galaxies within a similar $N$(\hi) range.  It should be noted that all our galaxies within this $N$(\hi) range have super-solar metallicities (i.e. Z/\Zsol$\,>$\,1), whereas the models of \citet{Sembach:2000} are for Z/\Zsol$\sim$\,0.3.


\paragraph{``Classical" Ionization Corrections in the Neutral Gas}

The \textsc{CLOUDY} models were also used to check the magnitude of the ``classical" ionization effects due to the lack of sampling of absorption lines relative to higher ionization stages of a certain element in the neutral gas. This was accomplished by looking at the ratio of ions in the radial plots of the models and by checking the relative column densities of a higher ionization state compared to the dominant lower ionization state of a certain element in the neutral gas only. Since the total amount of a certain element X is given by the following equation:

\begin{equation}
N(X) = N(X^i)+N(X^{i+1}), \label{eq:2}
\end{equation}

\noindent
where $N(X^i)$ is the dominant stage of ionization in the neutral gas for element $X$, the ionization correction is defined as the following in logarithmic scale:

\begin{equation}
ICF(X)_{neutral}=log[1+\frac{N(X^{i+1})}{N(X^i)}].\label{eq:icf2}
\end{equation}

\begin{figure*}
\center
\includegraphics[scale=1]{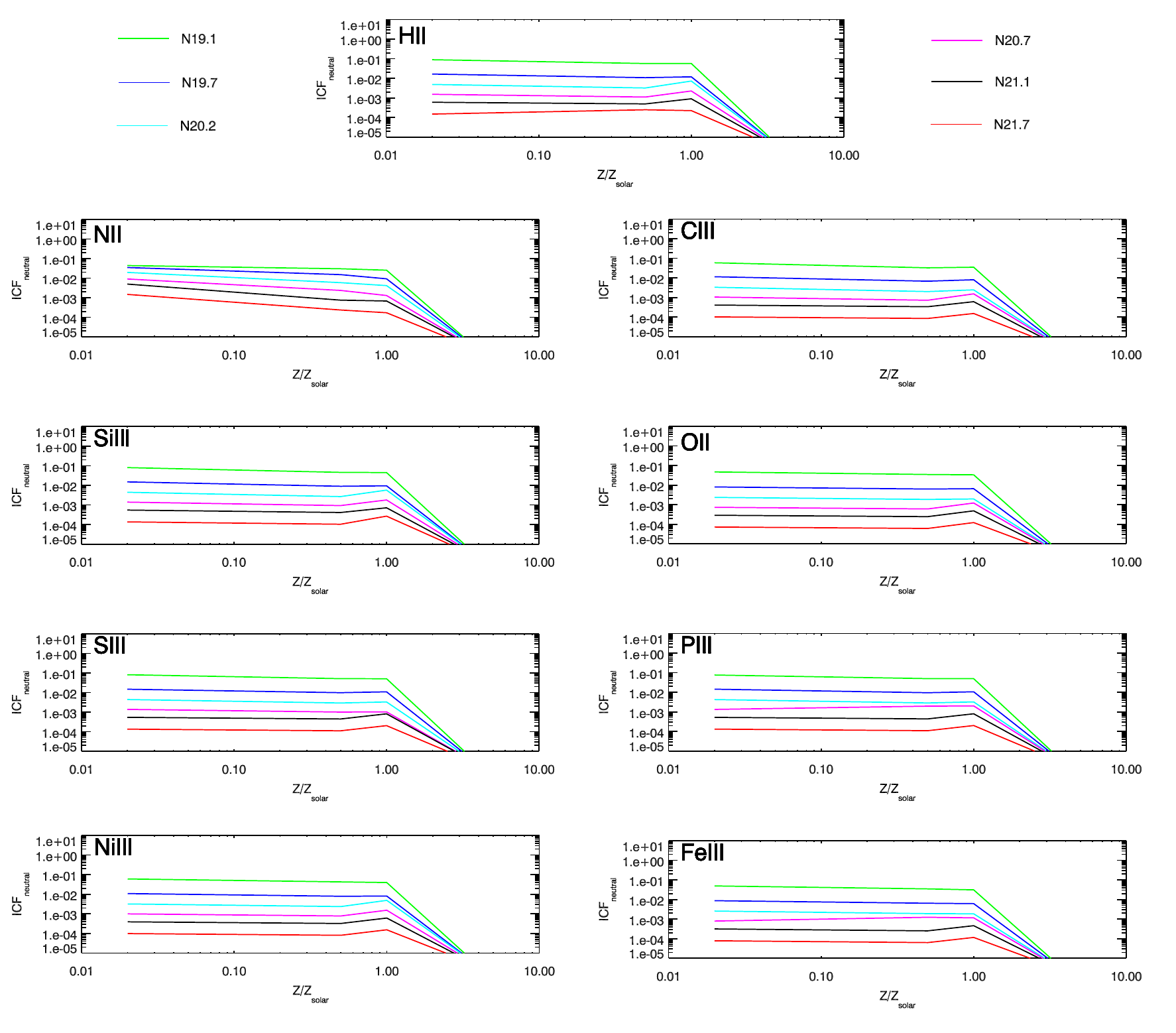}
\vspace{0.2cm}
\caption{``Classical" ionization corrections due to the lack of sampling of absorption lines relative to higher ionization states of a certain element in the neutral gas. The ICF$_{neutral}$ values represent the amount of a certain species locked into higher ionization states compared to the ones that dominate the cold gas in logarithmic scale.  These values were calculated as the average between the `worst-case scenario' of maximum amount of ionized gas and the `best-case scenario' of no ionization along the line of sight for each combination of $N$(\hi) and metallicity by using the \textsc{CLOUDY} models described in Section~\ref{sec:ioncorr} and equation~\ref{eq:icf2}. Results are shown for models with metallicities of 0.02, 0.5, 1.0, and 4.0~Z/\Zsol\, and neutral hydrogen column densities of log[$N$(\hi)/cm$^{-2}]\,$=  19.1, 19.7, 20.2, 20.7, 21.1, and 21.7 (the models with log[$N$(\hi)/cm$^{-2}$] = 18.2 and 18.7 were fully ionized in the `worst-case scenario' and were not considered further). Only the species affected by this correction are represented in this plot.}
\label{fig:cloudy_models_neutral}
\end{figure*}

Similarly to the ionization corrections due to contaminating gas along the line of sight, we run the `worst-case scenario' models corresponding to the maximum amount of ionized gas for each combination of metallicity and $N$(\hi), and averaged these upper limits with the values from the `best-case scenario' of no ionization (i.e., no ICF corrections). In Figure~\ref{fig:cloudy_models_neutral} we plot these  `average' ICF$_{neutral}$ values as a function of metallicity and for specific values of log[$N$(\hi)]. Only the higher-ionization states of the dominant ions within the neutral gas phase are considered in these plots. By interpolating between the grid of `average' ICFs in metallicity and $N$(\hi) space, neutral-gas ICFs specific to each galaxy were derived. In this case, the errors on the final ICFs are given by the difference between the `average' and the `maximum' ionization ICFs and are equal to the average. The final neutral-gas ICFs and associated errors adopted in our analysis are listed in the second part of Table~\ref{tab:ICFs}.
We find that no corrections are needed for the majority of the targets, except for those where log\,[$N$(\hi)/cm$^{-2}$]\,$\,\lesssim\,$\,20. The PDR is defined as the point at which the ionic fraction of \hi~is the same as the ionic fraction of \hii, and by this point onwards all the photons able to photoionize, e.g., \feii, \sii, or \sIii, have already been absorbed, so that the transition from, e.g., \sIiii~to \sIii~occurs just before the PDR, leaving negligible column densities of \sIiii~within the neutral gas.  As for the ionized-gas ICFs, since the models at log[$N$(\hi)/cm$^{-2}$] = 18.2 and 18.7 are fully ionized, they are not plotted in Fig.~\ref{fig:cloudy_models_neutral} and no neutral-gas ICFs are given for position 2 of M83 in Table~\ref{tab:ICFs}.

\subsubsection{Dust Depletion}
The presence of dust in the ISM can represent another serious complication in the interpretation of the metal abundances.  Refractory elements (e.g., Fe and Si) are more easily locked into dust grains than non-refractory ones (e.g., O and N).  This can clearly alter the relative heavy-element abundances.  Observations of the local ISM provide hints of selective depletions acting in dense clouds within our Galaxy \citep{Savage:1996}.  There is evidence of dust also in the DLAs, systems that many of our star-forming galaxies resemble with their relatively high \hi\, column densities.

In order to quantify this effect, following the methodology used in DLAs (e.g., see \citet{Cooke:2011} and references therein), the relative abundances of a refractory element and a volatile element coming from the same stellar processes (in order to avoid biases introduced by the specific star-formation history of a certain stellar system) need to be compared to the expected intrinsic nucleosynthetic ratio.  The best species for this kind of comparison are the Fe-peak elements Zn and Cr from SNe\,Ia, since the first element is only mildly depleted, while the latter is more refractory. Unfortunately there are no Cr\,{\sc ii}  or Zn\,{\sc ii}  lines within the COS wavelength range for this kind of comparison. We could then resort to Si and S, which are both $\alpha$ elements from SNe\,II known to be depleted to slightly different degrees (Si is more easily incorporated into dust grains than S). However, since all of our \sIii\, absorption profiles suffer from saturation effects, and thus only provide lower limits on $N$(\sIii), we can only provide an upper limit for [S/Si]. An upper limit in the range $0.3-0.7$ was estimated for the [S/Si] in our sample, with an average upper limit of $\sim 0.5$. For comparison, \citet{Savage:1996} measure depletion within the cool diffuse clouds to be [S/Si]=1.21 towards $\zeta$-Oph and [S/Si]\,$>\,$0.07 towards $\psi$-Persei, whereas towards the warm halo or the cool dense clouds the [S/Si] ratio ranges from 0.3 to 1.3.  Additionally, \citet{Prochaska:2002} suggest that lightly depleted regions of the ISM would have [S/Si]$\, \sim \,$0.2. Due to the fact that we only have an upper limit for [S/Si], we are unable to draw any conclusions from such comparisons. Whilst we expect dust to be an issue within our galaxies, we cannot constrain the amount of dust depletion present and therefore cannot make any corrections for it.


\subsection{Fine-Structure Lines}\label{sec:fine_structure}
Within our spectra, we observe a selection of fine-structure lines, which arise from fine-structure to ground state transitions (such as \oi*, \sIii*, and \cii*) .  The excitation of fine-structure levels can occur via three main mechanisms: collisions with electrons, radiative pumping by the local radiation field or excitation by photons from cosmic microwave background radiation \citep[although the latter mechanism is not so important for the Si$^+$ fine-structure levels as they are too far apart; ][] {Silva:2002}.  Much attention has been paid to these transitions as ratios with their ground-state counterparts (e.g., $N$(\sIii*)/$N$(\sIii)) can be used to determine the physical conditions of the gas from which the lines originate \citep{Bahcall:1968,Silva:2002,Savaglio:2004,Berger:2005,Howk:2005,Jenkins:2011}.  According to \citet{Bahcall:1968}, if observations suggest that the higher fine-structure states of a ground-state multiplet are occupied, then there are two conditions that the gas will meet.  Either the density of the absorbing material is large enough that rates for collisional excitation of fine-structure levels are greater than the photon decay rates, or the absorbing material is being subjected to a strong photon flux which populates the higher fine-structure states.  The presence of such lines is therefore an important indicator of the physical conditions of the gas. Here we describe the information gained from three such lines that could be studied within our COS spectra, i.e., \cii*, \oi* and \sIii*.

The \oi*~$\lambda$1304.85 absorption line originates from the middle ground-state fine-structure level of \oi, connected to the bottom fine-structure level by the 63.2~$\mu$m \foi\, line.  To get the middle level populated enough to see this line requires fairly dense, relatively warm (100-200~K) gas \citep[e.g., ][]{Morton:1975}.  Due to its blend with \sIii~$\lambda$1304.37, it is important to determine if \oi* is present as it must be taken into consideration when constraining the \sIii\, line.  One way to do this is assessing the presence of  \oi**~$\lambda$1306.03, which is connected to the \oi* line. \oi** originates from the topmost of the three ground state fine-structure levels of \oi\, and is connected to the middle fine-structure line by the 146~$\mu$m \foi\, line.  The additional excitation to get level three populated is not much more than that required to get level two populated, from which the \oi* line originates, and thus the two lines are often seen as a pair.  For example in \citet{Dinerstein:2006} both \oi* and \oi** are seen in absorption in a photo-dissociated region around a planetary nebula. Similarly \citet[][their Appendix B]{Jenkins:2011} see both lines towards the star HD~210839 and use them to deduce a gas temperature of $\sim$400~K.  On inspection of our data we see no evidence of \oi**~$\lambda$1306, which lies in an uncontaminated region of the spectra in all cases.  In addition to this, if we were to attribute the absorption around 1304~\AA\, to this line, the line would be systematically offset from the other absorption lines by $\sim$100--150~\kms.  The lack of \oi* and \oi** is to be expected as they are rarely detected in Galactic ISM sightlines. In particular, ruling out the presence of the \oi* line is extremely important. This line, if present, would contaminate \sIii~$\lambda$1304, the weakest \sIii\, line within our wavelength range and thus the most suited to better constrain an upper limit for the \sIii\, column density, considering that all the other \sIii\, lines are much stronger and suffer even more from saturation effects. 

\sIii* is detected in all of our targets, apart from the three most metal-poor targets (e.i., I~Zw~18, SBS~0335-052, SBS~1415+437) and NGC~3690, where it cannot be constrained due to contamination.  Whilst its detection is no surprise, having been detected in varying strengths in UV spectra of gamma-ray bursts \citep{Savaglio:2004}, damped \Lya\, systems \citep[e.g., ][]{Howk:2005} and Lyman Break Galaxies \citep[e.g.,][]{Pettini:2002}, its strength compared to \sIii\, is of upmost importance.  \sIii* absorption is associated with the local environment of the starburst, since the presence of this state requires large densities or intense IR/UV radiation fields, which are not common in interstellar environments \citep{Berger:2005}. E.g., ionized gas along the line of sight of QSOs has been associated to a relative strength of the \sIii* and \sIii\, column densities of $-2.36 < N$(\sIii*)/$N$(\sIii)$ <-1.58$ for a covering factor in the range 0.3--1.0 \citep{Srianand:2001}. A value of  $N$(\sIii*)/$N$(\sIii)$\,\sim -1.7$ has instead been associated to high-density neutral gas in a GRB afterglow \citep{Vreeswijk:2004}. Only an upper limit to the relative strength of the \sIii* and \sIii\, column densities can be inferred for each galaxy in our sample due to the lower limit values affecting the determination of $N$(\sIii) because of saturation effects in the \sIii\, lines. No conclusive results can thus be drawn related to the real physical conditions of the gas from which \sIii* originates in our case. An upper limit in the range between $-2.28$ and $-0.87$ was estimated for the $N$(\sIii*)/$N$(\sIii) in our sample, with an average upper limit of $\sim 1.5$. 



Unlike \sIii*, \cii* may arise in a cold ($T\sim100$~K ) or warm ($T\sim8000$~K )  neutral medium or in the ionized gas ($T\sim 10,000$~K ) \citep{Wolfe:2003,Srianand:2005}.  This is because the upper fine-structure levels in \sIii\, and \cii\, have excitation energies that differ by a factor of 4.  As with \sIii, the ratio of the excited- to ground-state column densities of C$^+$ can also be used to probe the physical properties of the absorbing gas. However, the measurement of $N$(\cii) in all the galaxies of our sample is not possible as \cii~$\lambda$1344 is highly saturated, deeming the $N$(\cii*)/$N$(\cii) ratio unusable for the purpose of probing the real conditions of the gas from which \cii* originates. 



\section{Results}\label{sec:results}
For each ion considered in the spectra of each galaxy of our sample, Table~\ref{tab:Nvals}  lists the average redshift of the transitions identified, the measured column density and $b$ parameter, and the lines used in the line-profile fitting (a non entry for the error in the $b$ parameter of a certain ion indicates that this parameter could not be constrained from the fit and needed to be assumed from another ion with similar kinematics properties).

For each galaxy, Table~\ref{tab:abundances} summarizes the column densities of ions representative of the total amount of an element in the neutral phase of the ISM, and the corresponding abundances relative to $N$(H) in logarithmic scale. Abundances are listed before and after applying the ICFs listed in Table~\ref{tab:ICFs}  with the errors propagated in quadrature. Abundances without and with ICF corrections  (except for position 2 in M83 where no entries corrected for ionization have been considered) are also listed relative to the solar abundances, by adopting the standard solar photospheric abundances published by \citet{Asplund:2009}, and the formula $[X/H]=$log$(X/H)-$log$(X/H)_{\odot}$. 
Errors on [X/H] reflect measurement errors in the column density $N$ as inferred from the line-profile fitting, since errors due to the continuum determination and/or the presence of stellar absorption in \Lya\, are negligible, errors on the solar abundance determinations,  and errors on the ICF corrections when relevant.  Column densities and abundances derived from absorption lines thought to be affected by saturation (both `classical' and `hidden', as discussed in Section~\ref{sec:saturation}) are actually to be considered as lower limits. 

\section{Individual Objects}\label{sec:objects}
In this section we briefly comment on the properties of each star-forming galaxy analyzed in this paper and on the characteristics of its neutral ISM as inferred from the absorption lines in its COS spectrum. 

\subsection{I~Zw~18}

With a metallicity of $\sim$\,1/30, I~Zw~18 is an object well-known as having the lowest metal content measured in the local Universe.  The system comprises two separate stellar components that share the same \hi\, envelope, the so-called main and secondary body \citep{Zwicky:1966}. The main body itself consists of two star-forming regions, with a brighter northwest component (containing the bright UV source targeted within this study) separated from a fainter southeast component by $\sim8''$.  Due to its distance of $\sim$\,18 Mpc, much debate has surrounded I~Zw~18 regarding its true age, with ages $<$40~Myr being predicted due to its record low oxygen abundance \citep{Izotov:1999}.  However, the clearest insight into the age of I~Zw~18 has been through \textit{HST} resolved stellar populations studies.  The most recent HST/ACS imaging by \citet{Aloisi:2007} confirmed the presence of a Red Giant Branch (RGB) with stars older than $\sim 1$ Gyr and thus ruled out the possibility that I~Zw~18 is a truly primordial galaxy formed recently in the local Universe.

\begin{figure*}
\center
\includegraphics{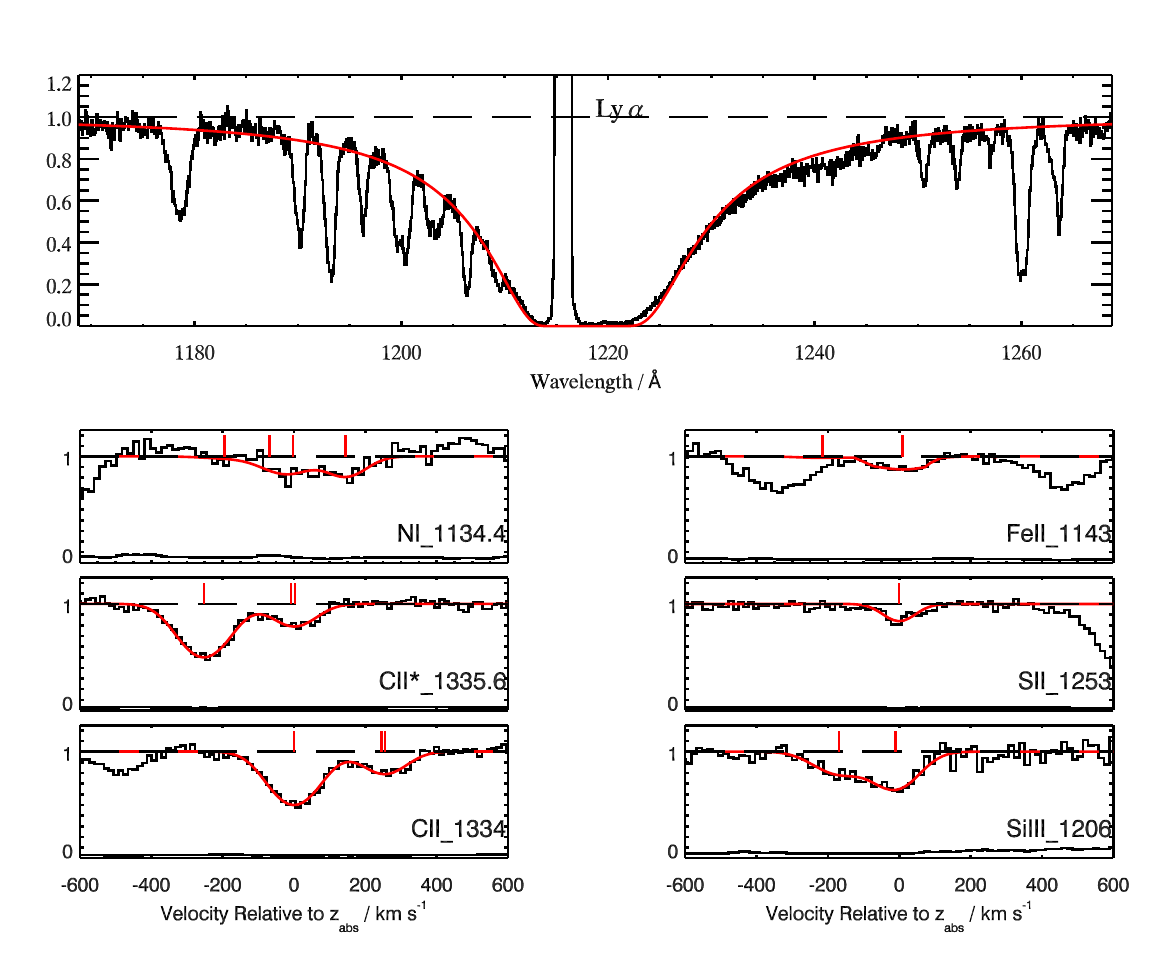}
\vspace{0.2cm}
\caption{Absorption-line profiles in I~Zw~18: observed spectrum (black histogram) and theoretical line profile (red solid line). The top panel shows the \Lya\, profile at $z_{abs}=0.0025575$ ($v = 767$ \kms).  The remaining panels display a selection of metal lines from within the spectrum, overlaid with theoretical line profiles whose model parameters are given in Table~\ref{tab:Nvals}. For all panels, the $y$-axis scale is the residual intensity after continuum normalization. The normalized continuum is shown instead by the black dashed line. The noise per pixel is also shown by the thin-line histogram at the bottom of the spectra (except for \Lya~where the errors are not plotted). The red tick marks above the normalized continuum indicate the locations of all the absorption lines considered and their multiple components when more were deemed necessary, with the label within each panel referring to the absorption of interest at zero velocity as inferred from the observed wavelength from the line-profile fitting. The strongest component of this absorption is plotted at zero velocity when multiple components are fitted.}
\label{fig:IZW18_spec}
\end{figure*}

I~Zw~18 has been studied extensively in the UV, firstly using GHRS  \citep{Kunth:1994} and more recently using FUSE \citep{Aloisi:2003}.  Both of the previous UV studies considered a foreground \hi\, column density in the direction of I~Zw~18 due to the Milky Way of log[$N$(\hi)] = 20.30  cm$^{-2}$ \citep{Stark:1992}.  In addition to the interstellar absorptions at the systemic velocity of I~Zw~18 of $\sim 760$~\kms, both studies witnessed absorptions at velocities around $\sim -160$~\kms corresponding to a \hi\, high-velocity cloud (HVC) in the direction of I~Zw~18 detected by \citet{Hulsbosch:1988}, \citet{Stark:1992}, and \citet{Hartmann:1994}. Both studies assumed a \hi\ column density of log[$N$(\hi)] =19.32 cm$^{-2}$ for this cloud. Figure~\ref{fig:IZW18_spec} (top panel) shows the \Lya\, absorption-line profile for I~Zw~18.  Whilst we consider the same column density adopted in previous studies for the HVC component, here we adopt a MW column density of log[$N$(\hi)] = 20.49 cm$^{-2}$ based on the latest \hi\ Leiden/Dwingeloo survey \citep{Hartmann:1997}.  Keeping the column density of the MW profile fixed, we estimate a column density of log[$N$(\hi)] = $21.28\pm0.03$ cm$^{-2}$ for the \hi\, within I~Zw~18.  

\begin{figure*}
\center
\includegraphics{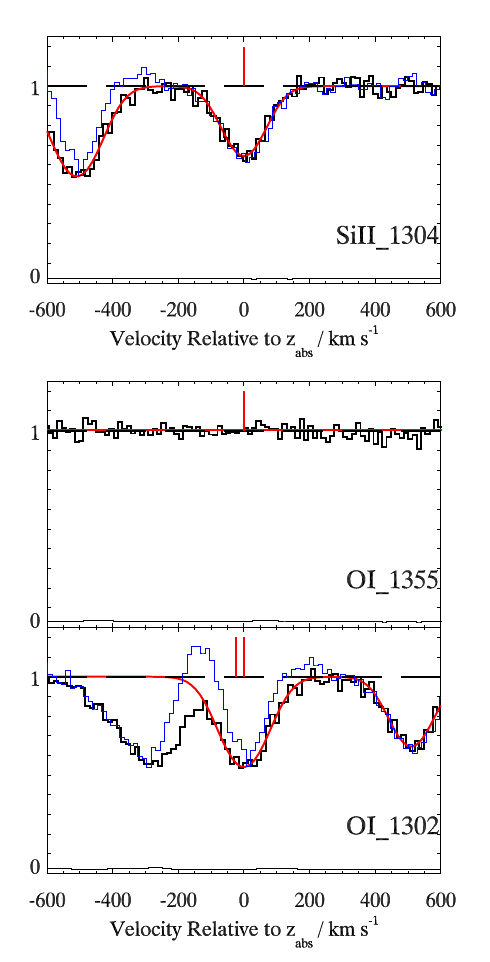}
\vspace{0.2cm}
\caption{Absorption-line profiles of \sIii~$\lambda$1304 (top panel) and \oi~$\lambda$1302 (bottom panel) in I~Zw~18 in a spectral region affected by geocoronal contamination. The region of the extremely faint \oi~$\lambda$1355 not affected by geocoronal contamination is also shown (middle panel). The thin blue histogram is the original spectrum, while the thick black histogram is the observed 'night' spectrum with geocoronal contamination removed. The theoretical line profile as fitted in the 'night' spectrum is plotted as the red solid line (the \oi~$\lambda$1302 profile also includes contamination by \pii~$\lambda$1301, which is deemed negligible due to the faintness of the line). For all panels, the $y$-axis scale is the residual intensity after continuum normalization. The normalized continuum is shown instead by the black dashed line. The noise per pixel is also shown by the thin-line histogram at the bottom of the spectra (except for \Lya~where the errors are not plotted). The red tick marks above the normalized continuum indicate the locations of all the absorption lines considered and their multiple components when more were deemed necessary, with the label within each panel referring to the absorption of interest at zero velocity as inferred from the observed wavelength from the line-profile fitting. The strongest component of this absorption is plotted at zero velocity when multiple components are fitted.} 
\label{fig:IZW18_OIspec}
\end{figure*}

A selection of the metal lines seen within the COS spectrum of I~Zw~18 are shown below the \Lya\, profile in Fig.~\ref{fig:IZW18_spec}.  Due to the low metallicity of this object, some species that are common to other galaxies within the sample (such as \pii, \mnii\,, and \nIii) were not present.  However, we were able to constrain all other high S/N lines, each best fitted with a single velocity component profile at an average redshift  z$_{abs}=0.0025464$ ($v\sim$760~\kms).  This is in agreement with the radial velocity of the stellar population (as measured from the \ciii~$\lambda$1176 photospheric absorption line), which also lies at  $v\sim$760~\kms. The higher ionization transition \sIiii~$\lambda$1206 which originates predominantly in the ionized gas, was detected in I~Zw~18 with two separate velocity components. The first velocity component (the only one listed in Table~\ref{tab:Nvals}) aligns with that of the ions arising from the neutral ISM (e.g., \oi\ and \Ni), suggesting a mix of ionized and neutral gas within the same absorbing cloud. Since \sIii* was not detected within the spectrum, we are not able to better constrain the physical conditions of this more highly ionized gas along the sightline. The second velocity component of \sIiii\, at a similar $b$ parameter and slightly lower column density, is instead blueshifted by $\sim 160$~\kms, suggesting an outflow of ionized gas or an additional \hii\, region along the sightline to I~Zw~18. All $b$ parameters from the measured absorption line profiles are in good agreement with one another ($\sim$80-90~\kms), suggesting that turbulent motion is dominating within this gas compared to thermal broadening.  Some amount of hidden saturation was affecting the strongest \feii\,$\lambda$1144 line, and one of the weakest lines (\feii\,$\lambda$1143) was used to constrain the ion column density, which turned out to be higher by 0.29~dex (Table~\ref{tab:saturation}). Due to geocoronal contamination in the region of \oi~$\lambda$1302 and \sIii~$\lambda$1304, it was necessary to fit these two lines in the `night' spectra  (see Fig.~\ref{fig:IZW18_OIspec}).

Comparing our results with those reported by \citet{Lebouteiller:2013} and obtained from our (HST GO program 11579; PI A.~Aloisi) and additional (HST GTO programs 11523 and 12028; PI J.~Green) COS observations, we find that on average our column densities are in agreement with those published in the above mentioned paper within the uncertainties (e.g., \sIii, \sii, \cii*, \Ni, and \feii), except for those ions that have only one saturated line in our COS spectra and for which we have only estimated a lower limit (i.e., \cii~and \oi).  

\subsection{SBS~0335$-$052}

Along with I~Zw~18, SBS~0335-052 is known as one of the most metal-poor galaxies in the local Universe.  SBS~0335-052 was found by \citet{Pustilnik:2001} to consist of two prominent \hi\, peaks separated in the east-west direction by 84$''$ (22~kpc at an assumed distance of $\sim$54~Mpc), a separation large enough for them to be treated as two distinct BCDs: SBS~0335-052E (considered here and simply called SBS~0335-052 hereafter for simplicity) and SBS~0335-052W.  The combination of SBS~0335-052 low metallicity ($\sim$\,1/25~\Zsol\,) and undetectable underlying old stellar population from HST imaging  suggests that this may be a young galaxy undergoing its first burst of star formation \citep{Thuan:1996}.  However, at the distance of this galaxy the detection of an RGB is almost impossible with current imaging capabilities and thus its `young' status remains unconfirmed.  

\begin{figure*}
\center
\includegraphics{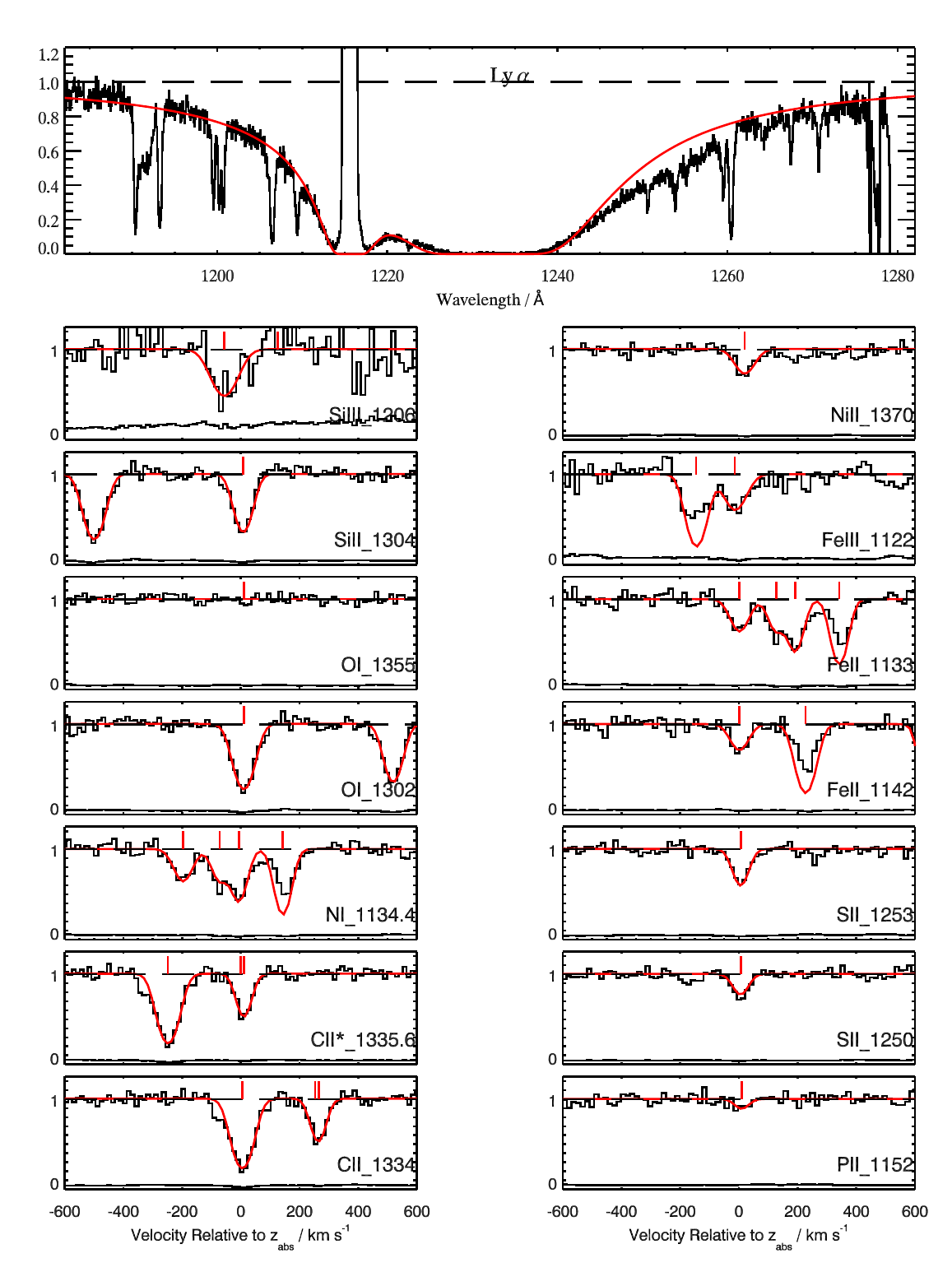}
\vspace{0.2cm}
\caption{Same as Fig.~\ref{fig:IZW18_spec} but for absorption-line profiles in SBS~0335-052.  The top panel shows the \Lya\, profile at $z_{abs}=0.0134685$ ($v = 4,038$ \kms).} 
\label{fig:SBS0335_spec}
\end{figure*}

The relatively high redshift of SBS~0335-052 ($z=0.0134685$ or $v\sim4038$~\kms, the highest value among the galaxies covered within this study), allowed for the separation of the Milky Way and galactic \Lya\, absorption-line profiles, as shown in the top-panel of Figure~\ref{fig:SBS0335_spec}.   We estimate log[$N$(\hi)] = $21.70\pm0.05$ cm$^{-2}$ for \hi\, within SBS~0335-052 and log[$N$(\hi)$_{MW}$] = $20.4\pm0.1$ cm$^{-2}$ for the MW \hi\, contribution along the same sightline. The \hi\, envelope of SBS~0335-052 has been previously studied using the large ($30''\times30''$) FUSE aperture by \citet{Thuan:2005sbs}, which found a slightly larger \hi\, column density of log[$N$(\hi)] = $21.86^{+0.08}_{-0.05}$ cm$^{-2}$.  The velocity of the stellar population, $\sim$4,070~\kms (as measured from the \ciii~$\lambda$1176 photospheric absorption line), is slightly redshifted from that of the \hi\, gas by $\sim$\,30~\kms, implying that the gas is outflowing compared to the UV background source.

A selection of metal lines seen within the COS spectrum of SBS~0335-052 are shown below the \Lya\, profile in Fig.~\ref{fig:SBS0335_spec}.  Each absorption line was best fitted with a single velocity component profile (overlaid).  The average redshift of all absorption profiles is z$_{abs}=$0.0134907 ($v\sim$4,044~\kms). This was sufficient enough to shift  \oi~$\lambda$1302 and \sIii~$\lambda$1304 absorption lines out of the geocoronal contamination and we were therefore able to use the original `day+night' spectrum to fit these lines.  We were able to constrain all the expected ions (i.e., those seen within the other spectra in this study), with the exception of \sIii* (its strongest line \sIii*~$\lambda$1264 was redshifted into the spectral break, while the next available strongest line \sIii*~$\lambda$1194 was not detected).  All $b$ parameters are in relatively good agreement with each other, at $b \sim$ 30-40~\kms.  \sIiii, which originates predominantly in ionized gas, displays a slightly higher than average $b$ parameter at $\sim$\,55~\kms, along with a velocity blueshifted by $\sim$\,55~\kms, which suggests that the ionized gas cloud is separate from the neutral ISM.   On account of the large redshift of this target, \feiii~$\lambda$1122 is visible within our COS wavelength range.  As with \sIiii\, its velocity is blueshifted with respect to velocities of the species that reside in the neutral ISM.  A considerable amount of hidden saturation (Section~\ref{sec:hidden_sat}) was present within this spectrum for both the \feii\, and \Ni\, lines.  In both cases, the weakest lines available were used to constrain the ion column densities, which amounted to a column density higher by 0.97 and 0.25~dex in the case of \feii\,and \Ni\, respectively (Fig.~\ref{fig:feii_saturation}, Table~\ref{tab:saturation}).

Comparing our results with those reported by \citet{Thuan:2005sbs} from FUSE observations of the ions in common, we find that on average our column densities are in agreement with the FUSE values within the uncertainties (e.g., \Ni, \pii, and \feii), except for those ions that have only one saturated line in our COS spectra and for which we can only estimate a lower limit (i.e., \sIii, \cii~and \oi). 



\subsection{SBS~1415+437}

SBS~1415+437 is classified as a metal-deficient blue compact dwarf galaxy, with an oxygen metallicity of $\sim$1/12~\Zsol. Initially thought to have formed stars only $\lesssim100$~Myr ago \citep{Thuan:1999}, SBS~1415+437  was considered as one of the best candidate primeval galaxies in the local universe. However, as with several other local metal-poor galaxies, deep \textit{HST}/ACS imaging was able to reveal a population of much older stars, in this case older than $\sim$1.3~Gyr \citep{Aloisi:2005}. 

\begin{figure*}
\center
\includegraphics{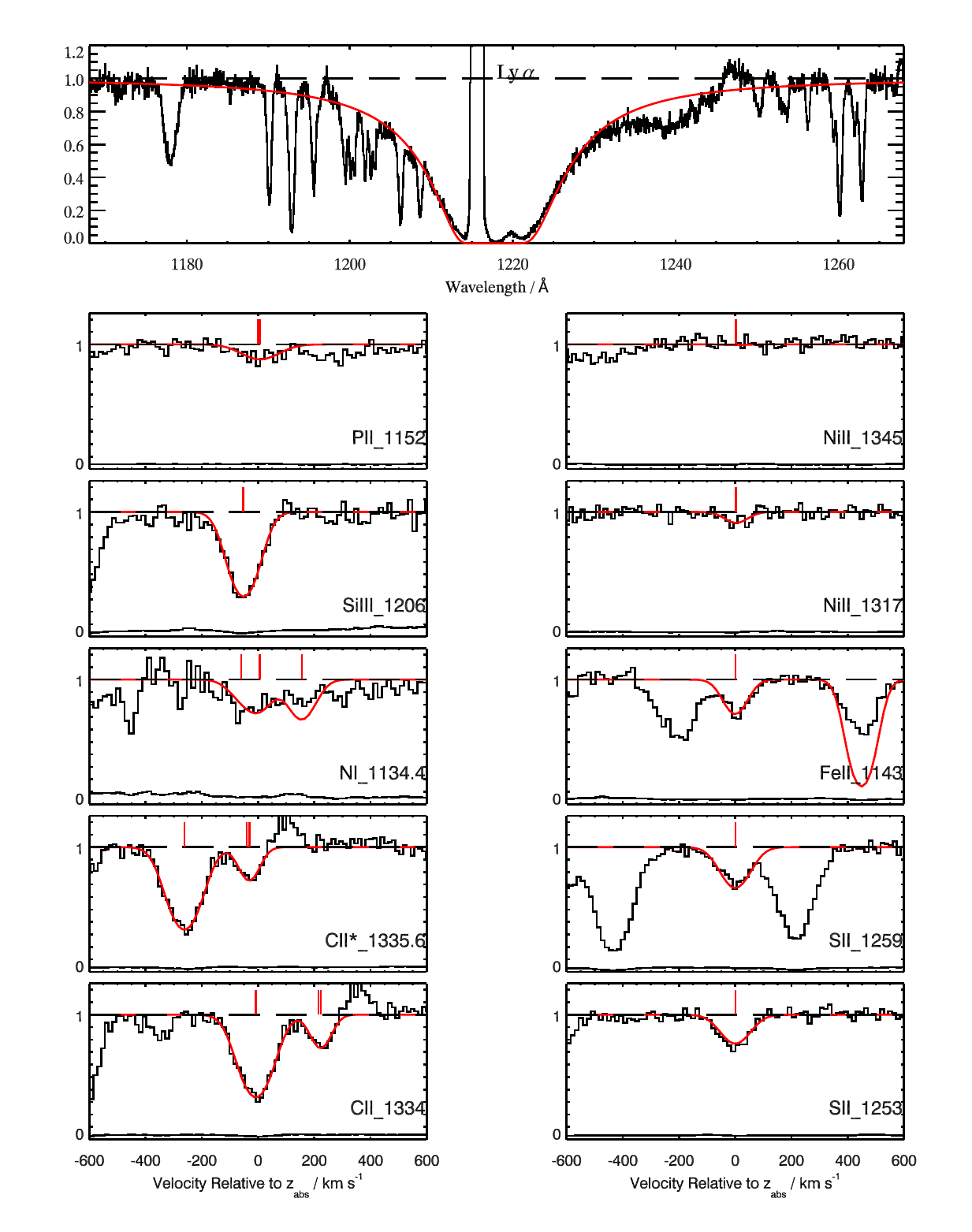}
\vspace{0.2cm}
\caption{Same as Fig.~\ref{fig:IZW18_spec} but for absorption-line profiles in SBS~1415+437.  The top panel shows the \Lya\, profile at $z_{abs}=0.0019413$ ($v = 582$ \kms).} 
\label{fig:SBS1415_spec}
\end{figure*}

With a redshift sufficiently high enough to clearly separate the centroid of its \Lya\, absorption profile from that of the Milky Way, we estimate log[$N$(\hi)] = $21.09\pm0.03$ cm$^{-2}$ for the \hi\, within SBS~1415+437 along with a Milky Way component of log[$N$(\hi$_{MW}$)] = 20.29 cm$^{-2}$ in its blue wing (see top panel of Fig.~\ref{fig:SBS1415_spec}). The Milky Way $N$(\hi) component was adopted from the latest \hi\ Leiden/Dwingeloo survey \citep{Hartmann:1997} and kept fixed during the fitting of the \Lya~absorption line.  The velocity of the stellar population ($\sim$ 610~\kms as measured from the \ciii~$\lambda$1176 photospheric absorption line) is slightly redshifted compared to that of the \hi\, gas by $\sim$ 30~\kms.

\begin{figure*}
\center
\includegraphics{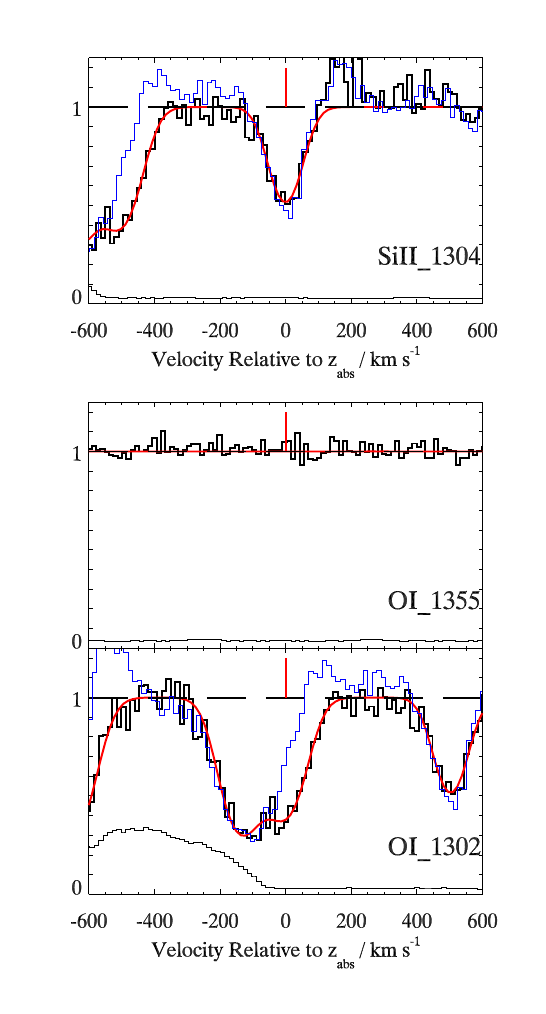}
\vspace{0.2cm}
\caption{Same as Fig.~\ref{fig:IZW18_OIspec} but for absorption-line profiles of \sIii~$\lambda$1304, \oi~$\lambda$1355, and \oi~$\lambda$1302 in SBS~1415+437.  The fit of \oi~$\lambda$1302 in the bottom panel includes contamination by \sIii~$\lambda1304_{MW}$, whose parameters were constrained by using the blue wing of the composite absorption.}
\label{fig:SBS1415_OIspec}
\end{figure*}

In the lower panels of Figure~\ref{fig:SBS1415_spec} we present a selection of the metal lines observed within the COS spectrum of SBS~1415+437.  The lines were best fitted with a single velocity component at an average redshift of $z_{abs}=0.0019377$ ($v\sim$ 580~\kms). This average excludes ions that typically originate from ionized gas, such as \sIiii, which is blueshifted by $\sim$50~\kms\, relative to the neutral ISM. \cii* is also blueshifted by $\sim$40~\kms, suggesting that in this case it is associated to the ionized more than the neutral gas. Rather than the Doppler width $b$ scaling inversely with ion mass, the $b$ parameters of all ions are mostly in agreement (within errors) around $\sim$\,60-70~\kms.  This suggests that turbulent broadening, rather than thermal broadening, is in affect within the neutral gas of SBS~1415+437.  Similarly to SBS~0335-052 (and many other galaxies within this sample), hidden saturation (Section~\ref{sec:hidden_sat}) was present in the \feii~$\lambda$1144 and \Ni~$\lambda$1134.9 absorption lines, which gave rise to column densities that were lower by 0.50 and 0.28 dex, respectively, compared to the column densities inferred when only using the weakest \feii\, and \Ni\, lines (Table~\ref{tab:saturation}).  Unlike most other galaxies within our sample (but similar to  I~Zw~18, another low-metallicity galaxy), the strongest \sIii* line within our wavelength range, \sIii*~$\lambda$1264, was not present.  Due to geocoronal contamination in the region of \oi~$\lambda$1302 and \sIii~$\lambda$1304 it was necessary to fit these two lines in the `night' spectra (see Fig.~\ref{fig:SBS1415_OIspec}).  

\subsection{NGC~4214}

NGC~4214 is a nearby ($\sim$ 3 Mpc) dwarf-barred irregular galaxy that consists of two main star-forming complexes containing hundreds of O stars, as well as a super star cluster.  It has a metallicity which is 1/3~\Zsol, placing it in the center of the metallicity scale covered within this study, along with NGC~5253 and NGC~4670.  This galaxy has been studied extensively in the last several years, mainly owing to its nearby proximity that allows  \textit{HST} to spatially resolve individual stars within it.  Morphologically, NGC~4214 harbors multiple sites of active star formation along a central bar-like structure, with bright stellar clusters and large cavities blown clear of gas by stellar winds (as revealed by recent \textit{HST}/WFC3 images by R. O'Connell and the WFC3 Scientific Oversight Committee), surrounded by a large disk of \hi\, gas \citep{Allsopp:1979}.

\begin{figure*}
\center
\includegraphics{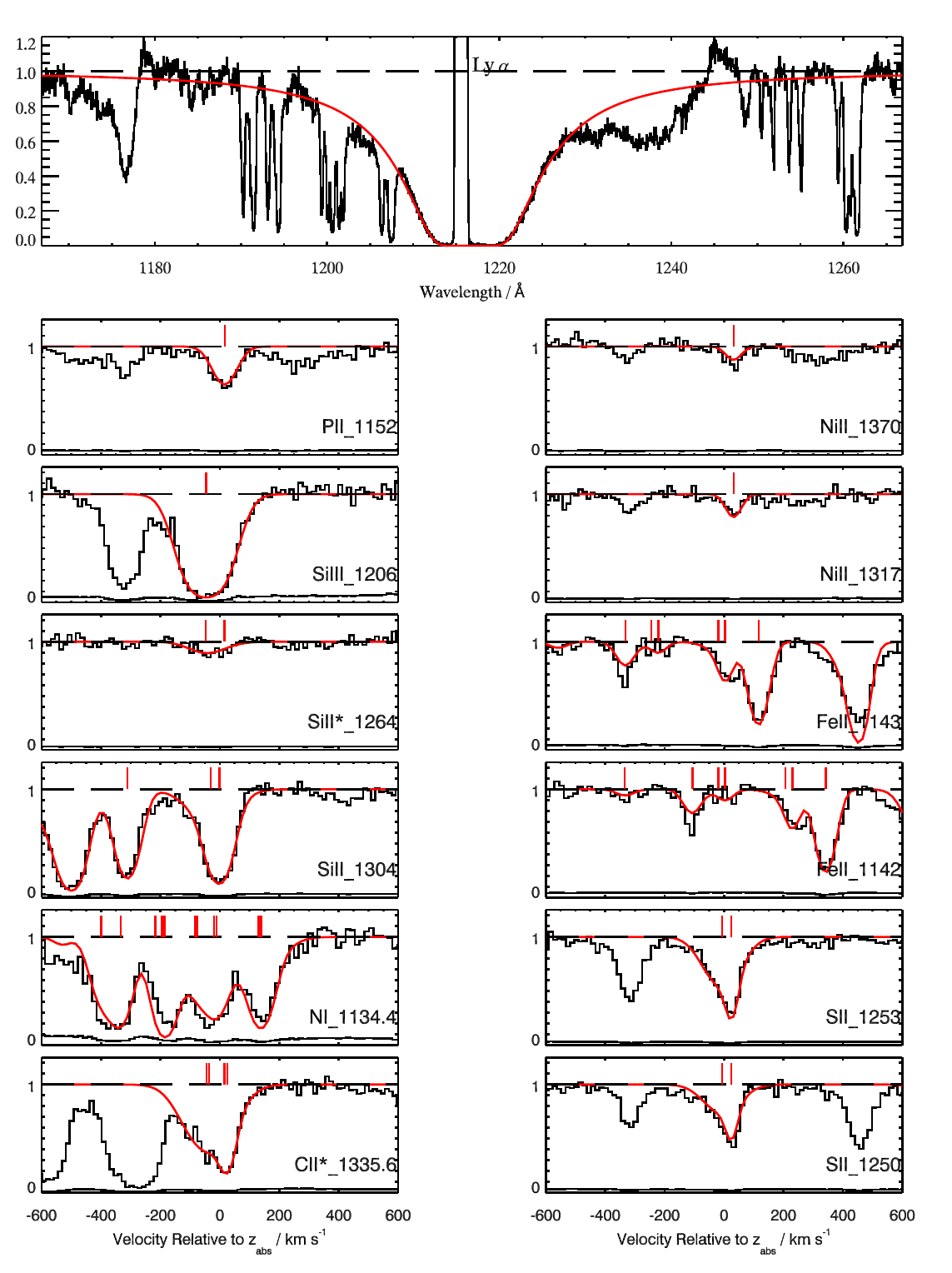}
\vspace{0.2cm}
\caption{Same as Fig.~\ref{fig:IZW18_spec} but for absorption-line profiles in NGC~4214.  The top panel shows the \Lya\, profile at $z_{abs}=0.0009544$ ($v = 286$ \kms).} 
\label{fig:NGC4214_spec}
\end{figure*}

The \Lya\, absorption-line region of NGC~4214 is shown in the top panel of Fig.~\ref{fig:NGC4214_spec}. The overlaid best-fit profile has a measured column density of log[$N$(\hi)] = $21.12\pm0.03$ cm$^{-2}$ combined with a Milky Way column density of log[$N$(\hi$_{MW}$)] = 20.08 cm$^{-2}$.  The latter was adopted from the latest \hi\ Leiden/Dwingeloo survey \citep{Hartmann:1997} and was kept fixed during the fitting.  The previously mentioned strong stellar winds within this galaxy are evident in the red and blue wings of the \Lya\, absorption (e.g., the large \Nv\, P-Cygni profile at 1243~\AA).  The stellar population velocity (as measured from the \ciii~$\lambda$1176 stellar absorption line) of $\sim$ 370 \kms\, is redshifted by $100$~\kms\, relative to the \hi\, gas.  However, since the \ciii~stellar line has a P-Cygni profile, this velocity offset should be regarded as a lower limit.

\begin{figure*}
\center
\includegraphics{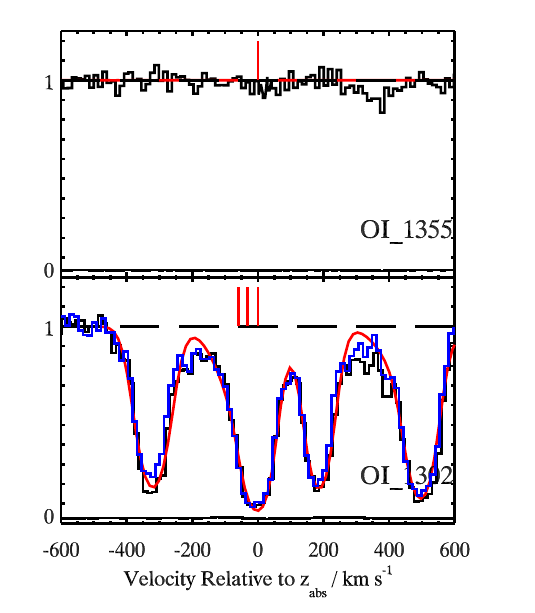}
\vspace{0.2cm}
\caption{Same as Fig.~\ref{fig:IZW18_OIspec} but for absorption-line profiles of \oi~$\lambda$1355, and \oi~$\lambda$1302 in NGC~4214 (\sIii~$\lambda$1304 was measured in the 'day+night' spectrum, so it is shown in the previous figure). The fit of \oi~$\lambda$1302 in the bottom panel shows the contamination by  \pii~$\lambda$1301, whose parameters were fixed from \pii~$\lambda$1152 whilst obtaining the model fits for \oi~$\lambda$1302.} 
\label{fig:NGC4214_OIspec}
\end{figure*}

A selection of the metal lines within the NGC~4214 spectrum are shown below the \Lya\, profile in Fig.~\ref{fig:NGC4214_spec}.  
Similarly to all the other galaxies within our sample, hidden saturation was present for the \feii~$\lambda$1144 absorption line and amounted to 0.19 dex compared to the other two weaker \feii~$\lambda$1143 and \feii~$\lambda$1142 lines that were fitted together. On the other hand,  \Ni~$\lambda$1134.9 did not show any hidden saturation and we were able to fit all the lines of the  \Ni~triplet around 1134 \AA~together. In this galaxy we were unable to constrain the fit to the \cii~$\lambda$1134 profile due to blending with the \cii\, absorption from the Milky Way. Because of geocoronal contamination around 1302~\AA, the \oi~$\lambda$1302 line was fitted on a `night' spectrum and can be seen in Fig.\ref{fig:NGC4214_OIspec}.  \sIii~$\lambda$1304 was unaffected by geocoronal contamination and was fitted on the `day+night' spectrum.

The ions fitted within this spectrum can be placed into three different groups, based on their average redshift and $b$ parameters.  Firstly a number of ions were best fitted using a 2-component line structure (\Ni, \oi, \sIii, \sii\,, and \feii), the first with an average redshift of $z_{abs,1}=0.0009808$ ($v_1\sim294$~\kms) and the second with $z_{abs,2}=0.0008936$ ($v_2\sim268$~\kms) blueshifted compared to the first one.  Although the radial velocities differ by only $\sim 25$~\kms, the two velocity components have very different $b$ parameters with $b_1=30$-40~\kms\, and $b_2=80$-90~\kms, suggesting that the two clouds of gas along the line of sight have quite different physical properties. Secondly, for the weaker absorption lines (i.e., \pii\,, and \nIii), we were able only to resolve a single velocity component.  Based on the average redshift of these lines ($z_{abs}=0.0010380$, $v\sim$\,311~\kms) and relatively low $b$ parameter of $\sim 40$~\kms, we may assume that we detected these ions only in the gas associated with the first high-redshift velocity component. The third and final group refers to the higher ionization ions from nebular gas (i.e., \sIii* and \sIiii) which have an average redshift of $z_{abs}=0.0007989$ ($v\sim240$~\kms) and a $b$ parameter of $\sim 90$ \kms. These ions are blueshifted by an additional 30~\kms\, from the second velocity component of the lower ionization ions while sharing similar kinematics properties (i.e., $b$ parameter).  This may mean that the second cloud of gas ($v_2$), whilst containing only low-ionization species, may be more aligned with regions of active star-formation, where a larger turbulent driven $b$ parameter is a result of stellar winds and outflows.

The \cii* ion has also two velocity components, one associated to the first component of the neutral gas with a redshift of $z_{abs}=0.0010080$ ($v\sim302$~\kms) and a similar $b$ parameter of $\sim 30$ \kms, and the other with a redshift of $z_{abs}=0.0008039$ ($v\sim241$~\kms) and a $b$ parameter of $\sim 100$ \kms associated instead to the more highly ionized blueshifted gas.

Since we were unable to resolve the \hi\, line profile into two velocity components, all abundances listed in Table~\ref{tab:abundances} are calculated from the total column density of the two velocity components, where applicable.
\subsection{NGC~5253}

NGC~5253 is a nearby starburst galaxy, at a distance of $\sim$3.8~Mpc and a metallicity of $\sim$\,1/3~\Zsol.  This starburst galaxy is famous for hosting the first case of observed localized enrichment in a \hii\, region, with \citet{Welch:1970},\citet{Walsh:1987,Walsh:1989}, and \citet{Kobulnicky:1997} each reporting the presence of a strong nitrogen overabundance in the starbursting nucleus of the galaxy.  The presence of Wolf-Rayet (WR) stars within the nucleus is firmly established and thought to be the source of the nitrogen-enriched ionized gas \citep[][and references therein]{Lopez-Sanchez:2007}.

\begin{figure*}
\center
\includegraphics[scale=0.5,angle=90]{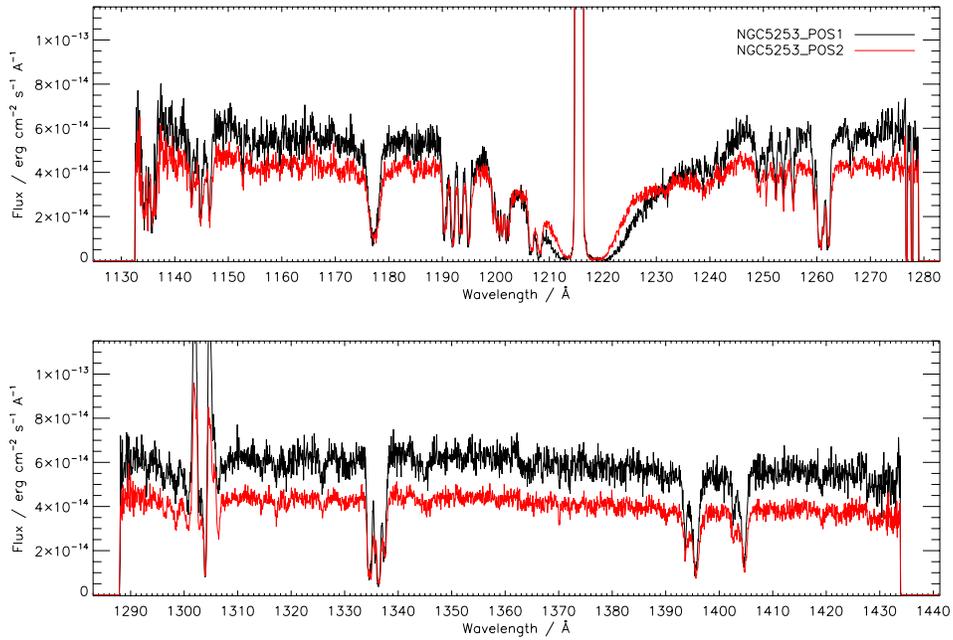}
\vspace{0.2cm}
\caption{Overlaid HST/COS G130M spectra of the two pointings obtained for NGC~5253 (position~1 in black and position~2 in red).} 
\label{fig:NGC5253_overlaid}
\end{figure*}

\begin{figure*}
\center
\includegraphics{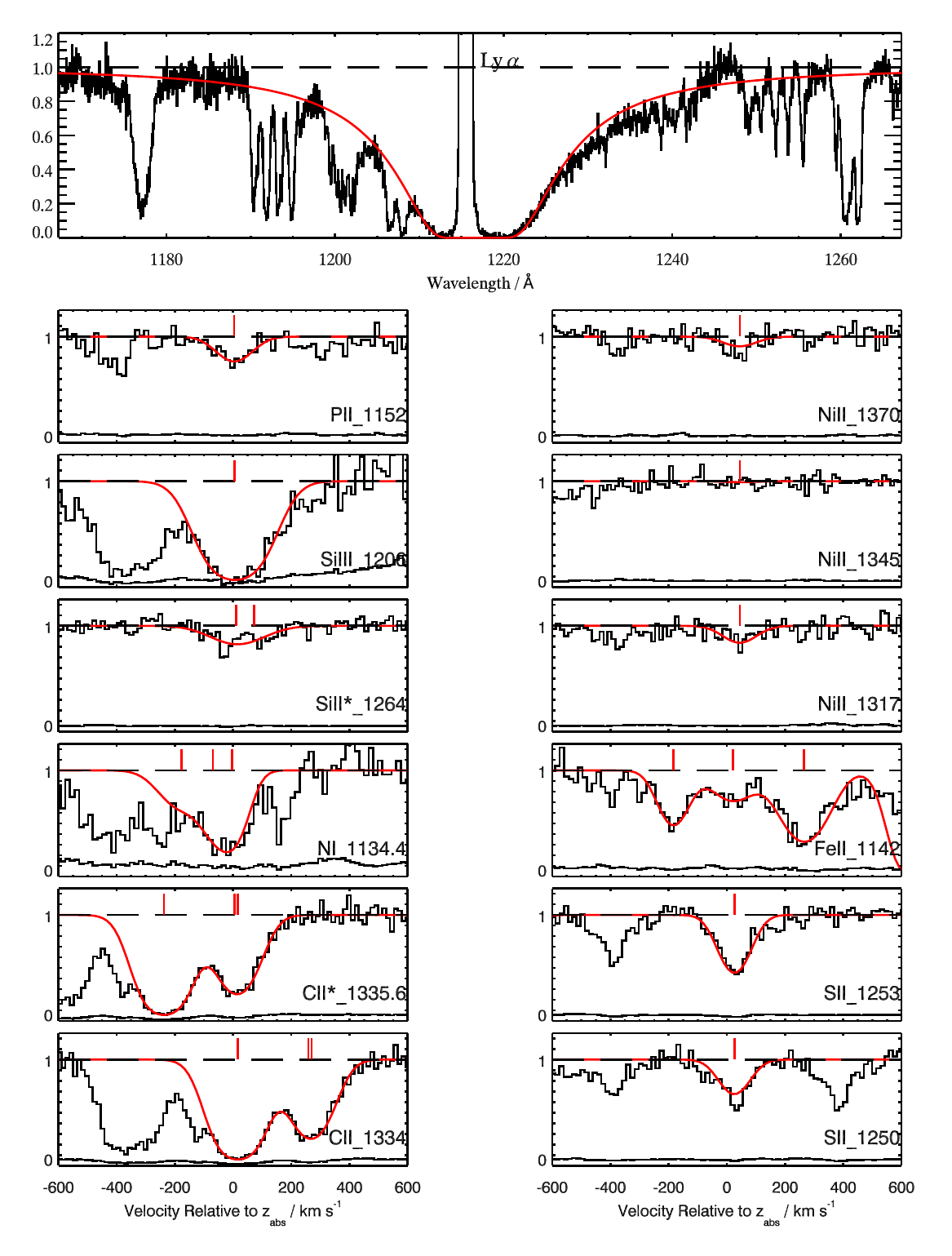}
\vspace{0.2cm}
\caption{Same as Fig.~\ref{fig:IZW18_spec} but for absorption-line profiles in position~1 of NGC~5253.  The top panel shows the \Lya\, profile at $z_{abs}=0.0012623$ ($v = 378$ \kms).} 
\label{fig:NGC5253-1_spec}
\end{figure*}

\begin{figure*}
\center
\includegraphics{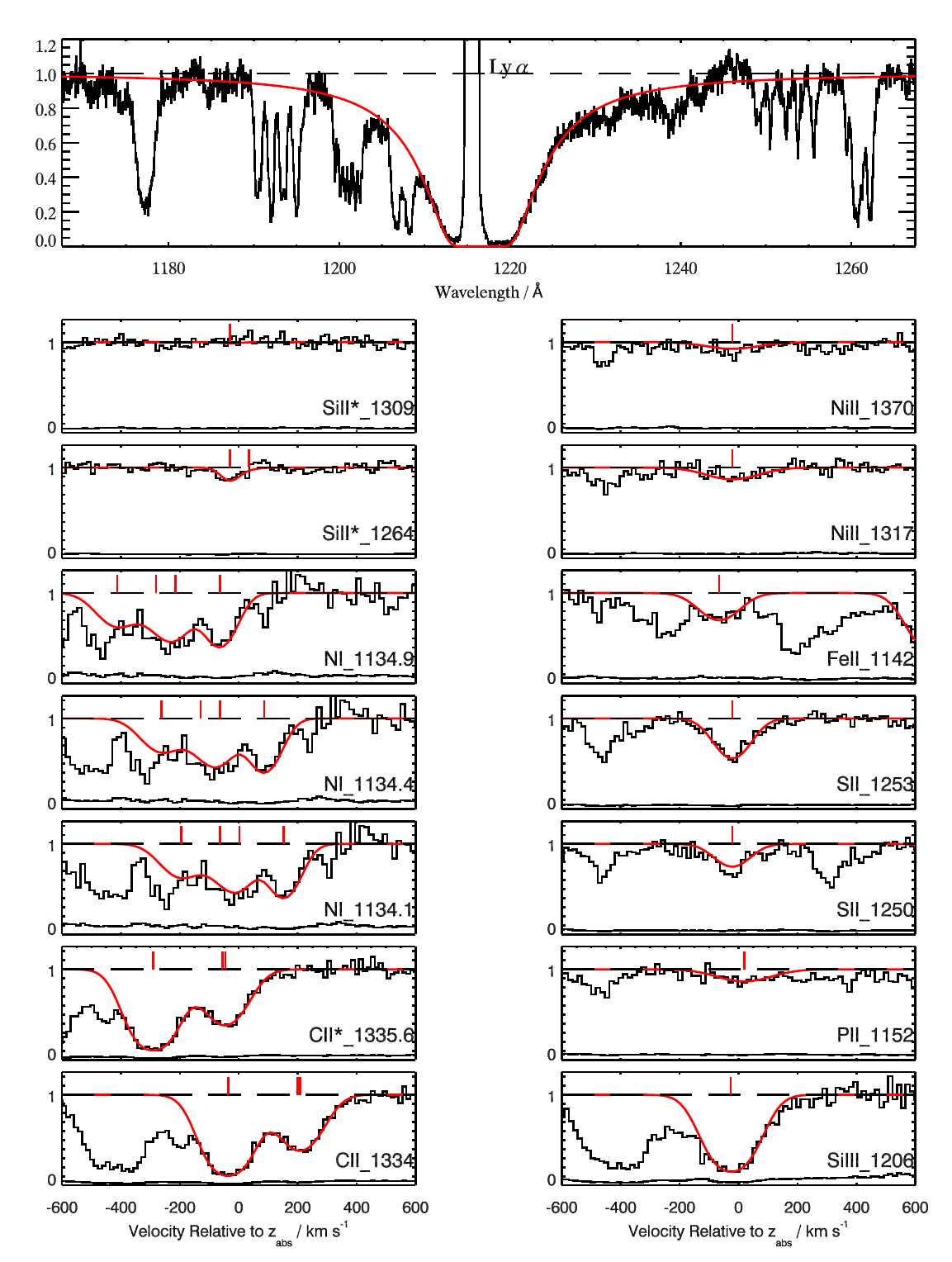}
\vspace{0.2cm}
\caption{Same as Fig.~\ref{fig:IZW18_spec} but for absorption-line profiles in position~2 of NGC~5253.  The top panel shows the \Lya\, profile at $z_{abs}=0.0015163$ ($v = 455$ \kms).} 
\label{fig:NGC5253-2_spec}
\end{figure*}

We obtained spectra from two separate pointings within NGC~5253, both located in the nucleus of the galaxy and separated by $\sim$7.1$''$ (a projected distance of $\sim$130~pc).  The spectra from the two pointings are overlaid in Fig.~\ref{fig:NGC5253_overlaid}.  The presence of a stellar population with strong winds (e.g., O and WR-stars) can be clearly seen in the spectra of both pointings in the form of strong P-Cygni profiles at $\sim$1248~\AA. Whilst there is relatively no difference in the velocity of the absorption lines, there is a difference in the strength of the profiles.  This is most clearly seen in the \Lya\, absorption profiles, where we estimate log[$N$(\hi)] = $21.20\pm0.01$ cm$^{-2}$ for position~1 (top panel of Fig.~\ref{fig:NGC5253-1_spec}) and log[$N$(\hi)] = $20.68\pm0.01$ cm$^{-2}$ for position~2 (top panel of Fig.~\ref{fig:NGC5253-2_spec}). A Milky Way column density of log[$N$(\hi$_{MW}$)] = 20.68 cm$^{-2}$ was adopted in both cases from the latest \hi\ Leiden/Dwingeloo survey \citep{Hartmann:1997} and was kept fixed during the fitting. 

\begin{figure*}
\center
\includegraphics{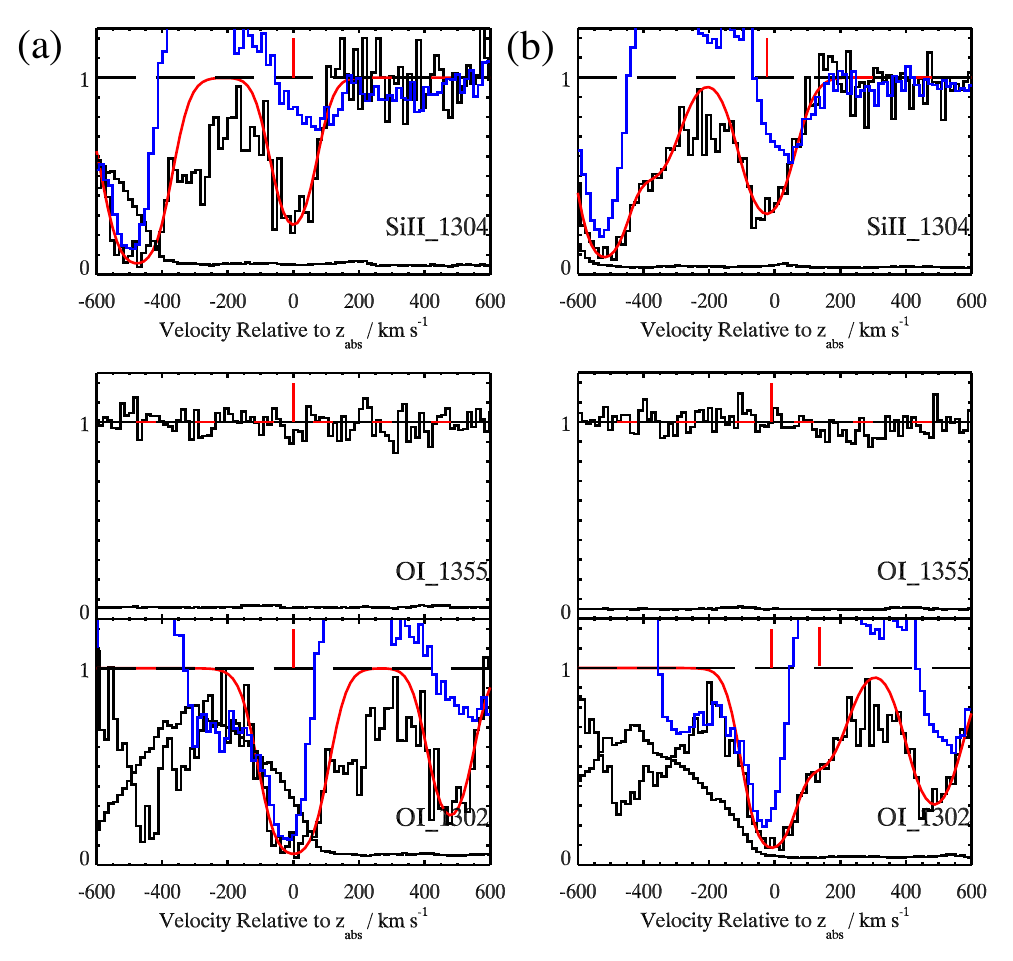}
\vspace{0.2cm}
\caption{Same as Fig.~\ref{fig:IZW18_OIspec} but for absorption-line profiles of \sIii~$\lambda$1304, \oi~$\lambda$1355, and \oi~$\lambda$1302 in position~1 (a) and position~2 (b) of NGC~5253.  The fit of \oi~$\lambda$1302 for position~2 in the right (b)  bottom panel includes contamination by \sIii~$\lambda1304_{MW}$, whose parameters were constrained by using the red wing of the composite absorption.} 
\label{fig:NGC5253-12_OIspec}
\end{figure*}

\textit{NGC~5253 Position 1:}
In Fig.~\ref{fig:NGC5253-1_spec} we present a selection of the lines observed within the COS spectrum of position~1 in NGC~5253.  All lines were best fitted with a single velocity component, at an average redshift of $z_{abs}=0.0013217$ ($v\sim$\,400~\kms). This average redshift  is in agreement with the velocity of the stellar population that also lies at $v\sim$\,400~\kms\,(as measured from the \ciii~$\lambda$1176 photospheric absorption line).  As is the case for previously discussed galaxies, the $b$ parameters of the different species do not scale inversely with the mass of the ion, suggesting that turbulent broadening is at play here, rather than thermal broadening.  Indeed, all ions have a $b$ parameter that is relatively constant around $\sim$\,80-100~\kms. The more highly ionized species \sIii* and \sIiii, are also in agreement with the lower ionization species, indicating a mix of neutral and ionized gas along the sightline.  A large amount of hidden saturation was present in both the \feii~$\lambda$1144 and \Ni~$\lambda$1134.9 absorption profiles (Section~\ref{sec:hidden_sat}), since an increase in column density of 0.92 and 0.46~dex was found for \feii\, and \Ni, respectively, when fitting the weakest lines (Table~\ref{tab:saturation}).  The \oi~$\lambda$ 1302 and \sIii~$\lambda$1304 absorption lines are shown in Fig.~\ref{fig:NGC5253-12_OIspec}a, and were fitted on `night' spectra due to geocoronal contamination within the 1304~\AA\, region.

\textit{NGC~5253 Position 2:}
The main absorption lines observed within the second NGC~5253 pointing are presented in Fig.~\ref{fig:NGC5253-2_spec}.  All lines were best fitted with a single velocity component, at an average redshift of $z_{abs}=0.0014118$ ($v\sim$\,420~\kms) which is in agreement with the radial velocity of the stellar population ($\sim$\,420~\kms, as measured from the \ciii~$\lambda$1176 photospheric absorption line). The $b$ parameters are reasonably constant around 80-90~\kms with a relatively large scatter around the mean mainly due to the large errors incurred from fits to the weakest lines, such as \nIii\, and \pii.  \sIiii~$\lambda$1206, which is thought to originate predominantly in ionized gas, is in line with the lower ionization species, indicating a mix of neutral and ionized gas along the sightline. \sIii* (also typically found in ionized gas) has instead a lower $b$ parameter of $\sim$\,40~\kms, suggesting that we only detect this species in some of the strongest unresolved velocity components that contribute to  \sIiii. As with Position 1, a $\sim$\,0.96~dex increase in column density was found when measuring one of weakest \feii~lines, rather than the strongest \feii\,line.  However, no hidden saturation was found in \Ni~$\lambda$1134.9 and all lines within the \Ni~triplet were fitted together satisfactorily.  The \oi~$\lambda$ 1302 and \sIii~$\lambda$1304 absorption lines are shown in Fig.~\ref{fig:NGC5253-12_OIspec}b, and were fitted on `night' spectra due to geocoronal contamination around 1304~\AA.

\subsection{NGC~4670}
NGC~4670 is an amorphous, metal-deficient (Z\,$\sim$\,1/3~\Zsol) blue compact dwarf galaxy.  It has often been compared to NGC~4214 as it displays a very similar optical spectrum suggestive of a vigorous of star formation, with the young stars superposed on a much older stellar population \citep{Kinney:1993}. 

\begin{figure*}
\center
\includegraphics{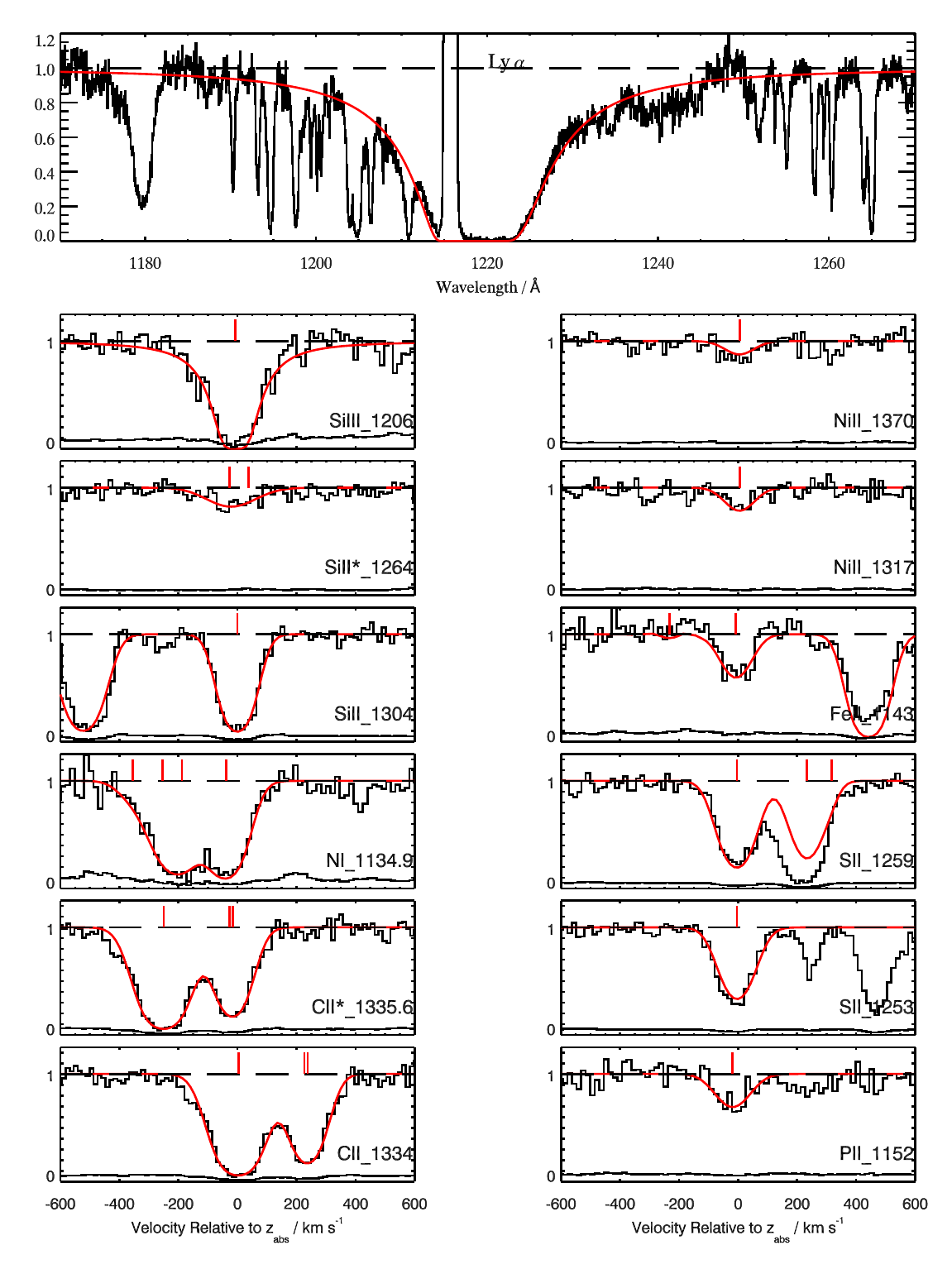}
\vspace{0.2cm}
\caption{Same as Fig.~\ref{fig:IZW18_spec} but for absorption-line profiles in NGC~4670.  The top panel shows the \Lya\, profile at $z_{abs}=0.0036438$ ($v = 1,092$ \kms).} 
\label{fig:NGC4670_spec}
\end{figure*}

The upper panel of Figure~\ref{fig:NGC4670_spec} displays the \Lya\, absorption region of NGC~4670 and, similarly to NGC~4214 and NGC~5253 discussed above, large wind profiles can be seen in the red wing of \Lya.  A best fit profile is overlaid, corresponding to log[$N$(\hi)] = $21.07\pm0.08$ cm$^{-2}$ which was constrained by the red wing of the profile, along with a Milky Way component of log[$N$(\hi)] = $20.25\pm0.12$ cm$^{-2}$ constrained by the blue wing of the profile and in excellent agreement with log[$N$(\hi)] = 20.24 cm$^{-2}$ measured in the latest \hi\ Leiden/Dwingeloo survey \citep{Hartmann:1997}.

\begin{figure*}
\center
\includegraphics{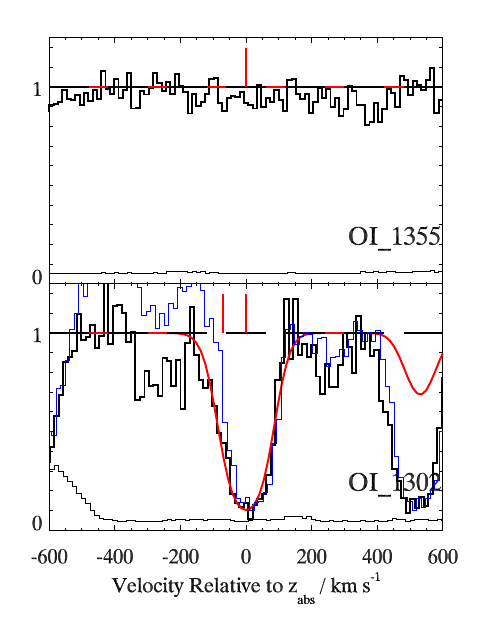}
\vspace{0.2cm}
\caption{Same as Fig.~\ref{fig:IZW18_OIspec} but for absorption-line profiles of \oi~$\lambda$1355, and \oi~$\lambda$1302 in NGC~4670 (\sIii~$\lambda$1304 was measured in the 'day+night' spectrum, so it is shown in the previous figure). The fit of \oi~$\lambda$1302 in the bottom panel shows the contamination by  \pii~$\lambda$1301, whose parameters were fixed from \pii~$\lambda$1152 whilst obtaining the model fits for \oi~$\lambda$1302.} 
\label{fig:NGC4670_OIspec}
\end{figure*}

The heavy element absorption lines (a selection of which are shown in Fig.~\ref{fig:NGC4670_spec}) were all fitted with a single velocity component at an average redshift of $z_{abs}=0.0036046$ ($v\sim1080$~\kms).  The velocity of the stellar component (as measured from the \ciii~$\lambda$1176 photospheric absorption line) is blueshifted by $\sim$\,35~\kms\, relative to the \hi\, gas.  
There is no overall decrease in $b$ parameter with ionic mass, with all ions displaying a $b$ parameter around  $\sim$ 70-80~\kms, which suggests that turbulent broadening (rather than thermal broadening) is occurring within the neutral gas. \sIiii~$\lambda$1206, which is thought to arise in the ionized gas, shows signs of a two-component velocity profile; however, a satisfactory fit to the profile could only be obtained with one component and by fixing the redshift $z$ and the $b$ parameter from the fit to \sIii~$\lambda$1304.  \sIii*, also arising from ionized gas, is in line with the lower ionization species once the large error on its $b$ parameter is taken into account, indicating a mix of neutral and ionized gas along the sightline. A relatively small amount of hidden saturation was seen within the \feii~triplet, with a 0.22~dex increase in $N$(\feii) when using the weakest lines (Section~\ref{sec:hidden_sat}).  No hidden saturation was seen from the strongest line within the \Ni~1134 triplet.  Due to geocoronal contamination around 1302~\AA, a `night' spectrum was used to fit the \oi~$\lambda$1302 line as shown in Fig.~\ref{fig:NGC4670_OIspec}. Due to the high redshift of this target, the \sIii~$\lambda$1304 was not affected by geocoronal contamination and was fitted in the `day+night' spectrum instead.

\subsection{NGC~4449}
NGC~4449 is a barred Magellanic-type irregular galaxy \citep{deVaucouleurs:1991} with star formation occurring throughout the galaxy at a rate almost twice that of the Large Magellanic Cloud \citep{Thronson:1987,Hunter:1999}.  It is a highly studied galaxy, mainly due to its nearby proximity ($\sim$\,3.8 Mpc) which provides excellent spatial resolution for extragalactic star cluster studies \citep[see][and references therein]{Reines:2008} and resolved stellar population studies \citep[see, e.g.,][]{Annibali:2008}.  \hi\, studies of NGC~4449 reveal a rather perturbed structure, with `extended streamers wrapping around the galaxy and counter-rotating (inner and outer) gas systems' \citep[see, e.g.,][]{Hunter:1998}.  Its \hii\, region metallicity of $\sim$\,1/2~\Zsol, places it in the mid--high metallicity range of galaxies covered within this study.

\begin{figure*}
\center
\includegraphics{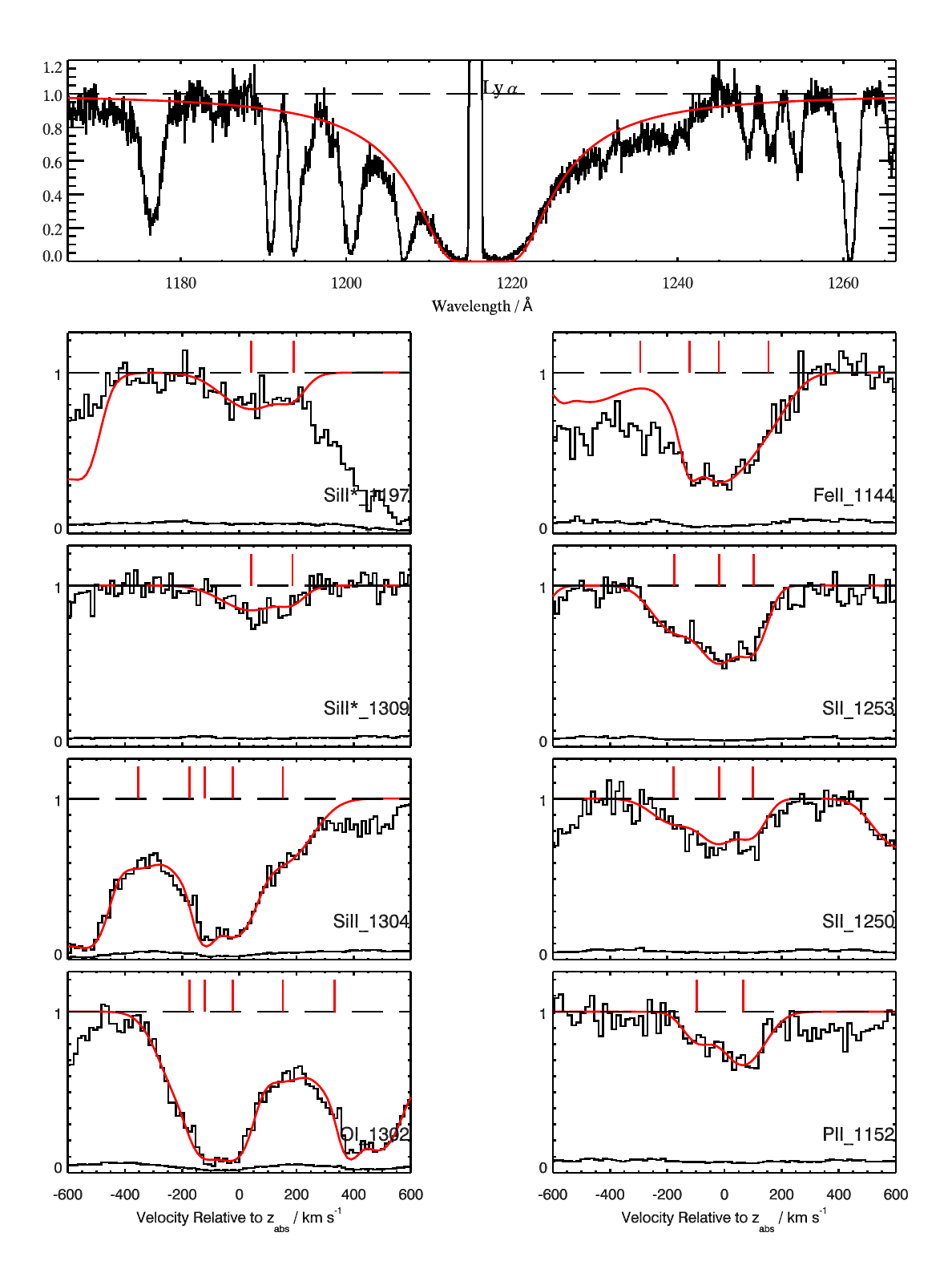}
\caption{Same as Fig.~\ref{fig:IZW18_spec} but for absorption-line profiles in NGC~4449.  The top panel shows the \Lya\, profile at $z_{abs}=0.0008061$ ($v = 242$ \kms).}
\label{fig:NGC4449_spec}
\end{figure*}

The top panel of Figure~\ref{fig:NGC4449_spec} displays the \Lya\, absorption region of NGC~4449.  A best fit profile is overlaid, corresponding to log[$N$(\hi)] = $21.14\pm0.03$ cm$^{-2}$ combined with an absorption from within the Milky Way of log[$N$(\hi)] = 20.21 cm$^{-2}$. Since the two absorption profiles are heavily blended with one another, the NGC~4449 \hi\, component was fitted by fixing the Milky Way component at the value measured in the latest \hi\ Leiden/Dwingeloo survey \citep{Hartmann:1997}, whilst leaving the profile fit parameters for the galaxy to vary freely.

The heavy element absorption lines in NGC~4449 (a selection of which are shown in Fig.~\ref{fig:NGC4449_spec}) were fitted with up to three velocity components. In \oi, \sIii, and \feii, three velocity components were visible, at average redshifts of 0.0001732 ($v_1 \sim$ 52~\kms), 0.0005096 ($v_2 \sim 153$~\kms) and 0.0010903 ($v_3 \sim 327$~\kms), respectively.  For \pii\, the lowest velocity component was clearly separated, while the two highest velocity components were blended into one, at z=0.0007982 ($v_{2+3} \sim 240$~\kms), probably due to the faintness of the line profiles. On the other hand, for \sii\, only the two highest velocity components were present likely due to the faintness of its lines. The $b$ parameters are somewhat less clearly defined between the three absorbing components; the lowest velocity component has $b_1 \sim$\,40-50~\kms, whereas the two higher velocity components are more in agreement with each other having $b_2 \sim$\,90-110~\kms and $b_3 \sim$\,~120-140\kms.  \sIii* (which is thought to arise predominantly from ionized gas) also has a two-component velocity structure with one component aligned with $v_2$ and a second slightly redshifted by $\sim 35$~\kms\, relative to $v_3$ but with a consistent $b$ parameter. The radial velocity of the stellar component is in perfect agreement with $v_{2+3}$, lying at $v\sim$240~\kms\, (as measured from the \ciii~$\lambda$1176 photospheric absorption line). The multi-component velocity structure observed in our COS data may be the result of the above mentioned complex gas system seen in the \hi\, imaging studies.  There is no overall decrease in $b$ parameter with ionic mass, which suggests that turbulent broadening is occurring within the neutral gas.

We were unable to constrain a \Ni\, column density for this galaxy because the two \Ni\, triplets (at $\sim$1134~\AA\, and $\sim$1200~\AA) suffer from blending due to the galaxy multiple velocity components and the Milky Way \Ni\, absorption profiles.  For the same reasons we were unable to constrain profile fits for \cii, \cii*, and \sIiii.   Another consequence of this multi-velocity component structure is that we were unable to constrain $N$(\feii) using the weaker $\lambda\lambda$1142, 1143 lines due to blending.  Therefore, the \feii\, column density and abundances are obtained from the \feii~$\lambda$1144 line and are deemed lower limits due to hidden saturation (Section~\ref{sec:hidden_sat}).  
Despite the mid--high metallicity of NGC~4449 within our COS sample, \nIii\, was not strong enough to be constrained above the noise of the spectrum.  There was no geocoronal contamination experienced during the observation of NGC~4449 and thus measurements of the \oi~$\lambda$1302 and \sIii~$\lambda$1304 absorption profiles were made on the original `day+night' spectrum.  As we were unable to constrain multiple velocity components within the \Lya\, profile, all abundances were calculated from the total column densities of all velocity components.

\subsection{NGC~3690}

NGC~3690 is the galaxy within our sample with the second highest metallicity ($Z \sim $~\Zsol) and the second highest spectroscopic redshift (see below). It is known as an extremely disturbed system that is interacting/merging with the companion galaxy IC~694 \citep{Gehrz:1983} located in the eastern part of the system.  Additional interest in this object arises from the fact that it is one of the most extreme known cases of an extended burst of star formation \citep{Augarde:1985}.  \hi\, studies of this system reveal that there is no disk-like structure, but instead an amorphous collection of bright star-forming regions and dust lanes \citep{Nordgren:1997}.  

\begin{figure*}
\center
\includegraphics{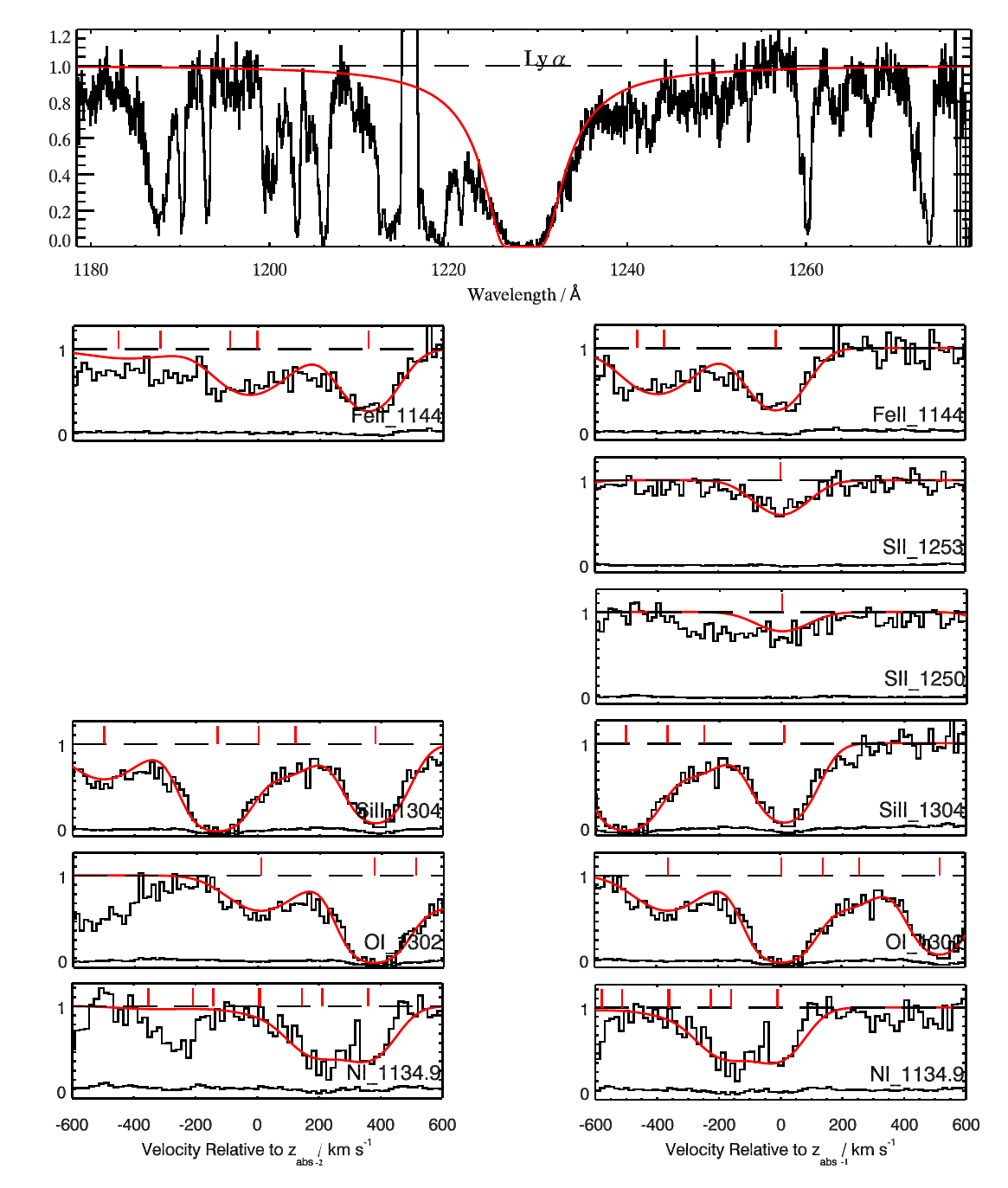}
\vspace{0.2cm}
\caption{Same as Fig.~\ref{fig:IZW18_spec} but for absorption-line profiles in NGC~3690.  The top panel shows the \Lya\, profile with two velocity components at $z_{abs,1}=0.0106978$ ($v_1 = 3,207$ \kms) and $z_{abs,2}=0.0095165$ ($v_2 = 2,853$ \kms) . The right column displays a selection of metal lines belonging to velocity component 1 and plotted relative to  $z_{abs,1}$, whereas the left column shows metal lines from velocity component 2 relative to  $z_{abs,2}$.} 
\label{fig:NGC3690_spec}
\end{figure*}

The \Lya\, absorption profile of NGC~3690 is shown in the top panel of Fig.~\ref{fig:NGC3690_spec}.  A positive consequence of this galaxy high redshift is that the intrinsic \Lya\, is completely separated from that of the Milky Way and is therefore not contaminated by it (however, the downside is that its COS spectrum has the lowest signal-to-noise of our sample with S/N$\sim$10).  We were able to constrain a two-component fit to the \Lya\, profile using in this case the uncontaminated red wing of the absorption line and the redshifts fixed from the average $z_{abs,1}$ and $z_{abs,2}$ measured from well constrained ions, such as \oi\, and \feii.  The best-fit profile was found to be log[$N$(\hi)] = $20.62\pm0.02$ cm$^{-2}$ at $z_{abs,1}=0.0106978$ ($v_1 \sim $\,3,210~\kms) and log[$N$(\hi)] = $19.82\pm0.08$ cm$^{-2}$ at $z_{abs,2}=0.0095165$ ($v_2 \sim $\,2,850~\kms). The radial velocity of the stellar population (as measured from the \ciii~$\lambda$1176 photospheric absorption line) lies at $\sim$\,3,050~\kms, which is the approximate midpoint between the two velocity components of the ISM.  Since NGC~3690 is the only target where we were able to separate two \hi\, velocity components in the \Lya\, profile, we calculated separate abundances for those ions where the column density of the two components was measured independently.

Unfortunately, due to the low S/N of this spectrum and the blending of its multiple velocity components, we were unable to constrain several of the ions available in other COS spectra within our sample, such as \cii, \cii*, \nIii, \pii, and \sIii*.  In addition, the \sIiii~$\lambda$1206 absorption line could not be constrained because redshifted into the Milky Way \Lya\, absorption-line profile. A consequence of the high redshift presented by NGC~3690, is that geocoronal contamination did not affect the \oi~$\lambda$1302 and \sIii~$\lambda$1304 lines, allowing the original `day+night' spectrum to be used for their fitting. In the lower panels of Fig.~\ref{fig:NGC3690_spec} we show a selection of the heavy element ions that we were able to constrain within the spectrum of NGC~3690.  With the exclusion of \sii, all of the ions displayed a clear two-component velocity structure in alignment with the \Lya\, profile velocities; one at an average redshift of $z_{abs,1}=0.0106992$ ($v_1\sim$3,210~\kms) and another at $z_{abs,2}=0.0094576$ ($v_2\sim$2,835~\kms).  This velocity structure could be a signature of NGC~3690 interaction with its neighboring galaxy IC~694.

A number of assumptions were made during the fitting process for those ions that we were able to constrain. These assumptions are as follows.  Component 2 ($v_2$ ) of  \oi\,$\lambda1302$~\AA\, and component 1 ($v_1$) of \sIii$\lambda$1304~\AA\, are well isolated and were thus used to get independent measurements of the b parameter and redshift of the individual $v_1$ and $v_2$ components. Component 1  ($v_1$ ) of  \oi\,$\lambda1302$~\AA\, and component 2 ($v_2$) of \sIii$\lambda$1304~\AA\, are instead merged together, and were thus constrained using the redshift and $b$ parameter determined from their corresponding isolated components.  Component 1 in  the couple of lines fitted for \sii\, and in \feii$\lambda1144$ is well isolated and was fitted independently, giving $b$ parameters and redshift values that agree within the errors with those found for component 1 of \sIii\, and \oi. This suggests that the gas from which $v_1$  originates is dominated by turbulent motions (i.e., the $b$ parameters are independent of ionic mass). We therefore decided to assume the same for $v_2$ and to use the average $b$ parameters from the separate components ($b_1\sim104$~\kms\, and $b_2\sim138$~\kms) to constrain both \Ni\, components and component 2 of \feii.  We were unable to detect component 2 for \sii, likely because a much lower column density made this component too faint to be detected.

Unfortunately, the complicated velocity structure described above prevented us from constraining $N$(\feii) from the weaker \feii\, lines alone and we were instead forced to also use \feii~$\lambda$1144, which is known to suffer from hidden saturation (Section~\ref{sec:hidden_sat}).  However, since the fit of this line for both velocity components was done in conjunction with the weaker \feii~$\lambda$1143 absorption and both lines agree quite well with the same fit, hidden saturation should not be at play.

\subsection{M83}
M83 is described in the literature as being a `remarkable grand-design spiral with Hubble type SAB(s)c and strong dust lanes that are offset along an oval distortion that contains a central starburst' \citep[][and references within]{Elmegreen:1998}.  It is from within this central starburst that our COS spectra were obtained, using two centrally located UV sources separated by approximately 6.5$''$ ($\sim$150~pc, see Fig.~\ref{fig:COS_targets}).  M83 is the highest metallicity galaxy within our sample, with an oxygen abundance $Z \sim 3$~\Zsol. Inclusion of this object within our sample meant we could fully probe depletion patterns at the higher end of the metallicity scale.  

\begin{figure*}
\center
\includegraphics[scale=0.5,angle=90]{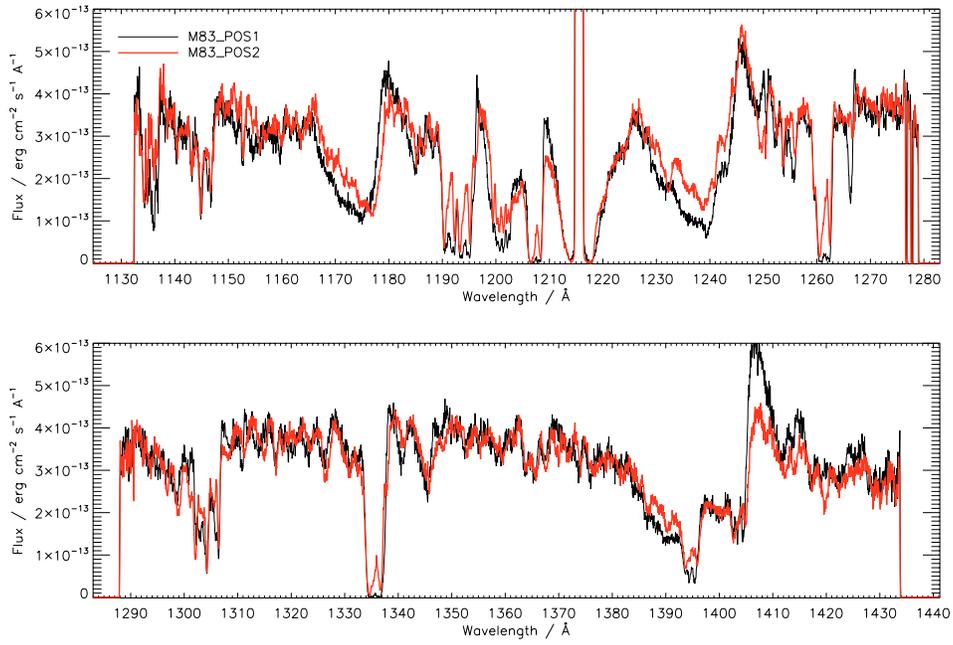}
\vspace{0.2cm}
\caption{Overlaid HST/COS G130M spectra of the two pointings obtained for M83 (position~1 in black and position~2 in red). The spectrum of position~1 has been scaled up to that of position~2 to allow for a better comparison of the line velocities.}
\label{fig:M83_overlaid}
\end{figure*}

The COS spectra from the two pointings are shown overlaid in Fig.~\ref{fig:M83_overlaid}, where the spectrum of position~1 has been scaled up to that of position~2.  This scaling allows us to highlight the relatively similar \hi\, column density between the two positions, and also a similar redshift for the \Lya\, and  metal absorption lines.  However, there is a velocity offset in the wind lines within the two spectra (e.g., \ciii~$\lambda$1176 and \sIiv~$\lambda$1402), suggesting that whilst the absorbing gas clouds may be spatially associated with each other, the stellar populations of the two UV background sources are not directly related to each other. In particular, the centroid of the \ciii~$\lambda$1176 photospheric line within each spectrum shows that the UV sources are offset from one another by $\sim$270~\kms.  The two targeted sources are within the eye of the central spiral \citep[our two pointings correspond to `hot spot' number 3 and 8 within][]{Elmegreen:1998} at different points of the inner arm, where a differential rotation would be expected.  The expanse of the wind lines within the M83 spectra are greater than within any other galaxy in our sample. P-Cygni profiles are visible in  \ciii~$\lambda$1176,  \Nv~$\lambda$1245, and \sIiv~$\lambda$1402. 

\begin{figure*}
\center
\includegraphics{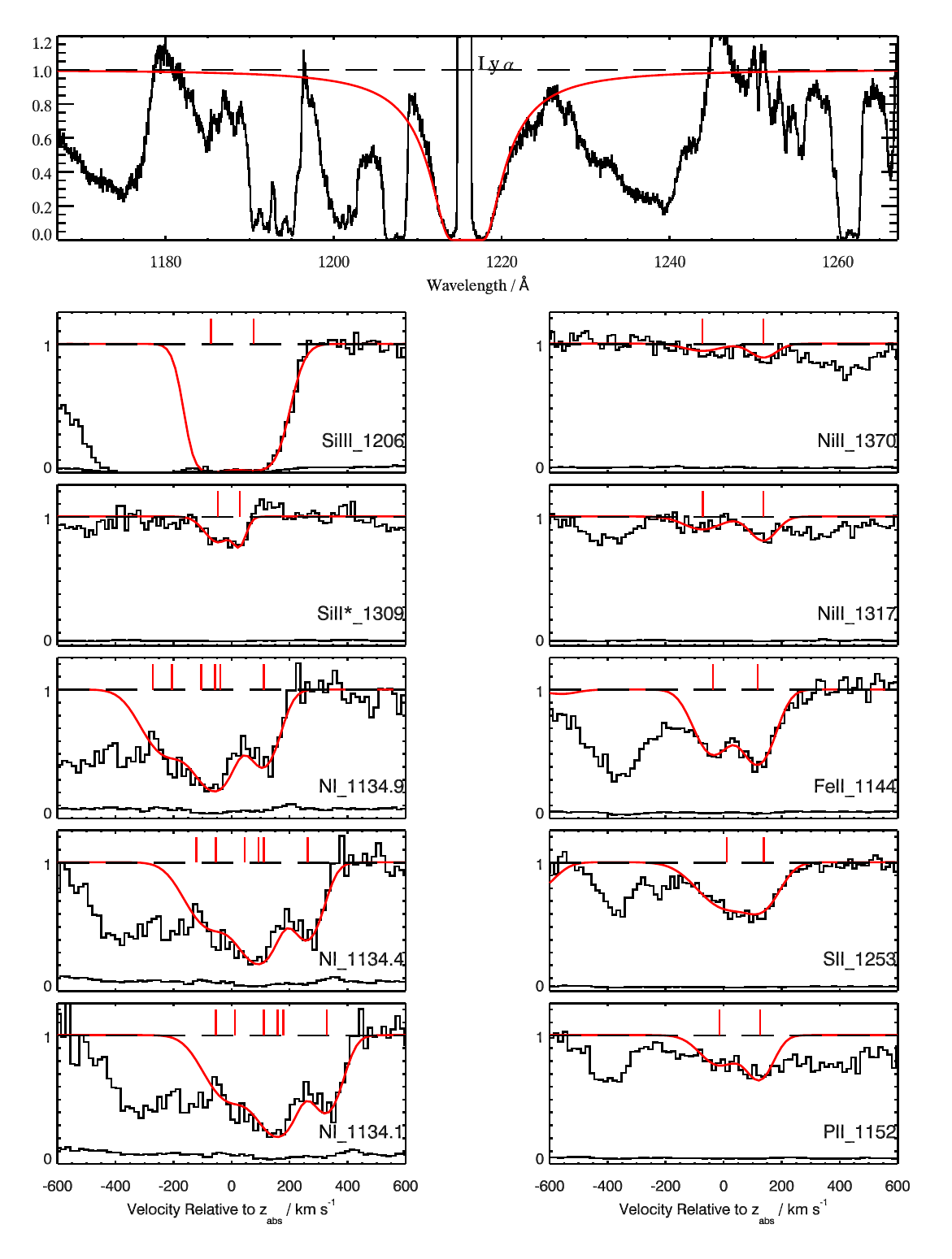}
\vspace{0.2cm}
\caption{Same as Fig.~\ref{fig:IZW18_spec} but for absorption-line profiles in position~1 of M83.  The top panel shows the \Lya\, profile at $z_{abs}=0.0011619$ ($v = 348$ \kms). Absorption profiles for the other lines fitted with two velocity components, are shown relative to component 2 at an average $z_{abs,2}=0.0011107$ ($v_2 = 333$ \kms).}
\label{fig:M83-1_spec}
\end{figure*}

\textit{M83 Position 1:}
In the top panel of Fig.~\ref{fig:M83-1_spec} we present the \Lya\, region in the COS spectrum of position~1 of M83, and a best-fit profile corresponding to an M83 component of log[$N$(\hi)] = $19.60\pm0.32$ cm$^{-2}$ and a Milky Way component of log[$N$(\hi)] = $20.57\pm0.05$ cm$^{-2}$. The profile was fitted by fixing the Milky Way component at the fit parameters found for position~2 of M83 (see below) and by allowing the parameters of the galaxy component to vary freely. A strong P-Cygni profile from \sIiii~$\lambda$1206 exists within the \Lya\, profile. This is attributable to winds from B0-1 supergiants \citep{Prinja:2002}.  However, the population of B0-1 stars is not large enough to contribute to the \Lya\, absorption-line profile, as detailed in Section~\ref{sec:hi_colDens}.

In the lower panels of Fig.~\ref{fig:M83-1_spec} we present a selection of metal absorption lines and their profile fits.  All metal absorption lines within the spectrum were fitted with two velocity components at average redshifts of $z_{abs,1}=0.0016155$ ($v_1\sim480$~\kms) and $z_{abs,2}=0.0011107$ ($v_2\sim330$~\kms).  The velocity of the stellar population (as measured by the \ciii~$\lambda$1176 photospheric line)  lies at $v\sim$150~\kms, i.e., a blueshift of up to $\sim 330$~\kms relative to the ISM.  In Fig.~\ref{fig:M83-1_spec} absorption profiles are displayed relative to  $z_{abs,2}$.   The $b$ parameters for the two components are rather different, with $b_1 \sim 60$-70~\kms\, and $b_2 \sim 90$-100~\kms, suggesting that the two components of absorbing gas are physically rather different. \sIiii\, and \sIii*~which are thought to arise in more highly ionized gas, also show a two-velocity component structure that is blueshifted relative to $v_1$ and $v_2$ by $\sim$70 and $\sim$25~\kms, suggesting that these species are indeed situated in separate clouds of higher-ionization absorbing gas (as also indicated by slightly different $b$ parameters). 

\begin{figure*}
\center
\includegraphics{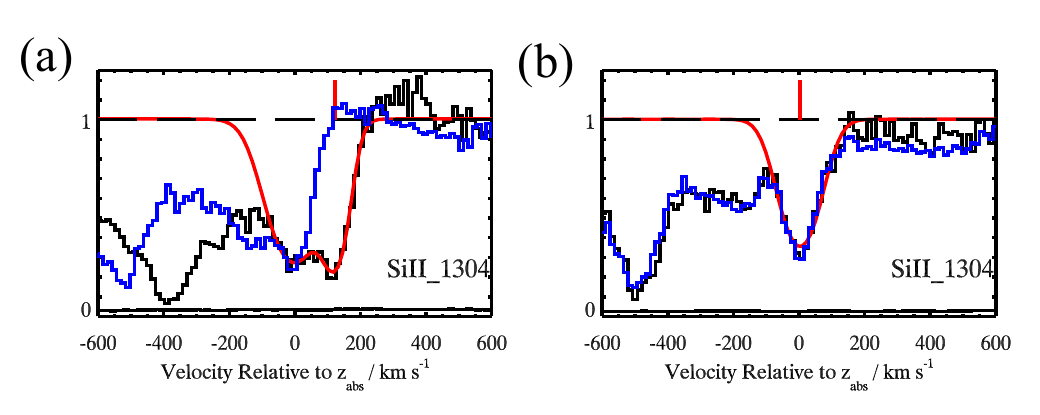}
\vspace{0.2cm}
\caption{Same as Fig.~\ref{fig:IZW18_OIspec} but for absorption-line profiles of \sIii~$\lambda$1304 in position~1 (a) and position~2 (b) of M83 (\oi~$\lambda$1302 was not measured). Two velocity components were used for position~1, while a single velocity component was used for position~2 (see Section~\ref{sec:objects} for more details).}
\label{fig:M83_OIspec}
\end{figure*}

As with other spectra within this sample that contain multi-velocity components, we were unable to constrain a number of lines due to blending (namely \cii, \cii*, and \oi).  $N$(\feii) could only be constrained from \feii~$\lambda1144$, which is known to suffer heavily from hidden saturation (see Section~\ref{sec:hidden_sat}). This is the reason why this column density and the inferred abundance are listed as lower limits in Tables~\ref{tab:Nvals}  and \ref{tab:abundances}. Due to the blending between the two velocity components, only the higher velocity component of \Ni~$\lambda$1134.9 was isolated enough to be fitted separately to constrain its parameters. The latter were then kept fixed, together with the redshift of the second component as inferred from an average of other ions, whilst fitting the second velocity component by using all the three lines of the \Ni~triplet at $\sim 1134$ \AA. Since we were able in this way to nicely fit this triplet, we do not think that hidden saturation is an issue. Due to geocoronal contamination, \sIii~$\lambda$1304 was fitted on a `night' spectrum and is shown in Fig.~\ref{fig:M83_OIspec}.

\begin{figure*}
\center
\includegraphics{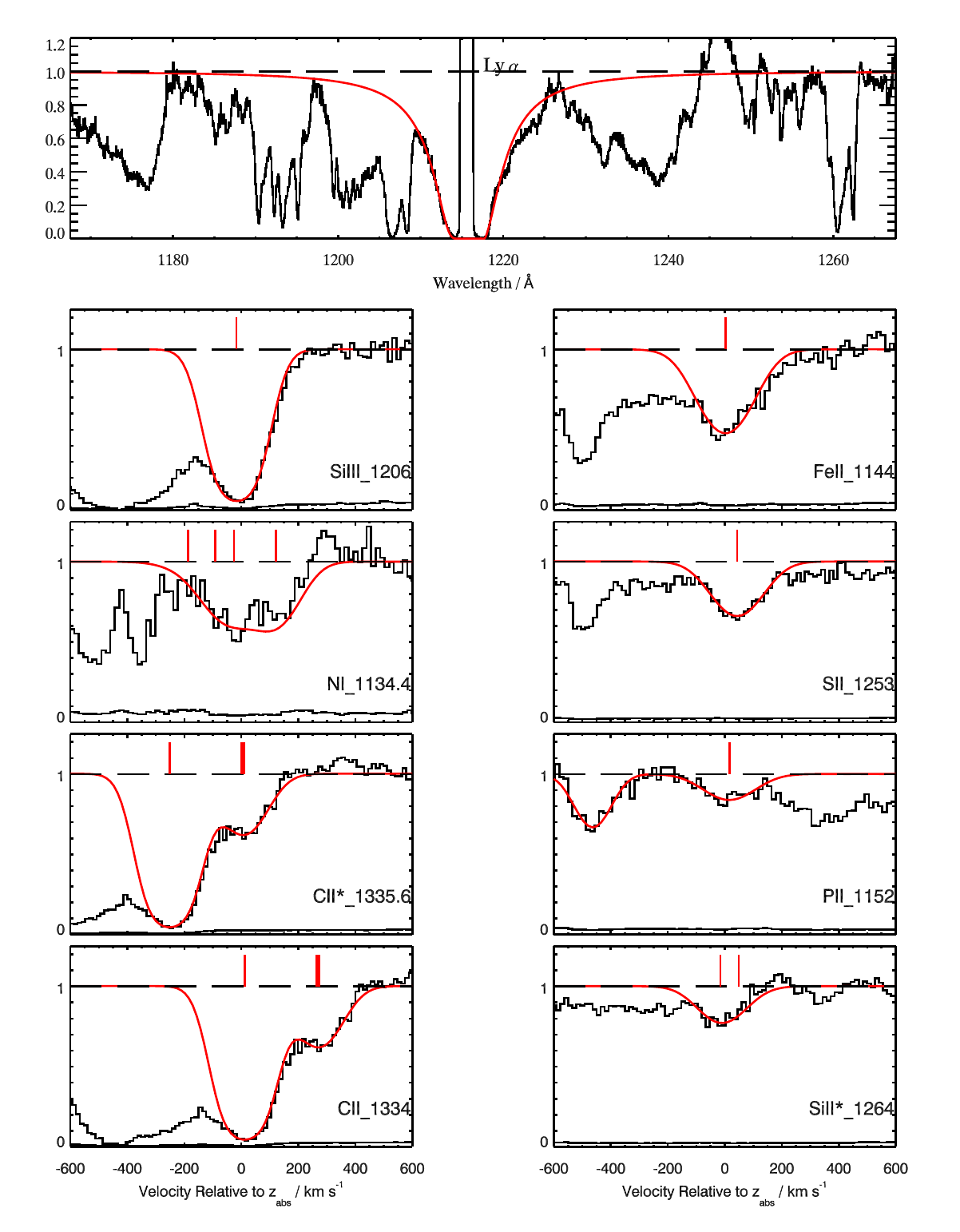}
\vspace{0.2cm}
\caption{Same as Fig.~\ref{fig:IZW18_spec} but for absorption-line profiles in position~2 of M83.  The top panel shows the \Lya\, profile at $z_{abs}=0.0015596$ ($v = 468$ \kms).}
\label{fig:M83-2_spec}
\end{figure*}

\textit{M83 Position 2}
The top panel of Fig.~\ref{fig:M83-2_spec} presents the \Lya\, region in the COS spectrum of position~2 of M83, and a best-fit profile corresponding to an M83 component of log[$N$(\hi)] = $18.44\pm0.30$ cm$^{-2}$ and a Milky Way component of log[$N$(\hi)] = $20.57\pm0.05$ cm$^{-2}$. The latter value is similar to the MW \hi\, column density of log[$N$(\hi)] = 20.8 as measured in the latest \hi\, Leiden/Dwingeloo survey \citep[see][]{Hartmann:1997}. The two \hi\,components were fitted simultaneously by fixing the MW redshift at 0.0 and the M83 redshift at the value of a well constrained ion, such as \feii.  The b parameter of the MW component was also fixed at the average value of 75 \kms as inferred from some isolated MW lines (\pii\,$\lambda1152$, the \Ni~triplet at $\sim 1134$, and \sii\,$\lambda1253$). The lower panels of Fig.~\ref{fig:M83-2_spec} show the main metal absorption lines fitted within the spectrum.  In contrast to position~1, the \sIiii~$\lambda$1206 stellar line does not show P-Cygni profile due to stellar winds in B0-1 stars affecting the \Lya\, profile fit. The spectrum of this COS target has the lowest \hi\, column density within our sample by almost $\sim 1$~dex. As a result, ICFs for the column densities measured from it are the highest compared to the other targets (see Table~\ref{tab:ICFs}).

Unlike M83 position~1, all lines are best fitted with a single velocity component at $z_{abs}=0.0015593$ ($v\sim470$~\kms) and with $b$ parameters of $\sim$\,100-110~\kms.  The \ciii~$\lambda$1176 photospheric line shows that the velocity of the stellar population lies at $\sim$\,420~\kms, a blueshift of $\sim$\,50~\kms\, relative to the \hi\, gas.  As a consequence of being only one velocity component within the absorbing gas, we were able to constrain more elements than in the spectrum of position~1, namely \cii\, and \cii*.  However, we were still unable to constrain \oi\ due to contamination from \sIii~$\lambda$1304 from the Milky Way, and \nIii\, due to noise in that region of the spectrum.  

As with M83 position~1, we were unable to use \feii~$\lambda$1142 and $\lambda$1143 to constrain $N$(\feii) due to severe blending with \feii\, absorption from the Milky Way, and \feii~$\lambda$1144 was instead used. Since this line is known to suffer from hidden saturation (Section~\ref{sec:hidden_sat}), the column density and abundance calculated from it are listed as lower limits in Tables~\ref{tab:Nvals}  and \ref{tab:abundances}.  With regards to contamination from ionized gas,  \sIiii\, and \sIii* (both thought to arise from more highly ionized gas) were found to have redshifts and $b$ parameters that are in agreement with the other species, suggesting that ionized gas may be present within the absorbing cloud.  
The \sIii~$\lambda$1304 absorption line is shown in Fig.~\ref{fig:M83_OIspec}, and was fitted on `night' spectra due to geocoronal contamination within the 1304~\AA\, region.

\section{An average spectrum of a $z=0$ Star-Forming Galaxy}\label{sec:avg_spec}

Averaging the column densities (in logarithmic scale) of each species as measured in the spectra of the sample, we have created a synthetic spectrum that represents the `average neutral ISM of a $z=0$ star-forming galaxy'.  This `average' spectrum does not include the following absorption lines:  (1) lines that are particular to the kinematical properties of a certain galaxy (e.g., multiple components); (2) lines that are known to originate in the ionized gas like, e.g., \sIiii, \feiii, \sIii*, and to some extend \cii*\,(only the dominant state of a certain species within the neutral ISM was included); (3) photospheric/stellar wind lines from the stellar population of the UV background source. 

Average column densities (and corresponding standard deviations) were derived only from individual values known to be unaffected by classical saturation, and were weighted by their errors.  For those ions with the strongest transitions affected by hidden saturation in some or all the targets of the sample, we also modeled these transitions by using the correct column density inferred from the weakest lines. As a consequence, the `saturated' transitions appear much stronger in the synthetic spectrum than what was actually observed. For species that are only available in saturated format  (e.g., \cii, \sIii, and \oi), we provided an average lower-limit profile. 
For those elements with multiple velocity components, we added the column densities within each spectrum before calculating the average, with the exception of NGC~3690 where two separate components of \hi~were measured.

The \Lya\, profile is represented by the average \hi~column density, log[$N$(\hi)/cm$^{-2}$] = $\,21.09\pm0.01$. For reference the average metallicity of  the sampled pointings as inferred from the literature is $Z \sim 1$ Z$_{\odot}$ (see Table~\ref{tab:ICFs} for the metallicity values of each pointing). The $b$ parameter for all lines was set at a nominal value of 100~\kms, while the average resolution of the sample, 25 \kms, was adopted. The average spectrum was created from column densities that have not been corrected for their ICFs, since this is what has been measured in the spectra. 

\begin{figure*}
\center
\includegraphics[scale=0.7,angle=90]{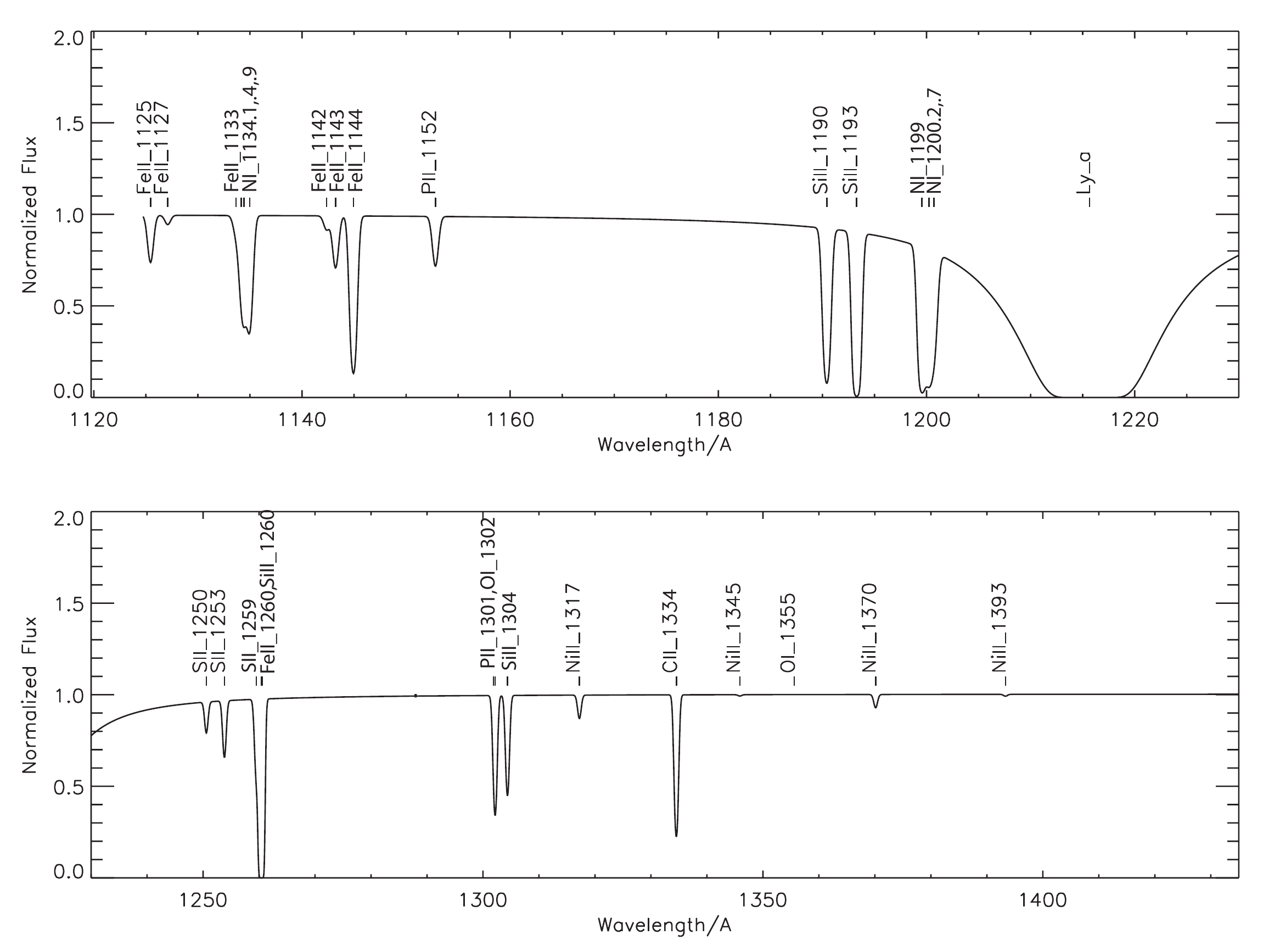}
\vspace{0.2cm}
\caption{Average flux-normalized absorption-line spectrum of the neutral gas in star-forming galaxies at $z=0$.  Synthetic line profiles were created using error-weighted average column densities of each species seen within our COS spectrum (excluding those known to originate in ionized gas or photosphere of stars), a nominal $b$ parameter of 100~\kms, and an average resolution of 25 \kms. The spectrum is also available in a machine-readable format in the online version of this publication.} 
\label{fig:avg_spec}
\end{figure*}

Figure~\ref{fig:avg_spec} shows the average flux-normalized absorption-line `synthetic' spectrum in the restframe wavelengths. The specific lines and average column densities adopted in our modeling are listed in Table~\ref{tab:avg_spec}, while the corresponding average abundances are listed in Table~\ref{tab:avg_abund}. No ICF corrections were considered, since these corrections are negligible (ICF$_{neutral}$) or comparable (ICF$_{ionized}$) to the errors at the average \hi~value of the `synthetic' spectrum (see Table~\ref{tab:ICFs}). The spectrum is also available in a machine-readable format in the online version and is designed to be used as a template for other star-forming galaxies within any redshift regime.

\section{Conclusion}\label{sec:discussion}
We have presented new \textit{HST}/COS FUV spectra of 9 nearby star-forming galaxies. For two galaxies within our sample, NGC~5253 and M83, two sets of spectroscopic data were obtained at different pointings.  The diversity in metallicity, galaxy type, and star formation activity covered by the galaxies in our sample is extensive, offering a unique opportunity to investigate the metallicity behavior of the neutral gas as a function of the galaxy properties at redshift $ z = 0$. The COS observations, with an average spectral resolution of $\sim$25~\kms, have S/N ranging from $\sim10$ to $30$ per 6-pixel resolution element and cover the wavelength region $\sim$1150--1450~\AA, providing an unparalleled view of the neutral gas in the local Universe.

Each of the 11 spectra were normalized using regions unaffected by absorption-line profiles, and column densities were inferred by using a multi-component line-profile fitting technique. Many interstellar absorption lines of neutral, singularly-, and doubly-ionized atoms of heavy elements were detected and analyzed in each spectrum.  

\hi~column densities were estimated by fitting \Lya. For each galaxy the intrinsic \Lya~is slightly blended with the Milky Way absorption component, so that a simultaneous fit of both components was required. The only exception is NGC~3690, where the higher redshift is enough to ensure complete separation of the intrinsic absorption from the Milky Way component. The \Lya\, absorption-line profile region was also carefully inspected for  possible stellar contamination. In order to exclude such contamination, model stellar spectra at the specific age and metallicity of the stellar population in each galaxy of the sample were produced, the corresponding stellar  \Lya\, absorption-line profiles were fitted and the resulting stellar column densities were compared to the interstellar ones. In each case, the stellar absorption contributed a negligible amount to the column density of \Lya\, and was not considered further.

\feii\, and \Ni\, are the ions with the best constrained column densities, as both are based on multiple lines with different oscillator strengths.  Column density determinations of other ions, like \cii, \cii*, \oi, \sIii, \sIii*, \sIiii, \pii, \sii, \feiii, and \nIii, were also made.  The single-, two-, and three-velocity component models used in the multi-component fitting technique turned out to be a good representation of the data when compared with the column densities inferred from the optical-depth method.

Several factors were taken into account when measuring column densities to ensure the most accurate measurements possible.  For observations affected by geocoronal emission around the 1302~\AA\, region,  \oi~$\lambda$1302 and \sIii~$\lambda$1304 absorption-line profiles were measured on `night-time' spectra. When multiple lines from the same ion were available giving inconsistent fitting results, column densities were measured from the line(s) with the lowest oscillator strength in order to minimize the effects of hidden saturation. This was the case, e.g., for \feii\, and \Ni\, since multiple lines with different oscillator strengths were available for these ions. The amount of hidden saturation was found to be a combination of the 
amount of a certain ion within the neutral gas  and the amount of gas along the line of sight. Hidden saturation effects occur when log[$N$(X)\,/\,cm$^{-2}$]$\,\gtrsim14.5$, with \feii~that is saturated when log[$N$(\hi)\,/\,cm$^{-2}$]$\,\gtrsim20.6$ and with \Ni~that is saturated when log[$N$(\hi)\,/\,cm$^{-2}$]$\,\gtrsim21.1$. Classical saturation effects were also assessed for each species by comparing our measurements with a theoretical curve of growth.

Abundances of several elements, i.e., C, N, O, Si, P, S, Fe, and Ni, were inferred from the column densities of the ions measured within each galaxy.  These abundances were corrected for classical ionization and for contamination of ionized gas along the line of sight according to ionization correction factors inferred from ad-hoc CLOUDY photoionization models specific to the metallicity and \hi\, column density of the galaxy. We were unable, however, to constrain the dust content of the sample due to the lack of appropriate dust depletion indicators in the wavelength range of our COS spectra.

Using in logarithmic scale the error-weighted average column densities of species thought to arise from the neutral ISM only, we derived an `average absorption-line spectrum of a $z=0$ star-forming galaxy'.  This synthetic spectrum is aimed to provide users with a `template' spectrum of the neutral ISM within star-forming galaxies.

Paper~II of this series will present a direct comparison of the metal abundances in the neutral and ionized gas (as inferred from the \hii~regions) within each galaxy of our COS sample and a discussion of the implications for the chemical (in)homogeneity of the multiphase ISM in these SFGs of the local Universe.  Paper~III will finally compare the interstellar abundances in the galaxies of our sample with those observed in the local and high-redshift Universe and discuss the implications of our findings within a cosmological context.

\acknowledgments
The authors give thanks to Edward Jenkins, Jason Tumlinson, Chris Thom, and Max Pettini for valuable discussions concerning various absorption-line related matters and to an anonymous referee for useful comments that greatly improved the paper. We also greatly appreciated discussions with Andrew Fox regarding photoionization modelling and Derck Massa regarding photospheric-line identification. We would like to sincerely thank Stephen McCandliss for his assistance in generating curves of growth and Vianney Lebouteiller for providing the model spectra used in \S~\ref{sec:hi_colDens}. Phil Hodge, Tom Ake, Stephane Beland, and Steve Penton are sincerely acknowledged for their advice and help with the removal of geocoronal contamination. AA would also like to thank Helene McLaughlin for collaboration on a related project and for help with the start of this project. We acknowledge the usage of the HyperLeda database (http://leda.univ-lyon1.fr). Some of the data presented in this paper were obtained from the Mikulsky Archive at the Space Telescope Science Institute (MAST). STScI is operated by the Association of Universities for Research in Astronomy, Inc., under NASA contract NAS5-26555. Support for MAST for non-HST data is provided by the NASA Office of Space Science via grant NNX09AF08G and by other grants and contracts. Support for Program number 11579 was provided by NASA through a grant from the Space Telescope Science Institute, which is operated by the Association of Universities for Research in Astronomy, Incorporated, under NASA contract NAS5-26555.

{\it Facilities:} \facility{HST (ACS/SBC; COS)}


\begin{table*}
\begin{scriptsize}
\caption{Global properties of the \emph{HST}/COS sample of nearby SFGs} 
\label{tab:sample}
\begin{tiny}
\begin{tabular}{lccccccccccc}
\tableline\tableline
Object Name   &    RA         &    DEC        & Type  &12+log(O/H) & Vel     & Distance        & E(B-V)   &  log(L$_{UV}+$L$_{FIR}$)  &  Reference        & Reference & Reference \\
              & (J2000)       & (J2000)       &       &            & (kms/s) & (Mpc)           &          &    (erg/s)              & L$_{UV}$,L$_{IR}$   & Distance  & 12+log(O/H)\\                                                                   
\tableline                                                                                                                                                 
I~Zw~18           	& 09 34 02.298 & +55 14 25.07     & BCD        & 7.2  	   & 753     &  18.2$\pm$1.5   & 0.032 	  & $<$41.12  	     	  & 1,2 	 	  & 10 	  & 17 	\\                                  
                                                                                                                                                                                                 
SBS~0335-052 	& 03 37 44.002 &$-$05 02 38.32   & BCD        & 7.3        & 4,043   &  53.7$\pm$3.8   & 0.047    & $<$43.63             & 3,4               & 11        & 17        \\                                       
                                                                                                                                                                                                
SBS~1415$+$437   & 14 17 1.406  & +43 30 4.75         & BCD        & 7.6        & 609     &  13.6$\pm$0.9   & 0.009    & 43.19                    & 5,6               & 12        & 18        \\                                     
                                                                                                                                                                                                
NGC~4214              & 12 15 39.413 & +36 19 35.17      & Irr           & 8.2        & 291     &  3.04$\pm$0.04  & 0.022    &  43.08                  & 1,7               & 13        & 19        \\                                     
                                                                                                                                                                                               
NGC~5253              & 13 39 56.976 &$-$31 38 27.01    & Irr           & 8.2        & 403     &  3.77$\pm$0.20  & 0.056    & 43.52                   & 1,7               & 14        & 20        \\                                    
                                                                                                                                                                                                 
NGC~4670 	       & 12 45 16.906 & +27 07 30.02       & BCD        & 8.2        & 1,069   &  23.1$\pm$1.6   & 0.015    & 44.20                   &  8                & 11        & 21        \\                                       
                                                                                                                                                                                                
NGC~4449             & 12 28 10.816 & +44 05 42.95       & Irr           & 8.3        & 203     &  3.82$\pm$0.27  & 0.019    &  43.65                  & 1,9               & 15        & 22        \\                                     
                                                                                                                                                                                                
NGC~3690             & 11 28 31.003 & +58 33 41.08       & Merger    & 8.8        & 3,119   &  48.5$\pm$3.4   & 0.017    & 46.23                   & 1,7               & 11        & 21        \\                                  
                                                                                                                                                                                                
M83                       & 13 37 0.515  &$-$29 52 00.48      & SAB(s)c   & 9.2        & 513     &  4.8$\pm$0.2  & 0.066    &  44.34                  & 1,7               & 16        & 22        \\

\tableline
\end{tabular}
\\[3.0mm]
Notes: Units of right ascension are hours, minutes, and seconds, and units of declination are degrees, arcminutes, and arcseconds.  The radial velocities, E(B-V) extinction and classifications of the objectes are cited from the NASA/IPAC Extragalactic Database (NED), which is operated by the Jet Propulsion Laboratory, California Institute of Techonology, under contract with the National Aeronautics and Space Administration.  Key for classification: BCD$=$Blue compact dwarf galaxy, Irr$=$Irregular galaxy and SAB(s)c$=$Weakly barred spiral galaxy with loosely wound arms (type c) and no ring-like structure. L$_{UV}$ and L$_{FIR}$ references: (1)=\citet{Kinney:1993}, (2)=\citet{Gezari:1999}, (3)=\citet{Leitherer:2011}, (4)=\citet{Dale:2001}, (5)=\textit{HST/FOS} archival spectra, (6)=\citet{ISOcat:2001}, 
(7)=\citet{IRAScat:1989}, (8)=This paper and (9)=\citet{Rush:1993}.  Distance references: (10)=\citet{Aloisi:2007}, (11)=\citet{Mould:2000}, (12)=\citet{Aloisi:2005}, (13)=\citet{Dalcanton:2009}, 
(14)=\citet{Sakai:2004}, (15)=\citet{Annibali:2008} and (16)=\citet{Radburn-Smith:2011}.  Metallicity references: (17)=\citet{Izotov:1999}, (18)=\citet{Thuan:1999_sbs},  (19)=\citet{Kobulnicky:1996}, 
(20)=\citet{Walsh:1989}, (21)=\citet{Heckman:1998}, (22)=\citet{Marble:2010}.
\end{tiny}
\end{scriptsize}
\end{table*}

\begin{table*}
\begin{center}
\begin{scriptsize}
\caption{Details of \textit{HST}/ACS observations from PID:11579, except for SBS~0335-052 which is archival data from PID:9470}\label{tab:obs_ACS}
\begin{tabular}{lccccccc}

\tableline \tableline

Target Name  &    RA        &    DEC        & Dataset & Obs. Date & Exp. Time & Aperture & Filter\\
                      &   (J2000)  &    (J2000)   &              &                &  (sec)            &                &         \\
\tableline                                  
                                            
I~Zw~18      & 09 34 02.298  & +55 14 25.07 &JB7H07010& 2009-02-07& 2900  & SBC-FIX &F125LP\\
                                                                   
SBS~0335-52  & 03 37 44.002 & -05 02 38.32 & J8F710020& 2002-10-25& 2700 & SBC-FIX & F140LP \\
                                                                   
SBS~1415+437 & 14 17 1.406  & +43 30 4.75  & JB7H08010& 2009-01-29& 2720  & SBC-FIX & F125LP\\
                                                                   
NGC~4214     & 12 15 39.413 & +36 19 35.17 & JB7H03010& 2009-02-08& 2680  & SBC-FIX & F125LP\\
                                                                   
NGC~5253     & 13 39 56.976  & -31 38 27.01& JB7H06010& 2009-03-07& 2660  & SBC-FIX & F125LP\\
                                                                   
                                                                   
NGC~4670     & 12 45 16.906 & +27 07 30.02 & JB7H05010& 2009-01-24& 2640  & SBC-FIX & F125LP\\
                                                                   
NGC~4449     & 12 28 10.816 & +44 05 42.95 & JB7H04010& 2009-02-03& 2720  & SBC-FIX & F125LP\\
                                                                   
NGC~3690     & 11 28 31.003 & +58 33 41.08 & JB7H02010& 2009-02-03& 2900  & SBC-FIX & F125LP\\
                                                                   
M83          & 13 37 0.515  & -29 52 00.48 & JB7H01010& 2009-01-03& 2640  & SBC-FIX & F125LP\\
             

\tableline    
\end{tabular}
\end{scriptsize}
\end{center}
\end{table*}

\begin{table*}
\begin{center}
\begin{scriptsize}
\caption{Details of \textit{HST}/COS observations from PID:11579}\label{tab:obs_COS}
\begin{tabular}{lccccccccc}
\tableline \tableline

Target Name          &    RA        &    DEC       & Dataset & Obs. Date & Exp. Time & Target Acq. Config. & Grating/Setting & FWHM$_{spec}$ & S/N \\
                     &    (J2000)   &    (J2000)   &         &           &  (s)      &                     &                 & (km\,s$^{-1}$) &    \\ 
\tableline                                         
                                                   
I~Zw~18              & 09 34 1.970  & +55 14 28.10& LB7H71010& 2011-01-24& 15523  & PSA/Mirror\,A & G130M/1291 &  22     &  31   \\
                                                                          
SBS~0335-052     & 03 37 43.980 & $-$05 02 38.90& LB7H91010& 2010-03-02& 9534  & PSA/Mirror\,A & G130M/1291  &  21     &   22  \\
                                                                                                                         
SBS~1415+437      & 14 17 1.420  & +43 30 5.16 & LB7H81010& 2010-04-16& 12767 & PSA/Mirror\,A & G130M/1291  &  32     &  29   \\ 
                                                                                                                         
NGC~4214             & 12 15 39.480 & +36 19 35.36& LB7H31010& 2011-05-04& 1968 & PSA/Mirror\,B & G130M/1291   &  25     &  26   \\
                                                                                                                         
NGC~5253-1        & 13 39 56.020  & $-$31 38 31.30&LB7H61010& 2010-07-03& 3038 & PSA/Mirror\,A & G130M/1291   &  34     &   16  \\
                                                                                                                         
NGC~5253-2        & 13 39 55.889 & $-$31 38 38.34&LB7H61020& 2010-07-03& 6296  & PSA/Mirror\,A & G130M/1291  &  26     &  18   \\
                                                                                                                         
NGC~4670             & 12 45 17.265 & +27 07 32.13& LB7H51010& 2009-11-26& 1544  & PSA/Mirror\,B & G130M/1291  &  23     &   18  \\
                                                                                                                         
NGC~4449             & 12 28 11.089 & +44 05 37.06& LB7H41010& 2010-05-26& 1736  & PSA/Mirror\,B & G130M/1291  &  23     &  17   \\
                                                                                                                         
NGC~3690             & 11 28 29.149 & +58 33 41.01& LB7H21010& 2010-09-25& 2276  & PSA/Mirror\,A & G130M/1291  &  22     &  11   \\
                                                                          
M83-1                   & 13 37 0.458  & $-$29 51 54.58& LB7H11010& 2010-08-06& 4093  & PSA/Mirror\,B & G130M/1291  &   21    &  26   \\
                                                                          
M83-2                   & 13 37 0.508  & $-$29 52 1.22 & LB7H11020& 2010-08-06& 2284  & PSA/Mirror\,B & G130M/1291  &   22    &  32   \\

\tableline
\end{tabular}
\end{scriptsize}
\end{center}
\end{table*}

\setlength{\tabcolsep}{3pt}

\begin{table*}
\begin{center}
\begin{scriptsize}
\caption{\hi\, column densities for each UV source targeted in our study and its Milky Way component}\label{tab:HIdensities}
\begin{tabular}{lccc|ccc|c}
\tableline \tableline

               & \multicolumn{3}{c}{Source}			                & \multicolumn{3}{c}{Milky Way}		                &  Stellar		\\
\cline{2-8}																								
Object	        &	log[$N$(\hi)]		& $\nu_{HI}$	& z$_{HI}^a$       &  log[$N$(\hi)]$_{MW}^{b,c}$    & $\nu_{HI}$	&   log[$N$(\hi)]$_{map}^d$   &  $\nu_{\lambda1176}^e$\\
	        &	(cm$^{-2}$)		&(km\,s$^{-1}$)	&               &	(cm$^{-2}$)	&(km\,s$^{-1}$)	&(cm$^{-2}$)     &(km\,s$^{-1}$)	        \\
\cline{1-8}																							
 I~Zw~18	&	21.28	$\pm$	0.03	&	767	&  0.0025575    &	20.49		&	0	&	20.49	&	763		\\
SBS~0335-052	&	21.70	$\pm$	0.05	&	4,038	&  0.0134685    &	20.4$\pm$0.1	&	0	&	20.15	&	4,073		\\
SBS~1415+437	&	21.09	$\pm$	0.03	&	582	&  0.0019413    &	20.29		&	0	&	20.29	&	610		\\
NGC~4214	&	21.12	$\pm$	0.03	&	286	&  0.0009544    &	20.08		&	$-$33	&	20.08	&	$>$369		\\
NGC~5253-1	&	21.20	$\pm$	0.01	&	378	&  0.0012623    &	20.68		&	0	&	20.68	&	399		\\
NGC~5253-2	&	20.65	$\pm$	0.05	&	455	&  0.0015613    &	20.68		&	27	&	$''$	&	422		\\
NGC~4670	&	21.07	$\pm$	0.08	&	1,092	&  0.0036438    &	20.25$\pm$0.12	&	0	&	20.24	&	1,045		\\
NGC~4449	&	21.14	$\pm$	0.03	&	242	&  0.0008061    &	20.21		&	0	&	20.21	&	236		\\
NGC~3690$_{v1}$	&	20.62	$\pm$	0.02	&	3,207	&  0.0106978    &	---		&	---	&	20.13	&	3,046		\\
NGC~3690$_{v2}$	&	19.82	$\pm$	0.08	&	2,853	&  0.0095165    &	---		&	---	&	$''$	&	$''$		\\
NGC~3690$_{v1+v2}$&	20.68	$\pm$	0.01	&	3,152	&  0.0105132    &	---	        &	 ---	&	$''$	&       $''$		\\
M83-1	        &	19.60	$\pm$	0.32	&	348	&  0.0011619    &	20.57$\pm$0.05$^f$	&	0	&	20.80	&       $>$150		\\
M83-2	        &	18.44	$\pm$	0.30	&	468	&  0.0015596    &	20.57$\pm$0.05	&	0	&	$''$	&       $>$420		\\
\tableline
\end{tabular}
\begin{description}
\vspace{0.3cm}
\begin{tiny}
\item[$^a$] Most of the redshifts were constrained from other species in the neutral gas and were kept fixed during the fit. 
\item[$^b$] $N$(\hi)$_{MW}$ values listed without errors correspond to cases where $N$(\hi)$_{map}$ was utilized and held fixed whilst deriving $N$(\hi) for the UV source in the galaxy, as described in Section~\ref{sec:hi_colDens}.
\item[$^c$] $N$(\hi)$_{MW}$ is not listed when the galaxy redshift is large enough and the strength of the Ly$\alpha$ absorption is small enough for $N$(\hi) to be uncontaminated by $N$(\hi)$_{MW}$.
\item[$^d$] $N$(\hi)$_{map}$ is estimated from the composite all-sky map of neutral hydrogen column density, formed from the Leiden/Dwingeloo survey data \citep{Hartmann:1997} and the composite $N$(\hi) map of \citet{Dickey:1990}. 
\item[$^e$] Systemic velocities of the stellar population within each UV source as derived from the \ciii~$\lambda$1176 photospheric absorption line.  These velocities are lower limits for NGC~4214, and for position~1 and position~2 of M83 due to the presence of P-Cygni absorption-line profiles at this wavelength.
\item[$^f$] This value of the MW H\,{\sc i} column density was assumed from position~2 of M83.
\end{tiny}
\end{description}
\end{scriptsize}
\end{center}
\end{table*}


\begin{table*}
\begin{center}
\begin{scriptsize}
\caption{List of lines identified in the COS spectra of this study and corresponding theoretical parameters}\label{tab:fvals}
\begin{tabular}{lccc|lccc}
\tableline \tableline
Line ID & $\lambda_{lab}^a$ (\AA) & $f^b$ & Reference& Line ID & $\lambda_{lab}^a$ (\AA) & $f^b$ & Reference \\
\tableline
\Lya\,   & 1215.6701 & 4.1640e-01 & 1 &             \sIii*~1197 &  1197.3938 &  1.4600e-01 & 7 \\ 
\ci~1139 & 1139.7930 & 1.3960e-02 &   1 &            \sIii*~1264 &  1264.7377 &  1.0600e-00 & 7\\              
\ci~1157 & 1157.9097 & 2.1780e-02 &   1 &            \sIii*~1265 &  1265.0020 &  1.1800e-01 & 7\\              
\ci~1188 & 1188.8332 & 1.2900e-02 &   2 &            \sIii*~1309  & 1309.2757 &  8.6700e-02 & 7\\   
\ci~1193.0 & 1193.0308 & 4.4470e-02 &   1&           \sIiii~1206 & 1206.5000 &  1.6690e-00 & 1\\         
\ci~1193.9 & 1193.9955 & 1.2800e-02 & 3&           \sIiv~1393 & 1393.7550 &  5.1400e-01 & 1\\         
\ci~1260 & 1260.7351 & 5.0700e-02 &   4 &             \sIiv~1402 & 1402.7700 &  2.5530e-01 & 1\\         
\ci~1277 & 1277.2452 & 8.5300e-02 &   4&               \pii~1152 &  1152.8180 &  2.3610e-01 & 1\\    
\ci~1280 & 1280.1353 & 2.4320e-02 &   1&              \pii~1301 &  1301.8740 &  1.2710e-02 & 8\\                                           
\ci~1328 & 1328.8333 & 7.5800e-02 &   4&               \sii~1250 &  1250.5840 &  5.4530e-03 & 1\\           
\cii~1344 & 1334.5323 & 1.27800e-01 & 1 &              \sii~1253 &  1253.8110 &  1.0880e-02 & 1\\         
\cii*~1335.6 & 1335.6627 & 1.2770e-02 &   1&          \sii~1259 &  1259.5190 &  1.6240e-02 & 1\\         
\cii*~1335.7 & 1335.7077 & 1.1490e-01 &   1&          \siii~1190 & 1190.2080 &  2.3000e-02 & 9\\                                            
\Ni~1134.1 & 1134.1653 &   1.3420e-02 & 1&            \feii~1121 & 1121.9748 &  2.0200e-02 &  10\\       
\Ni~1134.4 & 1134.4149 & 2.6830e-02 &  1&              \feii~1125 & 1125.4477 &  1.6000e-02 & 10\\                                              
\Ni~1134.9 & 1134.9803 &  4.0230e-02 &  1&             \feii~1127 & 1127.0984 &  2.8000e-03 & 10\\      
\Ni~1199 &   1199.5496 &  1.3280e-01 &  1&             \feii~1133 & 1133.6650 &  5.5000e-03 & 10\\      
\Ni~1200.2 & 1200.2233 &  8.8490e-02 &  1&             \feii~1142 & 1142.3656 &  4.2000e-03 & 10\\      
\Ni~1200.7 & 1200.7098 &  4.4230e-02 &  1&             \feii~1143 & 1143.2260 &  1.7700e-02 & 10\\      
\oi~1302 &  1302.1685 &  4.8870e-02 &  1&              \feii~1144 & 1144.9379 &  1.0600e-01 & 10\\      
\oi~1355 &   1355.5977 &  1.2480e-06 &  1&             \feii~1260 & 1260.5330 &  2.5000e-02 &   1\\      
\mgii~1239 & 1239.9253 &  5.9800e-04 & 5&        \feiii~1122 & 1122.5260 &  7.8840e-02 & 1\\                                                  
\mgii~1240 & 1240.3947 &  3.3220e-04 & 5&        \feiii~1207 & 1207.0500 &  4.4230e-06 & 1\\                                                  
\sIii~1190 & 1190.4158 &  2.9300e-01 & 2&       \feiii~1214 & 1214.5660 &  4.2730e-04 & 1\\          
\sIii~1193 & 1193.2897 &  5.8400e-01 & 2&        \nIii~1317 & 1317.2170 &  1.4580e-01 & 1\\          
\sIii~1260 & 1260.4221 &  1.1800e-00 & 2&        \nIii~1345 & 1345.8780 &  7.6900e-03 & 11$^c$\\          
\sIii~1304 & 1304.3702 &  8.6000e-02 & 6&         \nIii~1370 & 1370.1320 &  7.6900e-02 & 11$^c$\\               
\sIii*~1194 &  1194.5002 &  7.2900e-01 & 7&        \nIii~1393 & 1393.3240 &  1.0090e-02 & 11$^c$\\    
\sIii*~1197 &  1197.3938 &  1.4600e-01 & 7 & & & & \\   
 
\tableline
\end{tabular}
\begin{description}
\vspace{0.3cm}
\begin{tiny}
\item[$^a$] Vacuum wavelengths from \citet{Morton:1991}.
\item[$^b$] Oscillator strengths from references indicated in column (4).
\item[$^c$] f values scaled by 0.534 \citep[see][]{Fedchak:1999}.
\item \textsc{References.--} (1)=\citet{Morton:1991},(2)=\citet{Verner:1994},(3)=\citet{Wiese:1996},(4)=\citet{Morton:2003},(5)=\citet{Majumder:2002},(6)=\citet{Spitzer:1993},(7)=\citet{Verner:1996},(8)=\citet{Hibbert:1988},(9)=\citet{Tayal:1995},(10)=\citet{Howk:2000},(11)=\citet{Zsargo:1998}.


\end{tiny}
\end{description}
\end{scriptsize}
\end{center}
\end{table*}

\clearpage
\LongTables

\begin{deluxetable*}{lcccc}

\tablecolumns{5}
\tablewidth{0pc}
\tablecaption{Column Densities\label{tab:Nvals}}

\tablehead{
\colhead{Ion} & \colhead{z} & \colhead{log[$N$(X)/cm$^{-2}$]} & \colhead{$b$ (km\,s$^{-1}$)} & \colhead {Lines Used (\AA)} 
}


\startdata										
\cutinhead{I~Zw~18}

\cii	&	0.0025582	&	14.41	$\pm$	0.01	$^\dagger$&	93.38	$\pm$	3.23	&	1334	\\
\cii*	&	0.0025299	&	13.90	$\pm$	0.03	&	84.52	$\pm$	7.79	&	1135.6, 1135.7	\\
\Ni	&	0.0025479	&	14.38	$\pm$	0.03	&	76.08	$\pm$	9.84	&	1134.1,1134.4,1134.9	\\
\oi	&	0.0025575	&	14.80	$\pm$	0.02	$^\dagger$&	97.08	$\pm$	5.34	&	1302, 1355	\\
\sIii	&	0.0025667	&	14.36	$\pm$	0.01	$^\dagger$&	86.81	$\pm$	2.68	&	1304	\\
\sIiii &	0.0025239	&	13.10	$\pm$	0.02	$^\dagger$&	86.81	$\pm$	...	&	1206	\\
\sii	&	0.0025570	&	14.73	$\pm$	0.05	&	60.99	$\pm$	9.03	&	1253	\\
\feii	&	0.0025298	&	14.52	$\pm$	0.06	&	88.68	$\pm$	14.64	&	1143	\\
 
\cutinhead{SBS~0335-052}

\cii	&	0.0134870	&	14.41	$\pm$	0.01	$^\dagger$&	43.89	$\pm$	1.67	&	1334	\\
\cii*	&	0.0134691	&	13.93	$\pm$	0.03	&	32.20	$\pm$	2.70	&	1135.6, 1135.7	\\
\Ni	&	0.0134475	&	14.83	$\pm$	0.02	&	33.14	$\pm$	2.84	&	1134.1,1134.4	\\
\oi	&	0.0135029	&	14.77	$\pm$	0.01	$^\dagger$&	42.94	$\pm$	1.59	&	1302, 1355 	\\
\sIii	&	0.0134972	&	14.38	$\pm$	0.02	$^\dagger$&	38.27	$\pm$	1.74	&	1304	\\
\sIiii	&	0.0133030	&	13.15	$\pm$	0.08	$^\dagger$&	57.32	$\pm$	16.39	&	1206	\\
\pii	&	0.0135045	&	12.97	$\pm$	0.15	&	30.87	$\pm$	12.48	&	1152	\\
\sii	&	0.0134914	&	14.96	$\pm$	0.03	&	34.49	$\pm$	2.62	&	1250, 1253	\\
\feii	&	0.0134795	&	15.32	$\pm$	0.04	&	42.10	$\pm$	4.74	&	1133, 1142	\\
\feiii	&	0.0134180	&	14.24	$\pm$	0.07	&	41.43	$\pm$	9.50	&	1122	\\
\nIii	&	0.0135370	&	13.96	$\pm$	0.06	&	44.56	$\pm$	8.26	&	1345,1370	\\

\cutinhead{SBS~1415+437}

\cii	&	0.0019160	&	14.51	$\pm$	0.01	$^\dagger$&	75.64	$\pm$	1.91	&	1334	\\
\cii*	&	0.0018006	&	13.79	$\pm$	0.03	&	47.31	$\pm$	4.38	&	1135.6, 1135.7	\\
\Ni	&	0.0018782	&	14.57	$\pm$	0.04	&	62.60	$\pm$	8.30	&	1134.1,1134.4	\\
\oi	&	0.0019592	&	14.87	$\pm$	0.09	$^\dagger$&	76.27	$\pm$	7.72	&	1302, 1355 	\\
\sIii	&	0.0019453	&	14.42	$\pm$	0.01	$^\dagger$&	64.48	$\pm$	2.62	&	1304	\\
\sIiii	&	0.0017674	&	13.43	$\pm$	0.01	$^\dagger$&	68.00	$\pm$	2.52	&	1206	\\
\pii	&	0.0019467	&	13.43	$\pm$	0.08	&	90.24	$\pm$	18.41	&	1152	\\
\sii	&	0.0019381	&	14.96	$\pm$	0.02	&	69.82	$\pm$	3.81	&	1253, 1259	\\
\feii	&	0.0019686	&	14.83	$\pm$	0.04	$^\dagger$&	55.43	$\pm$	7.55	&	1143	\\
\nIii	&	0.0019492	&	13.26	$\pm$	0.10	&	54.07	$\pm$	13.58	&	1317, 1345	\\
 											
\cutinhead{NGC~4214}

\cii* -1	&	0.0010080	&	14.13	$\pm$	0.04	&	29.80	$\pm$	2.42	&	1135.6, 1135.7	\\
\cii* -3	&	0.0008039	&	14.59	$\pm$	0.02	&	100.64	$\pm$	2.97	&	$\prime\prime$	\\
\Ni -1	&	0.0009152	&	15.06	$\pm$	0.07	&	47.15	$\pm$	4.90	&	1134.1,1134.4,1134.9	\\
\Ni -2	&	0.0008850	&	14.65	$\pm$	0.20	&	100.00	$\pm$	...	&	$\prime\prime$	\\
\oi -1	&	0.0009841	&	15.07	$\pm$	0.02	$^\dagger$&	48.42	$\pm$	1.75	&	1302, 1355	\\
\oi -2	&	0.0008850	&	14.78	$\pm$	0.03	$^\dagger$&	100.00	$\pm$	...	&	$\prime\prime$	\\
\sIii -1	&	0.0009512	&	14.73	$\pm$	0.02	$^\dagger$&	49.19	$\pm$	1.45	&	1304	\\
\sIii -2	&	0.0008850	&	14.31	$\pm$	0.05	$^\dagger$&	100.00	$\pm$	...	&	$\prime\prime$	\\
\sIii*-3	&	0.0007951	&	12.62	$\pm$	0.15	&	79.80	$\pm$	27.60	&	1264, 1265, 1309	\\
\sIiii-3	&	0.0008026	&	13.96	$\pm$	0.01	$^\dagger$&	88.09	$\pm$	2.27	&	1206	\\
\pii-1	&	0.0010124	&	13.76	$\pm$	0.03	&	52.14	$\pm$	4.77	&	1152	\\
\sii -1	&	0.0010378	&	15.09	$\pm$	0.05	&	22.89	$\pm$	2.82	&	1250,1253	\\
\sii -2	&	0.0009290	&	15.38	$\pm$	0.03	&	80.76	$\pm$	4.44	&	$\prime\prime$	\\
\feii -1	&	0.0010156	&	14.66	$\pm$	0.06	&	31.77	$\pm$	4.28	&	1142, 1143	\\
\feii -2	&	0.0008840	&	14.23	$\pm$	0.20	&	61.77	$\pm$	18.99	&	$\prime\prime$	\\
\nIii-1	&	0.0010636	&	13.48	$\pm$	0.04	&	33.84	$\pm$	4.60	&	1317, 1370	\\

\cutinhead{NGC~5253-Pos.~1}

\cii	&	0.0013150	&	15.04	$\pm$	0.01	$^\dagger$&	98.56	$\pm$	4.00	&	1334	\\
\cii*	&	0.0012772	&	14.68	$\pm$	0.02	&	88.23	$\pm$	3.97	&	1135.6, 1135.7	\\
\Ni	&	0.0012530	&	15.24	$\pm$	0.03	&	63.95	$\pm$	7.69	&	1134.1, 1134.4	\\
\oi	&	0.0014092	&	15.44	$\pm$	0.14	$^\dagger$&	89.27	$\pm$	26.14	&	1302, 1355	\\
\sIii	&	0.0013121	&	14.79	$\pm$	0.02	$^\dagger$&	72.70	$\pm$	3.20	&	1304	\\
\sIii*	&	0.0012968	&	13.03	$\pm$	0.05	&	121.86	$\pm$	16.94	&	1264, 1265, 1309	\\
\sIiii	&	0.0012960	&	14.07	$\pm$	0.03	$^\dagger$&	124.23	$\pm$	12.91	&	1206 	\\
\pii	&	0.0012822	&	13.74	$\pm$	0.08	&	80.74	$\pm$	20.30	&	1152	\\
\sii	&	0.0013554	&	15.44	$\pm$	0.02	&	68.45	$\pm$	4.15	&	1250, 1253	\\
\feii	&	0.0013338	&	15.67	$\pm$	0.07	&	96.22	$\pm$	20.03	&	1142	\\
\nIii	&	0.0014081	&	13.68	$\pm$	0.08	&	76.67	$\pm$	20.04	&	1317, 1345, 1370	\\

\cutinhead{NGC~5253-Pos.~2}

\cii	&	0.0013976	&	14.92	$\pm$	0.01	$^\dagger$&	94.84	$\pm$	3.70	&	1334	\\
\cii*	&	0.0013344	&	14.55	$\pm$	0.02	&	92.79	$\pm$	4.53	&	1135.6, 1135.7	\\
\Ni	&	0.0013033	&	14.98	$\pm$	0.02	&	70.05	$\pm$	7.19	&	1134.1, 1134.4, 1134.9	\\
\oi	&	0.0014840	&	15.31	$\pm$	0.05	$^\dagger$&	79.58	$\pm$	10.73	&	1302, 1355	\\
\sIii	&	0.0014370	&	14.81	$\pm$	0.01	$^\dagger$&	90.74	$\pm$	3.88	&	1304	\\
\sIii*	&	0.0013754	&	12.53	$\pm$	0.08	&	41.24	$\pm$	12.02	&	1264, 1265, 1309	\\
\sIiii	&	0.0014290	&	13.80	$\pm$	0.02	$^\dagger$&	100.05	$\pm$	6.35	&	1206	\\
\pii	&	0.0015805	&	13.65	$\pm$	0.11	&	125.07	$\pm$	41.11	&	1152	\\
\sii	&	0.0014449	&	15.37	$\pm$	0.02	&	80.06	$\pm$	4.72	&	1250, 1253	\\
\feii	&	0.0012981	&	15.66	$\pm$	0.04	&	90.00	$\pm$	...	&	1142	\\
\nIii	&	0.0014452	&	13.77	$\pm$	0.13	&	121.22	$\pm$	48.73	&	1317, 1345, 1370	\\

\cutinhead{NGC~4670}

\cii	&	0.0036536	&	15.02	$\pm$	0.01	$^\dagger$&	91.22	$\pm$	1.95	&	1334	\\
\cii*	&	0.0035558	&	14.70	$\pm$	0.01	&	71.10	$\pm$	2.34	&	1135.6, 1135.7	\\
\Ni	&	0.0035143	&	15.42	$\pm$	0.01	&	75.35	$\pm$	3.44	&	1134.1,1134.4,1134.9	\\
\oi	&	0.0035890	&	15.26	$\pm$	0.02	$^\dagger$&	75.49	$\pm$	2.59	&	1302, 1355 	\\
\sIii	&	0.0036415	&	14.97	$\pm$	0.01	$^\dagger$&	65.62	$\pm$	1.76	&	1304	\\
\sIii*	&	0.0035529	&	12.99	$\pm$	0.06	&	108.91	$\pm$	20.21	&	1264, 1265, 1309	\\
\sIiii	&	0.0036415	&	13.98	$\pm$	0.03	$^\dagger$&	65.62	$\pm$	...	&	1206	\\
\pii	&	0.0035785	&	13.82	$\pm$	0.05	&	76.94	$\pm$	12.19	&	1152	\\
\sii	&	0.0036295	&	15.60	$\pm$	0.01	$^\dagger$&	72.47	$\pm$	2.06	&	1253, 1259	\\
\feii	&	0.0036338	&	15.07	$\pm$	0.05	&	67.25	$\pm$	9.89	&	1143	\\
\nIii	&	0.0036604	&	13.76	$\pm$	0.05	&	66.71	$\pm$	8.94	&	1317, 1345, 1370	\\

\cutinhead{NGC~4449}

\oi - 1	&	0.0001729	&	14.93	$\pm$	0.03	$^\dagger$&	56.92	$\pm$	5.48	&	1302, 1355 	\\
\oi - 2	&	0.0005062	&	15.21	$\pm$	0.02	$^\dagger$&	70.34	$\pm$	3.40	&	$\prime\prime$	\\
\oi - 3	&	0.0010891	&	15.00	$\pm$	0.07	$^\dagger$&	169.94	$\pm$	24.62	&	$\prime\prime$	\\
\sIii - 1	&	0.0001729	&	14.61	$\pm$	0.18	$^\dagger$&	42.47	$\pm$	9.47	&	1304	\\
\sIii - 2	&	0.0005062	&	15.00	$\pm$	0.14	$^\dagger$&	88.79	$\pm$	19.97	&	$\prime\prime$	\\
\sIii - 3	&	0.0010891	&	14.58	$\pm$	0.22	$^\dagger$&	127.58	$\pm$	44.71	&	$\prime\prime$	\\
\sIii* - 2	&	0.0007192	&	14.13	$\pm$	0.09	&	137.11	$\pm$	29.40	&	1197, 1309	\\
\sIii* - 3	&	0.0012061	&	13.43	$\pm$	0.35	&	57.95	$\pm$	23.00	&	$\prime\prime$	\\
\pii - 1	&	0.0002635	&	13.50	$\pm$	0.25	&	66.05	$\pm$	26.39	&	1152	\\
\pii - 2, 3	&	0.0007982	&	13.99	$\pm$	0.09	&	98.55	$\pm$	24.10	&	$\prime\prime$	\\
\sii - 2	&	0.0005213	&	15.43	$\pm$	0.09	&	88.04	$\pm$	16.35	&	1250, 1253	\\
\sii - 3	&	0.0009204	&	15.15	$\pm$	0.18	&	60.58	$\pm$	15.31	&	$\prime\prime$	\\
\feii - 1	&	0.0001739	&	13.93	$\pm$	0.17	$^\dagger$&	33.63	$\pm$	9.81	&	1144	\\
\feii - 2	&	0.0005163	&	14.93	$\pm$	0.06	$^\dagger$&	138.21	$\pm$	22.97	&	$\prime\prime$	\\
\feii - 3	&	0.0010928	&	14.17	$\pm$	0.20	$^\dagger$&	106.53	$\pm$	27.48	&	$\prime\prime$	\\

\cutinhead{NGC~3690}

\Ni - 1	&	0.0107248	&	15.11	$\pm$	0.02	&	104.00	$\pm$	...	&	1134.1,1134.4,1134.9	\\
\Ni - 2	&	0.0094680	&	13.63	$\pm$	0.71	&	138.00	$\pm$	...	&	$\prime\prime$	\\
\oi - 1	&	0.0107049	&	15.48	$\pm$	0.02	$^\dagger$&	104.00	$\pm$	...	&	1302, 1355	\\
\oi - 2	&	0.0094680	&	14.85	$\pm$	0.05	$^\dagger$&	138.39	$\pm$	22.80	&	$\prime\prime$	\\
\sIii - 1	&	0.0107248	&	15.09	$\pm$	0.02	$^\dagger$&	103.97	$\pm$	4.17	&	1304	\\
\sIii - 2	&	0.0094680	&	14.33	$\pm$	0.09	$^\dagger$&	138.00	$\pm$	...	&	$\prime\prime$	\\
\sii - 1	&	0.0106977	&	15.42	$\pm$	0.04	&	113.79	$\pm$	13.57	&	1250, 1253	\\
\feii - 1	&	0.0106426	&	14.84	$\pm$	0.03	&	110.30	$\pm$	8.60	&	1143, 1144	\\
\feii - 2	&	0.0094266	&	14.64	$\pm$	0.04	&	138.00	$\pm$	...	&	$\prime\prime$	\\

\cutinhead{M~83-Pos.~1}

\Ni - 1	&	0.0015711	&	14.95	$\pm$	0.03	$^\dagger$&	65.62	$\pm$	5.91	&	1134.9	\\
\Ni - 2	&	0.0010165	&	15.02	$\pm$	0.02	&	90.79	$\pm$	9.02	&	1134.1, 1134.4, 1134.9	\\
\sIii - 1	&	0.0016080	&	14.57	$\pm$	0.08	$^\dagger$&	49.89	$\pm$	4.05	&	1304	\\
\sIii - 2	&	0.0011923	&	14.88	$\pm$	0.05	$^\dagger$&	97.34	$\pm$	13.19	&	$\prime\prime$	\\
\sIii* - 1	&	0.0012922	&	13.62	$\pm$	0.19	&	29.14	$\pm$	7.97	&	1309	\\
\sIii* - 2	&	0.0010411	&	13.89	$\pm$	0.11	&	57.90	$\pm$	12.57	&	$\prime\prime$	\\
\sIiii - 1	&	0.0014288	&	14.15	$\pm$	0.03	$^\dagger$&	98.71	$\pm$	3.75	&	1206	\\
\sIiii - 2	&	0.0010047	&	13.88	$\pm$	0.06	$^\dagger$&	48.04	$\pm$	6.27	&	$\prime\prime$	\\
\pii - 1	&	0.0016196	&	13.80	$\pm$	0.03	&	61.00	$\pm$	...	&	1152	\\
\pii - 2	&	0.0011555	&	13.76	$\pm$	0.03	&	87.00	$\pm$	...	&	$\prime\prime$	\\
\sii - 1	&	0.0016429	&	15.08	$\pm$	0.21	&	75.61	$\pm$	14.00	&	1253	\\
\sii - 2	&	0.0012794	&	15.44	$\pm$	0.10	&	126.33	$\pm$	17.77	&	$\prime\prime$	\\
\feii - 1	&	0.0015954	&	14.55	$\pm$	0.04	$^\dagger$&	72.42	$\pm$	6.16	&	1144	\\
\feii - 2	&	0.0010640	&	14.50	$\pm$	0.05	$^\dagger$&	80.83	$\pm$	11.20	&	$\prime\prime$	\\
\nIii - 1	&	0.0016562	&	13.65	$\pm$	0.05	&	61.13	$\pm$	8.62	&	1317, 1370	\\
\nIii - 2	&	0.0009562	&	13.51	$\pm$	0.12	&	87.34	$\pm$	29.22	&	$\prime\prime$	\\

\cutinhead{M~83-Pos.~2}

\cii	&	0.0015950	&	15.09	$\pm$	0.01	$^\dagger$&	101.75	$\pm$	3.75	&	1334	\\
\cii*	&	0.0015537	&	14.30	$\pm$	0.02	&	106.40	$\pm$	6.09	&	1335.6, 1335.7	\\
\Ni 	&	0.0014658	&	14.90	$\pm$	0.03	&	110.00	$\pm$	...	&	1134.1, 1134.4	\\
\sIii	&	0.0015648	&	14.69	$\pm$	0.01	$^\dagger$&	78.50	$\pm$	2.00	&	1304	\\
\sIii*	&	0.0015012	&	13.13	$\pm$	0.06	&	111.93	$\pm$	20.26	&	1264, 1265	\\
\sIiii	&	0.0014956	&	13.99	$\pm$	0.01	$^\dagger$&	100.81	$\pm$	1.80	&	1206	\\
\pii 	&	0.0016062	&	13.73	$\pm$	0.04	&	123.25	$\pm$	15.18	&	1152	\\
\sii	&	0.0016918	&	15.39	$\pm$	0.02	&	120.26	$\pm$	7.66	&	1253	\\
\feii 	&	0.0015596	&	14.71	$\pm$	0.02	$^\dagger$&	125.24	$\pm$	9.04	&	1144	\\										

\enddata
											

\end{deluxetable*}

\begin{table*}
\begin{center}
\begin{scriptsize}
\caption{Amount of hidden saturation in the strongest \feii~and \Ni~absorption lines} \label{tab:saturation}
\begin{tabular}{lcccc}
\tableline\tableline		
 & \feii\, & \Ni\, & & \\		
Galaxy	&	log[$N$(1142 or 1143)]$-$log[$N$(1144)]			&	log[$N$(1134.1, 1134.4)]$-$log[$N$(1134.9)]			&	log[$N$(\hi)]	&	12+log(O/H)	\\
               &                        (dex)                                                     &                                      (dex)                                                     &
(cm$^{-2}$)     &                             \\
\tableline
I~Zw~18	&	0.29	$\pm$	0.06	&	0.02	$\pm$	0.05	&	21.28	&	7.2	\\
SBS~0335$-$052	&	0.97	$\pm$	0.04	&	0.25	$\pm$	0.04	&	21.70	&	7.3	\\
SBS~1415+437	&	0.50	$\pm$	0.04	&	0.28	$\pm$	0.12	&	21.09	&	7.6	\\
NGC~4214	&	0.19	$\pm$	0.08	&	0.08	$\pm$	0.08	&	21.12	&	8.2	\\
NGC~5253-1	&	0.92	$\pm$	0.07	&	0.46	$\pm$	0.06	&	21.20	&	8.2	\\
NGC~5253-2	&	0.96	$\pm$	0.04	&	0.03	$\pm$	0.04	&	20.65	&	8.2	\\
NGC~4670	&	0.22	$\pm$	0.05	&	0.03	$\pm$	0.03	&	21.07	&	8.2	\\
NGC~4449	&	---			&	---			&	21.14	&	8.3	\\
NGC~3690$_{v1}$	&	---			&	---			&	20.62	&	8.8	\\
NGC~3690$_{v2}$	&	---			&	---			&	19.82	&	8.8	\\
M83-1	&	--- 	&	---			&	19.60	&	9.2	\\
M83-2	&	---			&	0.07	$\pm$	0.06	&	18.44	&	9.2	\\								
\tableline
\end{tabular}
\begin{description}
\vspace{0.3cm}
\begin{tiny}
Note -- This table reports the amount of hidden saturation seen within the \feii~$\lambda$1144 and \Ni~$\lambda$1134.9 absorption lines.  Columns 2 (\feii) and 3 (\Ni) list the difference in the logarithm of the column densities $N_1$ and $N_2$ from lines with different oscillator strengths such as $f_1<f_2$.  For \feii\, this corresponds to $\lambda_1$=$\lambda$1142 or $\lambda$1143, $\lambda_2$=$\lambda$1144 and for \Ni\, $\lambda_1$=$\lambda\lambda$1134.1,1134.4 and $\lambda_2$=$\lambda$1134.9. Dashed lines indicate galaxies for which we were unable to constrain $N_1$ and/or $N_2$ due to blending. Columns 4 and 5 list the \hi\, column density (as listed in Table~\ref{tab:HIdensities}) and metallicity (as listed in Table~\ref{tab:sample}), respectively.  The values reported in this table can be seen in graphical format in Figures~\ref{fig:saturation_trends}--\ref{fig:saturation_all}. 
\end{tiny}
\end{description}
\end{scriptsize}
\end{center}
\end{table*}

\begin{sidewaystable*}
\vspace{9cm}
\begin{center}
\begin{scriptsize}
\caption{Ionization correction factors for the metallicity and \hi\, column density of the galaxies in the sample}\label{tab:ICFs}
\begin{tabular}{lccccccccccc}
\tableline \tableline
\multicolumn{12}{c}{Ionized-Gas ICFs, ICF$_{ionized}$}\\
Galaxy & log[$N$(\hi)] & Z/\Zsol &\hi& \cii & \Ni&  \oi&  \sIii&  \pii&  \sii& \feii& \nIii \\
\tableline
SBS~0335$-$052 &  21.70 &  0.04 & 
 0.00 &
 0.02$\pm$ 0.02 &
 0.00 &
 0.00 &
 0.00 &
 0.01$\pm$ 0.01 &
 0.00 &
 0.00 &
 0.00 \\
I~Zw~18 &  21.28 &  0.03 & 
 0.00 &
 0.05$\pm$ 0.05 &
 0.00 &
 0.00 &
 0.01$\pm$ 0.01 &
 0.01$\pm$ 0.01 &
 0.01$\pm$ 0.01 &
 0.00 &
 0.01$\pm$ 0.01 \\
NGC~5253-1 &  21.20 &  0.32 & 
 0.00 &
 0.05$\pm$ 0.05 &
 0.00 &
 0.00 &
 0.02$\pm$ 0.02 &
 0.01$\pm$ 0.01 &
 0.01$\pm$ 0.01 &
 0.00 &
 0.01$\pm$ 0.01 \\
NGC~4449 &  21.14 &  0.41 & 
 0.00 &
 0.06$\pm$ 0.06 &
 0.00 &
 0.00 &
 0.02$\pm$ 0.02 &
 0.02$\pm$ 0.02 &
 0.01$\pm$ 0.01 &
 0.01$\pm$ 0.01 &
 0.01$\pm$ 0.01 \\
NGC~4214 &  21.12 &  0.32 & 
 0.00 &
 0.06$\pm$ 0.06 &
 0.00 &
 0.00 &
 0.02$\pm$ 0.02 &
 0.02$\pm$ 0.02 &
 0.01$\pm$ 0.01 &
 0.00 &
 0.01$\pm$ 0.01 \\
SBS~1415+437 &  21.09 &  0.08 & 
 0.00 &
 0.07$\pm$ 0.07 &
 0.00 &
 0.00 &
 0.02$\pm$ 0.02 &
 0.02$\pm$ 0.02 &
 0.02$\pm$ 0.02 &
 0.00 &
 0.01$\pm$ 0.01 \\
NGC~4670 &  21.07 &  0.32 & 
 0.00 &
 0.07$\pm$ 0.07 &
 0.00 &
 0.00 &
 0.02$\pm$ 0.02 &
 0.02$\pm$ 0.02 &
 0.02$\pm$ 0.02 &
 0.01$\pm$ 0.01 &
 0.01$\pm$ 0.01 \\
NGC~5253-2 &  20.65 &  0.32 & 
 0.00 &
 0.14$\pm$ 0.14 &
 0.00 &
 0.00 &
 0.06$\pm$ 0.06 &
 0.05$\pm$ 0.05 &
 0.04$\pm$ 0.04 &
 0.01$\pm$ 0.01 &
 0.02$\pm$ 0.02 \\
NGC~3690$_{v1}$ &  20.62 &  1.29 & 
 0.00 &
 0.13$\pm$ 0.13 &
 0.00 &
 0.00 &
 0.08$\pm$ 0.08 &
 0.04$\pm$ 0.04 &
 0.03$\pm$ 0.03 &
 0.02$\pm$ 0.02 &
 0.02$\pm$ 0.02 \\
NGC~3690$_{v2}$ &  19.82 &  1.29 & 
 0.02$\pm$ 0.02 &
 0.41$\pm$ 0.41 &
 0.01$\pm$ 0.01 &
 0.02$\pm$ 0.02 &
 0.29$\pm$ 0.29 &
 0.19$\pm$ 0.19 &
 0.16$\pm$ 0.16 &
 0.12$\pm$ 0.12 &
 0.10$\pm$ 0.10 \\
M83-1 &  19.60 &  3.24 & 
 0.04$\pm$ 0.04 &
 0.47$\pm$ 0.47 &
 0.03$\pm$ 0.03 &
 0.04$\pm$ 0.04 &
 0.41$\pm$ 0.41 &
 0.26$\pm$ 0.26 &
 0.23$\pm$ 0.23 &
 0.18$\pm$ 0.18 &
 0.12$\pm$ 0.12 \\
M83-2 &  18.44 &  3.24 & 
NaN &
NaN &
NaN &
NaN &
NaN &
NaN &
NaN &
NaN &
NaN \\ 
\tableline
\multicolumn{12}{c}{Neutral-Gas ICFs, ICF$_{neutral}$}\\
Galaxy & log[$N$(\hi)] & Z/\Zsol &\hii & \ciii & \Nii&  \oii&  \sIiii&  \piii&  \siii& \feiii& \nIiii \\
\tableline
SBS~0335$-$052 &  21.70 &  0.04 & 
 0.00 &
 0.00 &
 0.00 &
 0.00 &
 0.00 &
 0.00 &
 0.00 &
 0.00 &
 0.00 \\
I~Zw~18 &  21.28 &  0.03 & 
 0.00 &
 0.00 &
 0.00 &
 0.00 &
 0.00 &
 0.00 &
 0.00 &
 0.00 &
 0.00 \\
NGC~5253-1 &  21.20 &  0.32 & 
 0.00 &
 0.00 &
 0.00 &
 0.00 &
 0.00 &
 0.00 &
 0.00 &
 0.00 &
 0.00 \\
NGC~4449 &  21.14 &  0.41 & 
 0.00 &
 0.00 &
 0.00 &
 0.00 &
 0.00 &
 0.00 &
 0.00 &
 0.00 &
 0.00 \\
NGC~4214 &  21.12 &  0.32 & 
 0.00 &
 0.00 &
 0.00 &
 0.00 &
 0.00 &
 0.00 &
 0.00 &
 0.00 &
 0.00 \\
SBS~1415+437 &  21.09 &  0.08 & 
 0.00 &
 0.00 &
 0.00 &
 0.00 &
 0.00 &
 0.00 &
 0.00 &
 0.00 &
 0.00 \\
NGC~4670 &  21.07 &  0.32 & 
 0.00 &
 0.00 &
 0.00 &
 0.00 &
 0.00 &
 0.00 &
 0.00 &
 0.00 &
 0.00 \\
NGC~5253-2 &  20.65 &  0.32 & 
 0.00 &
 0.00 &
 0.01$\pm$ 0.01 &
 0.00 &
 0.00 &
 0.00 &
 0.00 &
 0.00 &
 0.00 \\
NGC~3690$_{v1}$ &  20.62 &  1.29 & 
 0.00 &
 0.00 &
 0.00 &
 0.00 &
 0.00 &
 0.00 &
 0.00 &
 0.00 &
 0.00 \\
NGC~3690$_{v2}$ &  19.82 &  1.29 & 
 0.01$\pm$ 0.01 &
 0.01$\pm$ 0.01 &
 0.01$\pm$ 0.01 &
 0.01$\pm$ 0.01 &
 0.01$\pm$ 0.01 &
 0.01$\pm$ 0.01 &
 0.01$\pm$ 0.01 &
 0.01$\pm$ 0.01 &
 0.01$\pm$ 0.01 \\
M83-1 &  19.60 &  3.24 & 
 0.03$\pm$ 0.03 &
 0.01$\pm$ 0.01 &
 0.02$\pm$ 0.02 &
 0.01$\pm$ 0.01 &
 0.02$\pm$ 0.02 &
 0.02$\pm$ 0.02 &
 0.02$\pm$ 0.02 &
 0.02$\pm$ 0.02 &
 0.02$\pm$ 0.02 \\
M83-2 &  18.44 &  3.24 & 
NaN &
NaN &
NaN &
NaN &
NaN &
NaN &
NaN &
NaN &
NaN \\
\tableline
\multicolumn{12}{c}{Total ICFs, ICF$_{total}$}\\
Galaxy & log[$N$(\hi)] & Z/\Zsol & H & C & N & O & Si & P & S & Fe & Ni \\
\tableline
SBS~0335$-$052 &  21.70 &  0.04 & 
 0.00 &
 0.02$\pm$ 0.02 &
 0.00 &
 0.00 &
 0.00 &
 0.01$\pm$ 0.01 &
 0.00 &
 0.00 &
 0.00 \\
I~Zw~18 &  21.28 &  0.03 & 
 0.00 &
 0.05$\pm$ 0.05 &
 0.00 &
 0.00 &
 0.01$\pm$ 0.01 &
 0.01$\pm$ 0.01 &
 0.01$\pm$ 0.01 &
 0.00 &
 0.01$\pm$ 0.01 \\
NGC~5253-1 &  21.20 &  0.32 & 
 0.00 &
 0.05$\pm$ 0.05 &
 0.00 &
 0.00 &
 0.02$\pm$ 0.02 &
 0.01$\pm$ 0.01 &
 0.01$\pm$ 0.01 &
 0.00 &
 0.01$\pm$ 0.01 \\
NGC~4449 &  21.14 &  0.41 & 
 0.00 &
 0.06$\pm$ 0.06 &
 0.00 &
 0.00 &
 0.02$\pm$ 0.02 &
 0.02$\pm$ 0.02 &
 0.01$\pm$ 0.01 &
 0.01$\pm$ 0.01 &
 0.01$\pm$ 0.01 \\
NGC~4214 &  21.12 &  0.32 & 
 0.00 &
 0.06$\pm$ 0.06 &
 0.00 &
 0.00 &
 0.02$\pm$ 0.02 &
 0.02$\pm$ 0.02 &
 0.01$\pm$ 0.01 &
 0.00 &
 0.01$\pm$ 0.01 \\
SBS~1415+437 &  21.09 &  0.08 & 
 0.00 &
 0.07$\pm$ 0.07 &
 0.00 &
 0.00 &
 0.02$\pm$ 0.02 &
 0.02$\pm$ 0.02 &
 0.02$\pm$ 0.02 &
 0.00 &
 0.01$\pm$ 0.01 \\
NGC~4670 &  21.07 &  0.32 & 
 0.00 &
 0.07$\pm$ 0.07 &
 0.00 &
 0.00 &
 0.02$\pm$ 0.02 &
 0.02$\pm$ 0.02 &
 0.02$\pm$ 0.02 &
 0.01$\pm$ 0.01 &
 0.01$\pm$ 0.01 \\
NGC~5253-2 &  20.65 &  0.32 & 
 0.00 &
 0.14$\pm$ 0.14 &
-0.01$\pm$ 0.01 &
 0.00 &
 0.06$\pm$ 0.06 &
 0.05$\pm$ 0.05 &
 0.04$\pm$ 0.04 &
 0.01$\pm$ 0.01 &
 0.02$\pm$ 0.02 \\
NGC~3690$_{v1}$ &  20.62 &  1.29 & 
 0.00 &
 0.13$\pm$ 0.13 &
 0.00 &
 0.00 &
 0.08$\pm$ 0.08 &
 0.04$\pm$ 0.04 &
 0.03$\pm$ 0.03 &
 0.02$\pm$ 0.02 &
 0.02$\pm$ 0.02 \\
NGC~3690$_{v2}$ &  19.82 &  1.29 & 
 0.01$\pm$ 0.02 &
 0.40$\pm$ 0.41 &
 0.00$\pm$ 0.01 &
 0.01$\pm$ 0.02 &
 0.28$\pm$ 0.29 &
 0.18$\pm$ 0.19 &
 0.15$\pm$ 0.16 &
 0.11$\pm$ 0.12 &
 0.09$\pm$ 0.10 \\
M83-1 &  19.60 &  3.24 & 
 0.01$\pm$ 0.05 &
 0.46$\pm$ 0.47 &
 0.01$\pm$ 0.04 &
 0.03$\pm$ 0.04 &
 0.39$\pm$ 0.41 &
 0.24$\pm$ 0.26 &
 0.21$\pm$ 0.23 &
 0.16$\pm$ 0.18 &
 0.10$\pm$ 0.12 \\
M83-2 &  18.44 &  3.24 & 
NaN &
NaN &
NaN &
NaN &
NaN &
NaN &
NaN &
NaN &
NaN \\
\tableline
\end{tabular}
\end{scriptsize}
\end{center}
\end{sidewaystable*}
\clearpage
\LongTables
\begin{deluxetable*}{llccccccc}
\tablecolumns{6}
\tablewidth{0pc}
\tablecaption{Interstellar Abundances\label{tab:abundances}}

\tablehead{\colhead{Element}	&	\colhead{Ion}	&	\colhead{log[$N(X)$/cm$^{-2}$]}			&	\colhead{log(X/H)}&	\colhead{log(X/H)$_{ICF}^a$}			&	\colhead{log(X/H)$_{\odot}^b$}	&	\colhead{[X/H]$^c$}	&	\colhead{[X/H]$_{ICF}^d$}			}															
\cutinhead{I~Zw~18}															\\
																	
H	&	\hi	&	21.28	$\pm$	0.03	&				&				&				&								\\
C	$^\dagger$&	\cii	&	14.41	$\pm$	0.01	&	$-$6.87	$\pm$	0.03	&	$-$6.92	$\pm$	0.06	&	$-$3.57	$\pm$	0.05	&	$-$3.30	$\pm$	0.06	&	$-$3.35	$\pm$	0.08	\\
N	&	\Ni	&	14.38	$\pm$	0.03	&	$-$6.90	$\pm$	0.04	&	$-$6.90	$\pm$	0.04	&	$-$4.17	$\pm$	0.05	&	$-$2.73	$\pm$	0.06	&	$-$2.73	$\pm$	0.06	\\
O	$^\dagger$&	\oi	&	14.80	$\pm$	0.02	&	$-$6.48	$\pm$	0.04	&	$-$6.48	$\pm$	0.04	&	$-$3.31	$\pm$	0.05	&	$-$3.17	$\pm$	0.06	&	$-$3.17	$\pm$	0.06	\\
Si	$^\dagger$&	\sIii	&	14.36	$\pm$	0.01	&	$-$6.92	$\pm$	0.03	&	$-$6.93	$\pm$	0.03	&	$-$4.49	$\pm$	0.03	&	$-$2.43	$\pm$	0.04	&	$-$2.44	$\pm$	0.04	\\
S	&	\sii	&	14.73	$\pm$	0.05	&	$-$6.55	$\pm$	0.06	&	$-$6.56	$\pm$	0.06	&	$-$4.88	$\pm$	0.03	&	$-$1.67	$\pm$	0.07	&	$-$1.68	$\pm$	0.07	\\
Fe	&	\feii	&	14.52	$\pm$	0.06	&	$-$6.76	$\pm$	0.07	&	$-$6.76	$\pm$	0.07	&	$-$4.50	$\pm$	0.04	&	$-$2.26	$\pm$	0.08	&	$-$2.26	$\pm$	0.08	\\
\cutinhead{SBS~0335$-$052}															\\
H	&	\hi	&	21.70	$\pm$	0.05	&				&				&				&				&				\\
C	$^\dagger$&	\cii	&	14.41	$\pm$	0.01	&	$-$7.29	$\pm$	0.05	&	$-$7.31	$\pm$	0.05	&	$-$3.57	$\pm$	0.05	&	$-$3.72	$\pm$	0.07	&	$-$3.74	$\pm$	0.07	\\
N	&	\Ni	&	14.83	$\pm$	0.02	&	$-$6.87	$\pm$	0.05	&	$-$6.87	$\pm$	0.05	&	$-$4.17	$\pm$	0.05	&	$-$2.70	$\pm$	0.07	&	$-$2.70	$\pm$	0.07	\\
O	$^\dagger$&	\oi	&	14.77	$\pm$	0.01	&	$-$6.93	$\pm$	0.05	&	$-$6.93	$\pm$	0.05	&	$-$3.31	$\pm$	0.05	&	$-$3.62	$\pm$	0.07	&	$-$3.62	$\pm$	0.07	\\
Si	$^\dagger$&	\sIii	&	14.38	$\pm$	0.02	&	$-$7.32	$\pm$	0.05	&	$-$7.32	$\pm$	0.05	&	$-$4.49	$\pm$	0.03	&	$-$2.83	$\pm$	0.06	&	$-$2.83	$\pm$	0.06	\\
P	&	\pii	&	12.97	$\pm$	0.15	&	$-$8.73	$\pm$	0.16	&	$-$8.74	$\pm$	0.16	&	$-$6.59	$\pm$	0.03	&	$-$2.14	$\pm$	0.16	&	$-$2.15	$\pm$	0.16	\\
S	&	\sii	&	14.96	$\pm$	0.03	&	$-$6.74	$\pm$	0.06	&	$-$6.74	$\pm$	0.06	&	$-$4.88	$\pm$	0.03	&	$-$1.86	$\pm$	0.07	&	$-$1.86	$\pm$	0.07	\\
Fe	&	\feii	&	15.32	$\pm$	0.04	&	$-$6.38	$\pm$	0.06	&	$-$6.38	$\pm$	0.06	&	$-$4.50	$\pm$	0.04	&	$-$1.88	$\pm$	0.07	&	$-$1.88	$\pm$	0.07	\\
Ni	&	\nIii	&	13.96	$\pm$	0.06	&	$-$7.74	$\pm$	0.08	&	$-$7.74	$\pm$	0.08	&	$-$5.78	$\pm$	0.04	&	$-$1.96	$\pm$	0.09	&	$-$1.96	$\pm$	0.09	\\
																
\cutinhead{SBS~1415+437}															\\																	
H	&	\hi	&	21.09	$\pm$	0.03	&				&				&				&				&				\\
C	$^\dagger$&	\cii	&	14.51	$\pm$	0.01	&	$-$6.58	$\pm$	0.03	&	$-$6.65	$\pm$	0.08	&	$-$3.57	$\pm$	0.05	&	$-$3.01	$\pm$	0.06	&	$-$3.08	$\pm$	0.09	\\
N	&	\Ni	&	14.57	$\pm$	0.04	&	$-$6.52	$\pm$	0.05	&	$-$6.52	$\pm$	0.05	&	$-$4.17	$\pm$	0.05	&	$-$2.35	$\pm$	0.07	&	$-$2.35	$\pm$	0.07	\\
O	$^\dagger$&	\oi	&	14.87	$\pm$	0.09	&	$-$6.22	$\pm$	0.09	&	$-$6.22	$\pm$	0.09	&	$-$3.31	$\pm$	0.05	&	$-$2.91	$\pm$	0.10	&	$-$2.91	$\pm$	0.10	\\
Si	$^\dagger$&	\sIii	&	14.42	$\pm$	0.01	&	$-$6.67	$\pm$	0.03	&	$-$6.69	$\pm$	0.04	&	$-$4.49	$\pm$	0.03	&	$-$2.18	$\pm$	0.04	&	$-$2.20	$\pm$	0.05	\\
P	&	\pii	&	13.43	$\pm$	0.08	&	$-$7.66	$\pm$	0.09	&	$-$7.68	$\pm$	0.09	&	$-$6.59	$\pm$	0.03	&	$-$1.07	$\pm$	0.09	&	$-$1.09	$\pm$	0.09	\\
S	&	\sii	&	14.96	$\pm$	0.02	&	$-$6.13	$\pm$	0.04	&	$-$6.15	$\pm$	0.04	&	$-$4.88	$\pm$	0.03	&	$-$1.25	$\pm$	0.05	&	$-$1.27	$\pm$	0.05	\\
Fe	$^\dagger$&	\feii	&	14.83	$\pm$	0.04	&	$-$6.26	$\pm$	0.05	&	$-$6.26	$\pm$	0.05	&	$-$4.50	$\pm$	0.04	&	$-$1.76	$\pm$	0.06	&	$-$1.76	$\pm$	0.06	\\
Ni	&	\nIii	&	13.26	$\pm$	0.10	&	$-$7.83	$\pm$	0.10	&	$-$7.84	$\pm$	0.10	&	$-$5.78	$\pm$	0.04	&	$-$2.05	$\pm$	0.11	&	$-$2.06	$\pm$	0.11	\\

\cutinhead{NGC~4214}															\\
H	&	\hi	&	21.12	$\pm$	0.03	&				&				&				&				&				\\
N	&	\Ni	&	15.20	$\pm$	0.08	&	$-$5.92	$\pm$	0.09	&	$-$5.92	$\pm$	0.09	&	$-$4.17	$\pm$	0.05	&	$-$1.75	$\pm$	0.10	&	$-$1.75	$\pm$	0.10	\\
O	$^\dagger$&	\oi	&	15.25	$\pm$	0.02	&	$-$5.87	$\pm$	0.04	&	$-$5.87	$\pm$	0.04	&	$-$3.31	$\pm$	0.05	&	$-$2.56	$\pm$	0.06	&	$-$2.56	$\pm$	0.06	\\
Si	$^\dagger$&	\sIii	&	14.87	$\pm$	0.02	&	$-$6.25	$\pm$	0.04	&	$-$6.27	$\pm$	0.04	&	$-$4.49	$\pm$	0.03	&	$-$1.76	$\pm$	0.05	&	$-$1.78	$\pm$	0.05	\\
P	&	\pii	&	13.76	$\pm$	0.03	&	$-$7.36	$\pm$	0.04	&	$-$7.38	$\pm$	0.04	&	$-$6.59	$\pm$	0.03	&	$-$0.77	$\pm$	0.05	&	$-$0.79	$\pm$	0.05	\\
S	&	\sii	&	15.56	$\pm$	0.03	&	$-$5.56	$\pm$	0.04	&	$-$5.57	$\pm$	0.04	&	$-$4.88	$\pm$	0.03	&	$-$0.68	$\pm$	0.05	&	$-$0.69	$\pm$	0.05	\\
Fe	&	\feii	&	14.80	$\pm$	0.07	&	$-$6.32	$\pm$	0.08	&	$-$6.32	$\pm$	0.08	&	$-$4.50	$\pm$	0.04	&	$-$1.82	$\pm$	0.09	&	$-$1.82	$\pm$	0.09	\\
Ni	&	\nIii	&	13.48	$\pm$	0.04	&	$-$7.64	$\pm$	0.05	&	$-$7.65	$\pm$	0.05	&	$-$5.78	$\pm$	0.04	&	$-$1.86	$\pm$	0.06	&	$-$1.87	$\pm$	0.06	\\
																
\cutinhead{NGC~5253-Pos.~1}															\\																	
H	&	\hi	&	21.20	$\pm$	0.01	&				&				&				&				&				\\
C	$^\dagger$&	\cii	&	15.04	$\pm$	0.01	&	$-$6.16	$\pm$	0.01	&	$-$6.21	$\pm$	0.05	&	$-$3.57	$\pm$	0.05	&	$-$2.59	$\pm$	0.05	&	$-$2.64	$\pm$	0.07	\\
N	&	\Ni	&	15.24	$\pm$	0.03	&	$-$5.96	$\pm$	0.03	&	$-$5.96	$\pm$	0.03	&	$-$4.17	$\pm$	0.05	&	$-$1.79	$\pm$	0.06	&	$-$1.79	$\pm$	0.06	\\
O	$^\dagger$&	\oi	&	15.44	$\pm$	0.14	&	$-$5.76	$\pm$	0.14	&	$-$5.76	$\pm$	0.14	&	$-$3.31	$\pm$	0.05	&	$-$2.45	$\pm$	0.15	&	$-$2.45	$\pm$	0.15	\\
Si	$^\dagger$&	\sIii	&	14.79	$\pm$	0.02	&	$-$6.41	$\pm$	0.02	&	$-$6.43	$\pm$	0.03	&	$-$4.49	$\pm$	0.03	&	$-$1.92	$\pm$	0.04	&	$-$1.94	$\pm$	0.04	\\
P	&	\pii	&	13.74	$\pm$	0.08	&	$-$7.46	$\pm$	0.08	&	$-$7.47	$\pm$	0.08	&	$-$6.59	$\pm$	0.03	&	$-$0.87	$\pm$	0.09	&	$-$0.88	$\pm$	0.09	\\
S	&	\sii	&	15.44	$\pm$	0.02	&	$-$5.76	$\pm$	0.02	&	$-$5.77	$\pm$	0.02	&	$-$4.88	$\pm$	0.03	&	$-$0.88	$\pm$	0.04	&	$-$0.89	$\pm$	0.04	\\
Fe	&	\feii	&	15.67	$\pm$	0.07	&	$-$5.53	$\pm$	0.07	&	$-$5.53	$\pm$	0.07	&	$-$4.50	$\pm$	0.04	&	$-$1.03	$\pm$	0.08	&	$-$1.03	$\pm$	0.08	\\
Ni	&	\nIii	&	13.68	$\pm$	0.08	&	$-$7.52	$\pm$	0.08	&	$-$7.53	$\pm$	0.08	&	$-$5.78	$\pm$	0.04	&	$-$1.74	$\pm$	0.09	&	$-$1.75	$\pm$	0.09	\\
																
\cutinhead{NGC~5253-Pos.~2}												\\
H	&	\hi	&	20.65	$\pm$	0.05	&				&				&				&				&				\\
C	$^\dagger$&	\cii	&	14.92	$\pm$	0.01	&	$-$5.73	$\pm$	0.05	&	$-$5.87	$\pm$	0.15	&	$-$3.57	$\pm$	0.05	&	$-$2.16	$\pm$	0.07	&	$-$2.30	$\pm$	0.16	\\
N	&	\Ni	&	14.98	$\pm$	0.02	&	$-$5.67	$\pm$	0.05	&	$-$5.66	$\pm$	0.05	&	$-$4.17	$\pm$	0.05	&	$-$1.50	$\pm$	0.07	&	$-$1.49	$\pm$	0.07	\\
O	$^\dagger$&	\oi	&	15.31	$\pm$	0.05	&	$-$5.34	$\pm$	0.07	&	$-$5.34	$\pm$	0.07	&	$-$3.31	$\pm$	0.05	&	$-$2.03	$\pm$	0.09	&	$-$2.03	$\pm$	0.09	\\
Si	$^\dagger$&	\sIii	&	14.81	$\pm$	0.01	&	$-$5.84	$\pm$	0.05	&	$-$5.90	$\pm$	0.08	&	$-$4.49	$\pm$	0.03	&	$-$1.35	$\pm$	0.06	&	$-$1.41	$\pm$	0.09	\\
P	&	\pii	&	13.65	$\pm$	0.11	&	$-$7.00	$\pm$	0.12	&	$-$7.05	$\pm$	0.13	&	$-$6.59	$\pm$	0.03	&	$-$0.41	$\pm$	0.12	&	$-$0.46	$\pm$	0.13	\\
S	&	\sii	&	15.37	$\pm$	0.02	&	$-$5.28	$\pm$	0.05	&	$-$5.32	$\pm$	0.06	&	$-$4.88	$\pm$	0.03	&	$-$0.40	$\pm$	0.06	&	$-$0.44	$\pm$	0.07	\\
Fe	&	\feii	&	15.66	$\pm$	0.04	&	$-$4.99	$\pm$	0.06	&	$-$5.00	$\pm$	0.06	&	$-$4.50	$\pm$	0.04	&	$-$0.49	$\pm$	0.08	&	$-$0.50	$\pm$	0.08	\\
Ni	&	\nIii	&	13.77	$\pm$	0.13	&	$-$6.88	$\pm$	0.14	&	$-$6.90	$\pm$	0.14	&	$-$5.78	$\pm$	0.04	&	$-$1.10	$\pm$	0.15	&	$-$1.12	$\pm$	0.15	\\									
								
\cutinhead{NGC~4670}															\\																	
H	&	\hi	&	21.07	$\pm$	0.08	&				&				&				&				&				\\
C	$^\dagger$&	\cii	&	15.02	$\pm$	0.01	&	$-$6.05	$\pm$	0.08	&	$-$6.12	$\pm$	0.11	&	$-$3.57	$\pm$	0.05	&	$-$2.48	$\pm$	0.09	&	$-$2.55	$\pm$	0.12	\\
N	&	\Ni	&	15.42	$\pm$	0.01	&	$-$5.65	$\pm$	0.08	&	$-$5.65	$\pm$	0.08	&	$-$4.17	$\pm$	0.05	&	$-$1.48	$\pm$	0.09	&	$-$1.48	$\pm$	0.09	\\
O	$^\dagger$&	\oi	&	15.26	$\pm$	0.02	&	$-$5.81	$\pm$	0.08	&	$-$5.81	$\pm$	0.08	&	$-$3.31	$\pm$	0.05	&	$-$2.50	$\pm$	0.09	&	$-$2.50	$\pm$	0.09	\\
Si	$^\dagger$&	\sIii	&	14.97	$\pm$	0.01	&	$-$6.10	$\pm$	0.08	&	$-$6.12	$\pm$	0.08	&	$-$4.49	$\pm$	0.03	&	$-$1.61	$\pm$	0.09	&	$-$1.63	$\pm$	0.09	\\
P	&	\pii	&	13.82	$\pm$	0.05	&	$-$7.25	$\pm$	0.09	&	$-$7.27	$\pm$	0.09	&	$-$6.59	$\pm$	0.03	&	$-$0.66	$\pm$	0.09	&	$-$0.68	$\pm$	0.09	\\
S	$^\dagger$&	\sii	&	15.60	$\pm$	0.01	&	$-$5.47	$\pm$	0.08	&	$-$5.49	$\pm$	0.08	&	$-$4.88	$\pm$	0.03	&	$-$0.59	$\pm$	0.09	&	$-$0.61	$\pm$	0.09	\\
Fe	&	\feii	&	15.07	$\pm$	0.05	&	$-$6.00	$\pm$	0.09	&	$-$6.01	$\pm$	0.09	&	$-$4.50	$\pm$	0.04	&	$-$1.50	$\pm$	0.10	&	$-$1.51	$\pm$	0.10	\\
Ni	&	\nIii	&	13.76	$\pm$	0.05	&	$-$7.31	$\pm$	0.09	&	$-$7.32	$\pm$	0.09	&	$-$5.78	$\pm$	0.04	&	$-$1.53	$\pm$	0.10	&	$-$1.54	$\pm$	0.10	\\					

\cutinhead{NGC~4449}															\\
H	&	\hi	&	21.14	$\pm$	0.03	&				&				&				&				&				\\
O	$^\dagger$&	\oi	&	15.54	$\pm$	0.02	&	$-$5.60	$\pm$	0.04	&	$-$5.60	$\pm$	0.04	&	$-$3.31	$\pm$	0.05	&	$-$2.29	$\pm$	0.06	&	$-$2.29	$\pm$	0.06	\\
Si	$^\dagger$&	\sIii	&	15.25	$\pm$	0.10	&	$-$5.89	$\pm$	0.10	&	$-$5.91	$\pm$	0.10	&	$-$4.49	$\pm$	0.03	&	$-$1.40	$\pm$	0.10	&	$-$1.42	$\pm$	0.10	\\
P	&	\pii	&	14.11	$\pm$	0.09	&	$-$7.03	$\pm$	0.09	&	$-$7.05	$\pm$	0.09	&	$-$6.59	$\pm$	0.03	&	$-$0.44	$\pm$	0.09	&	$-$0.46	$\pm$	0.09	\\
S	&	\sii	&	15.61	$\pm$	0.09	&	$-$5.53	$\pm$	0.09	&	$-$5.54	$\pm$	0.09	&	$-$4.88	$\pm$	0.03	&	$-$0.65	$\pm$	0.09	&	$-$0.66	$\pm$	0.09	\\
Fe	$^\dagger$&	\feii	&	15.04	$\pm$	0.06	&	$-$6.10	$\pm$	0.07	&	$-$6.11	$\pm$	0.07	&	$-$4.50	$\pm$	0.04	&	$-$1.60	$\pm$	0.08	&	$-$1.61	$\pm$	0.08	\\

\cutinhead{NGC~3690$_{v1}$}															\\																
H	&	\hi	&	20.62	$\pm$	0.02	&				&				&				&				&				\\
N	&	\Ni	&	15.11	$\pm$	0.02	&	$-$5.51	$\pm$	0.03	&	$-$5.51	$\pm$	0.03	&	$-$4.17	$\pm$	0.05	&	$-$1.34	$\pm$	0.06	&	$-$1.34	$\pm$	0.06	\\
O	$^\dagger$&	\oi	&	15.48	$\pm$	0.02	&	$-$5.14	$\pm$	0.03	&	$-$5.14	$\pm$	0.03	&	$-$3.31	$\pm$	0.05	&	$-$1.83	$\pm$	0.06	&	$-$1.83	$\pm$	0.06	\\
Si	$^\dagger$&	\sIii	&	15.09	$\pm$	0.02	&	$-$5.53	$\pm$	0.03	&	$-$5.61	$\pm$	0.09	&	$-$4.49	$\pm$	0.03	&	$-$1.04	$\pm$	0.04	&	$-$1.12	$\pm$	0.09	\\
S	&	\sii	&	15.42	$\pm$	0.04	&	$-$5.20	$\pm$	0.04	&	$-$5.23	$\pm$	0.05	&	$-$4.88	$\pm$	0.03	&	$-$0.32	$\pm$	0.05	&	$-$0.35	$\pm$	0.06	\\
Fe	&	\feii	&	14.84	$\pm$	0.03	&	$-$5.78	$\pm$	0.04	&	$-$5.80	$\pm$	0.04	&	$-$4.50	$\pm$	0.04	&	$-$1.28	$\pm$	0.06	&	$-$1.30	$\pm$	0.06	\\
	&		&				&				&				&		&								\\															
																	
\cutinhead{NGC~3690$_{v2}$}															\\																	
H	&	\hi	&	19.82	$\pm$	0.08	&				&				&				&								\\
N	&	\Ni	&	13.63	$\pm$	0.71	&	$-$6.19	$\pm$	0.71	&	$-$6.18	$\pm$	0.71	&	$-$4.17	$\pm$	0.05	&	$-$2.02	$\pm$	0.71	&	$-$2.01	$\pm$	0.71	\\
O	$^\dagger$&	\oi	&	14.85	$\pm$	0.05	&	$-$4.97	$\pm$	0.09	&	$-$4.97	$\pm$	0.09	&	$-$3.31	$\pm$	0.05	&	$-$1.66	$\pm$	0.10	&	$-$1.66	$\pm$	0.10	\\
Si	$^\dagger$&	\sIii	&	14.33	$\pm$	0.09	&	$-$5.49	$\pm$	0.12	&	$-$5.76	$\pm$	0.31	&	$-$4.49	$\pm$	0.03	&	$-$1.00	$\pm$	0.12	&	$-$1.27	$\pm$	0.31	\\
Fe	&	\feii	&	14.64	$\pm$	0.04	&	$-$5.18	$\pm$	0.09	&	$-$5.28	$\pm$	0.15	&	$-$4.50	$\pm$	0.04	&	$-$0.68	$\pm$	0.10	&	$-$0.78	$\pm$	0.15	\\															

\cutinhead{M83-Pos.~1}															\\																	
H	&	\hi	&	19.60	$\pm$	0.32	&				&				&				&				&				\\
N	$^\dagger$&	\Ni	&	15.29	$\pm$	0.02	&	$-$4.31	$\pm$	0.32	&	$-$4.31	$\pm$	0.33	&	$-$4.17	$\pm$	0.05	&	$-$0.14	$\pm$	0.32	&	$-$0.14	$\pm$	0.33	\\
Si	$^\dagger$&	\sIii	&	15.05	$\pm$	0.04	&	$-$4.55	$\pm$	0.32	&	$-$4.93	$\pm$	0.52	&	$-$4.49	$\pm$	0.03	&	$-$0.06	$\pm$	0.32	&	$-$0.44	$\pm$	0.52	\\
P	&	\pii	&	14.08	$\pm$	0.02	&	$-$5.52	$\pm$	0.32	&	$-$5.75	$\pm$	0.41	&	$-$6.59	$\pm$	0.03	&	+1.07	$\pm$	0.32	&	+0.84	$\pm$	0.41	\\
S	&	\sii	&	15.60	$\pm$	0.09	&	$-$4.00	$\pm$	0.33	&	$-$4.20	$\pm$	0.41	&	$-$4.88	$\pm$	0.03	&	+0.88	$\pm$	0.33	&	+0.68	$\pm$	0.41	\\
Fe	$^\dagger$&	\feii	&	14.83	$\pm$	0.03	&	$-$4.77	$\pm$	0.32	&	$-$4.92	$\pm$	0.37	&	$-$4.50	$\pm$	0.04	&	$-$0.27	$\pm$	0.32	&	$-$0.42	$\pm$	0.37	\\
Ni	&	\nIii	&	13.89	$\pm$	0.06	&	$-$5.71	$\pm$	0.33	&	$-$5.80	$\pm$	0.35	&	$-$5.78	$\pm$	0.04	&	+0.07	$\pm$	0.33	&	$-$0.02	$\pm$	0.35	\\																

\cutinhead{M83-Pos.~2}															\\																	
H	&	\hi	&	18.44	$\pm$	0.30	&				&				&				&				&				\\
C	$^\dagger$&	\cii	&	15.09	$\pm$	0.01	&	$-$3.35	$\pm$	0.30	&	NaN		&	$-$3.57	$\pm$	0.05	& 	+0.22	$\pm$	0.30	&	NaN			\\
N	&	\Ni	&	14.90	$\pm$	0.03	&	$-$3.54	$\pm$	0.30	&	NaN			&	$-$4.17	$\pm$	0.05	&	+0.63	$\pm$	0.30	&	NaN		\\
Si	$^\dagger$&	\sIii	&	14.69	$\pm$	0.01	&	$-$3.75	$\pm$	0.30	&	NaN			&	$-$4.49	$\pm$	0.03	&	+0.74	$\pm$	0.30	&	NaN		\\
P	&	\pii	&	13.73	$\pm$	0.04	&	$-$4.71	$\pm$	0.30	&	NaN			&	$-$6.59	$\pm$	0.03	&	+1.88	$\pm$	0.30	&	NaN		\\
S	&	\sii	&	15.39	$\pm$	0.02	&	$-$3.05	$\pm$	0.30	&	NaN			&	$-$4.88	$\pm$	0.03	&	+1.83	$\pm$	0.30	&	NaN			\\
Fe	$^\dagger$&	\feii	&	14.71	$\pm$	0.02	&	$-$3.73	$\pm$	0.30	&	NaN			&	$-$4.50	$\pm$	0.04	&	+0.77	$\pm$	0.30	&	NaN		\\


\end{deluxetable*}

\begin{description}
\vspace{0.0cm}
\begin{tiny}
\item[$^a$]Ionization-corrected ratios in logarithmic scale, i.e. calculated after applying the ionization correction factors listed in Table~\ref{tab:ICFs} with Equation~\ref{eq:icf0}.
\item[$^b$]Solar photospheric abundances from \citet{Asplund:2009}.
\item[$^c$][X/H]=log(X/H)$-$log(X/H)$_{\odot}$.
\item[$^d$][X/H]$_{ICF}$=log(X/H)$_{ICF}$$-$log(X/H)$_{\odot}$.
\end{tiny}
\end{description}

\begin{table*}
\begin{center}
\begin{scriptsize}
\caption{List of lines measured in the individual COS spectra of our sample of nearby SFGs and average column densities} \label{tab:avg_spec}
\begin{tabular}{lcc|lcc}
\tableline \tableline
Line ID & $\lambda$ (\AA) & log($N$/cm$^{-2}$) & Line ID & $ \lambda$(\AA) & log($N$/cm$^{-2}$)\\
\tableline
\Lya            &     1215.67 &        21.09 $\pm$ 0.01 &                   &                    &                                        \\
\cii~1334   &     1334.53 & $>$14.77                     &  \sii~1250  &      1250.58 &      15.28 $\pm$       0.01 \\ 
\Ni~1134.1 &    1134.17 &        15.15 $\pm$ 0.01 &  \sii~1253  &      1253.81 &      15.28 $\pm$       0.01 \\ 
\Ni~1134.4 &    1134.41 &        15.15 $\pm$ 0.01 &  \sii~1259  &      1259.52 &      15.28 $\pm$       0.01 \\ 
\Ni~1134.9 &    1134.98 &        15.15 $\pm$ 0.01 &  \feii~1121 &      1121.97 &      15.05 $\pm$       0.02 \\
\Ni~1199    &    1199.55 &        15.15 $\pm$ 0.01 &  \feii~1125 &      1125.45 &      15.05 $\pm$       0.02 \\
\Ni~1200.2 &    1200.22 &        15.15 $\pm$ 0.01 &  \feii~1127 &      1127.10 &      15.05 $\pm$       0.02 \\
\Ni~1200.7 &    1200.71 &        15.15 $\pm$ 0.01 &  \feii~1133 &      1133.66 &      15.05 $\pm$       0.02 \\
\oi~1302    &    1302.17 &  $>$15.05                     &  \feii~1142 &      1142.37 &      15.05 $\pm$       0.02 \\
\oi~1355    &    1355.60 &  $>$15.05                     &  \feii~1143 &      1143.23 &      15.05 $\pm$       0.02 \\
\sIii~1190  &     1190.42 &  $>$14.68                    &  \feii~1144  &      1144.94 &      15.05 $\pm$       0.02 \\
\sIii~1193  &     1193.29 &  $>$14.68                    &  \feii~1260  &      1260.53 &      15.05 $\pm$       0.02 \\
\sIii~1260  &     1260.42 &  $>$14.68                    &  \nIii~1317  &      1317.22 &      13.68 $\pm$       0.02 \\
\sIii~1304  &     1304.37 &  $>$14.68                    &  \nIii~1345  &      1345.88 &      13.68 $\pm$       0.02 \\
\pii~1152  &      1152.82 &        13.90 $\pm$ 0.01 & \nIii~1370  &      1370.13 &      13.68 $\pm$       0.02 \\
\pii~1301  &      1301.87 &        13.90 $\pm$ 0.01 & \nIii~1393  &      1393.32 &      13.68 $\pm$       0.02 \\
\tableline
\end{tabular}
\end{scriptsize}
\end{center}
\end{table*}

\begin{table*}
\begin{center}
\begin{scriptsize}
\caption{Average abundances of the elements detected in the COS spectra of our sample of nearby SFGs} \label{tab:avg_abund}
\begin{tabular}{l|ccc}
\tableline \tableline
Element & log(X/H)                       &     log(X/H)$_{\odot}^a$ &           [X/H]$^b$             \\
\tableline
C  &   $>$ $-$6.32                      &      $-$3.57 $\pm$ 0.05     &     $>$ $-$ 2.75              \\
N  &           $-$5.94 $\pm$ 0.01  &      $-$4.17 $\pm$ 0.05     &     $-$1.77 $\pm$ 0.05   \\
O  &   $>$ $-$6.04                      &      $-$3.31 $\pm$ 0.05     &     $>$ $-$ 2.73              \\
Si  &   $>$ $-$6.41                      &      $-$4.49 $\pm$ 0.03     &     $>$ $-$ 1.92              \\
P   &           $-$7.19 $\pm$ 0.01  &      $-$6.59 $\pm$ 0.03     &     $-$0.60 $\pm$ 0.03   \\
S   &           $-$5.81 $\pm$ 0.01  &      $-$4.88 $\pm$ 0.03     &     $-$0.93 $\pm$ 0.03   \\
Fe &           $-$6.04 $\pm$ 0.02  &      $-$4.50 $\pm$ 0.04     &     $-$1.54 $\pm$ 0.04   \\
Ni &           $-$7.41 $\pm$ 0.02   &     $-$5.78 $\pm$ 0.04     &     $-$1.63 $\pm$ 0.04   \\
\tableline
\end{tabular}
\begin{description}
\vspace{0.0cm}
\begin{tiny}
\item[$^a$]Solar photospheric abundances from \citet{Asplund:2009}.
\item[$^b$][X/H]=log(X/H)$-$log(X/H)$_{\odot}$.
\end{tiny}
\end{description}
\end{scriptsize}
\end{center}
\end{table*}


\clearpage

\clearpage

\clearpage

\bibliographystyle{mn2e}
\bibliography{references.bib}


\end{document}